\documentclass[11pt,citesort]{article}

\usepackage{epsfig}

\usepackage{cite}
\usepackage{dsfont}
\usepackage{slashed}
\usepackage{amsfonts,color}
\usepackage{hyperref}
\usepackage[centertags]{amsmath}       
\usepackage{amssymb,amscd}
\usepackage{cancel}

\newcommand{\be}{\begin{eqnarray}}
\newcommand{\ee}{\end{eqnarray}}
\newcommand{\beq}{\begin{equation}}
\newcommand{\eeq}{\end{equation}}

\newcommand{\nn}{\nonumber}

\newcommand{\pio}{\mbox{$\pi^0$}}
\newcommand{\gzero}{g_0^\theta}
\newcommand{\gone}{g_1^\theta}

\newcommand{\btheta}{\bar\theta}

\newcommand{\mpi}{M_{\pi}}
\newcommand{\fpi}{F_{\pi}}
\newcommand{\Mstr}{\delta m_{np}^{\rm str}}
\newcommand{\Mpistr}{(\delta\mpi^2)^{\rm str}}

\newcommand{\NthreeLO}{{\sf N$^3$LO}}
\newcommand{\NtwoLO}{{\sf N$^2$LO}}

\newcommand{\NLO}{{\sf NLO}}

\newcommand{\he}{{}^3{\rm He}}
\newcommand{\hy}{{}^3{\rm H}}

\begin{document}
\begin{titlepage}
\renewcommand{\thefootnote}{\fnsymbol{footnote}}

\vspace{2.0cm}

\begin{center}
{\Large\bf P- and T-Violating Lagrangians in Chiral Effective Field Theory\\
[0.3cm]
 and Nuclear Electric Dipole Moments}
\vspace{1.2cm}

{\large \bf J. Bsaisou$^a$, Ulf-G. Mei{\ss}ner$^{a,b,c,d}$, A. Nogga$^{a,b,c}$,
and A. Wirzba$^{a,b,}$\footnotemark[1]} 

\vspace{0.5cm}

{\large 
$^a$ 
{\it Institute for Advanced Simulation, Institut f\"ur Kernphysik, 
and J\"ulich Center for Hadron Physics, Forschungszentrum J\"ulich, 
D-52425 J\"ulich, Germany}}

\vspace{0.25cm}
{\large
$^b$ \it{JARA -- Forces and Matter Experiments, Forschungszentrum J\"ulich, D-52425 J\"ulich, Germany}}

\vspace{0.25cm}
{\large
$^c$ \it{JARA -- High Performance Computing, Forschungszentrum J\"ulich, D-52425 J\"ulich, Germany}}

\vspace{0.25cm}
{\large 
$^d$ 
{\it Helmholtz-Institut f\"ur Strahlen- und Kernphysik and Bethe Center for 
Theoretical Physics, Universit\"at Bonn, D-53115 Bonn, Germany}}
\end{center}

\vspace{1.5cm}

\begin{abstract}
A scheme to derive hadronic interactions induced by effective multi-quark terms is presented within the framework of chiral effective field theory. It is employed to work out the list of parity- and time-reversal-symmetry-violating hadronic interactions that are relevant for the computation of nuclear contributions to the electric dipole moments of the hydrogen-2, helium-3 and hydrogen-3 nuclei. We also derive the scattering and Faddeev equations required to compute electromagnetic form factors in general and electric dipole moments in particular. 
\end{abstract}

\vfill
\footnotetext[1]{Corresponding author.\\ {\it E-mail addresses:} 
                           j.bsaisou@fz-juelich.de (J. Bsaisou),
                           meissner@hiskp.uni-bonn.de (Ulf-G. Mei{\ss}ner),\\ 
                           a.nogga@fz-juelich.de (A. Nogga),
                           a.wirzba@fz-juelich.de (A. Wirzba).}
\end{titlepage}


\section{Introduction}
Electric Dipole Moments (EDMs) of non-degenerate quantum systems with non-vanishing total spin and at least
one non-vanishing charge 
imply the simultaneous violation of parity ($P$) and time-reversal ($T$) symmetry and hence by the $CPT$ theorem also the violation of $CP$ ($C$: charge-conjugation symmetry). 
For a comprehensive review, see, {\it e.g.}, Ref.\,\cite{Bigi:2000yz}.
The non-vanishing charge can be of a more general nature such as the baryon, flavor or lepton number.
Possible candidates for these non-degenerate quantum systems are: all non-selfconjugate subatomic particles with
non-vanishing spin  --- whether elementary and stable such as the electron or unstable
such as the muon or even composite as the nucleon or nuclei --- but also specially chosen dia- or paramagnetic atoms or even molecules
with a non-degenerate ground state.
According to \cite{sakharov}, $CP$ violation is a necessary condition for an asymmetry between matter and anti-matter in the universe to be dynamically generated. However, $CP$ violation by the complex phase of the Cabibbo-Kobayashi-Maskawa (CKM) matrix of the Standard Model (SM) is insufficient to account for the observed matter-antimatter asymmetry \cite{Bennett:2012zja,Ade:2013zuv}. This indicates the existence of other mechanism(s) of $CP$ violation within and/or beyond the SM.

EDMs of non-degenerate  (non-selfconjugate and non-zero-spin) quantum systems
induced by the complex phase of the CKM matrix are significantly suppressed and experimentally inaccessible in the foreseeable future \cite{Pospelov_Ritz}. Other sources of $P$ and $T$ violation within and beyond the SM might give rise to considerably larger EDMs. Therefore, any measurement of a non-vanishing EDM of any of the above listed quantum
sytems in the foreseeable future would be evidence of $P$ and $T$ violation beyond the CKM matrix of the SM. 

This reasoning has led to the emergence of new experimental concepts to measure directly and/or indirectly the EDMs of the electron, single nucleons and light nuclei. The current upper bounds on the neutron and proton EDMs are $|d_{n}|\!<2.9\cdot 10^{-26}e\,{\rm cm}$\cite{Baker:2006ts} and 
$|d_{p}|\!<7.9\cdot 10^{-25}e\,\rm{cm}$\cite{Griffith:2009zz}, respectively, where the upper bound on the proton EDM has been inferred from a measurement of the diamagnetic ${}^{199}$Hg atom. The upper bound on the electron EDM, obtained indirectly from a measurement of the EDM of the ThO molecule, is currently at $|d_{e}|<8.7\cdot 10^{-29}e\,\rm{cm}$\,\cite{Baron:2013eja}. The leading contributions from the SM CKM matrix to the electron EDM involve four electroweak loops \cite{Pospelov_Ritz}, while the leading diagrams for $d_n$ and $d_p$ have only three loops \cite{dim4CKM1,Pospelov_Ritz}. The SM mechanism of \cite{Mannel:2012qk,Mannel:2012hb} leads to comparable predictions for $d_n$ and $d_p$. The current upper bound on the neutron EDM is therefore closer to its SM prediction than the current upper bound on the electron EDM.

The proposals to measure the EDMs of charged light nuclei directly at dedicated storage rings with an envisioned accuracy of $\sim 10^{-29}\,e\,{\rm cm}$ \cite{Semertzidis:2003iq,Orlov:2006su,Semertzidis:2011qv,Semertzidis:2011zz,Lehrach,Pretz:2013us,Rathmann:2013rqa,BNL_deuteron} are the primary motivation for this work. A single EDM measurement would not be sufficient to identify the source of $P$ and $T$ violation and EDM measurements of several light nuclei are in general required to impose constraints on the sources of $P$ and $T$ violation within and beyond the SM \cite{jbepja,deVries:2011an,Engel:2013lsa,Wirzba:2014mka,Yamanaka:2014nba,edmoln}. This paper provides a detailed explanation of the formal but necessary aspects of the investigation of the EDMs of light nuclei, {\it i.e.} the hydrogen-2 nucleus (deuteron), the helium-3 nucleus (helion) and the hydrogen-3 nucleus (triton). These formal aspects are the derivation of $P$- and $T$-violating hadronic interactions and the derivation of scattering (2N systems) and Faddeev (3N systems) equations to numerically compute nuclear form factors,
see also the thesis \cite{dissertation}. The former point is also applicable to calculations of $P$- and $T$-violating moments of heavier systems. 

Within the SM, the $\theta$-term of quantum chromodynamics (QCD) \cite{'tHooft:1976up} is the only source of $P$ and $T$ violation beyond the complex phase of the CKM matrix (and the phase(s) of its leptonic counterpart, the 
Pontecorvo-Maki-Nakagawa-Sakata matrix). 
It is given by the full contraction of the gluon field-strength tensor $G_{\mu\nu}^{a}$, $a=1,\cdots,8$, with its dual (we use the Einstein summation convention throughout this paper) and is parametrized by an angle $\theta$. The $\theta$-term can be removed by an axial $U(1)$ transformation via the axial $U(1)$ anomaly  at the price of picking up an extra complex phase of the quark-mass matrix 
in the QCD (quantum) action~\cite{Baluni,Crewther:1979pi,Mereghetti:2010tp}: 
\begin{eqnarray}
 \lefteqn{S_{{\rm QCD}}+S_{{\rm QCD}}^{\theta}
 = S_{{\rm QCD}}
    -\theta\frac{g_s^2}{64\pi^2}
     \int \epsilon^{\mu\nu\rho\sigma}G_{\rho\sigma}^{a}G_{\mu\nu}^{a}\ {\rm d}^4x}
    && \nonumber \\
 && \xrightarrow{\phantom{-}U_A(1)}
   S_{{\rm QCD}}+ \frac{\bar\theta\bar{m}}{2}  
      \int \left ( \bar{q}i\gamma_5q
      +\epsilon\bar{q}i\gamma_5\tau_3q\right) {\rm d}^4x \nonumber\\
&&\xrightarrow{SU_A(2)}
S_{{\rm QCD}}+\bar{\theta}m^{\ast}\int \bar{q}i\gamma_5q
 \ {\rm d}^4 x\ .\label{eq:lagtheta}
\end{eqnarray}
Here 
$g_s$ is the strong coupling constant. Since the ultimate interest of this work is in nucleons, light nuclei and light mesons, the strange quark is not retained as an explicit degree of freedom and we focus on QCD for the two lightest flavors. The doublet
$q=(u,d)$ is thus the isospin doublet of the two lightest quark flavors, up and down, $\tau_{i}$ denote the three isospin Pauli matrices and $\epsilon^{\mu\nu\rho\sigma}$ is the completely antisymmetric tensor with $\epsilon^{0123}=1$. \mbox{$\bar{\theta}=\theta-{\rm arg}\det(\mathcal{M})$} is the physical \mbox{$\theta$-parameter} \cite{Crewther:1979pi,Mereghetti:2010tp} and \mbox{$\mathcal{M}=\bar{m}\mathds{1}+\epsilon\bar{m}\tau_3$}, modulo an overall complex phase, is the quark mass matrix with 
\mbox{$\bar{m}=(m_u+m_d)/2$} and $\epsilon=(m_u-m_d)/(m_u+m_d)$ for the up- and down-quark masses $m_u$ and $m_d$, respectively. The second axial $SU(2)$ rotation in Eq.\,(\ref{eq:lagtheta}) moves the $\theta$-term completely into a $P$- and $T$-violating and isospin-conserving quark-mass term with reduced mass \mbox{$m^{\ast}\!=\!m_um_d/(m_u+m_d)$}.

Beyond the SM (BSM) theories such as supersymmetry, multi-Higgs scenarios, left-right symmetric models, etc.\  (see {\it e.g.} \cite{RamseyMusolf:2006vr,Weinberg:1989dx,Pati:1974yy,Mohapatra:1974hk,Mohapatra:1974gc,Senjanovic:1975rk,Minkowski:1977sc,Senjanovic:1978ev,Mohapatra:1979ia,Mohapatra:1980yp}) describe physics at high energies and involve degrees of freedom of masses significantly larger than the Higgs mass. Their induced interactions at the energy scale of the SM, $\Lambda_{{\rm SM}}\sim 250$ GeV, can be obtained by employing an effective field theory approach: all heavy BSM degrees of freedom are systematically integrated out to obtain a set of non-renormalizable effective operators composed of SM degrees of freedom \cite{Buchmuller:1985jz,dim6Weinberg,DeRujula:1990db,Pospelov_Ritz,Grzadkowski:2010es,RamseyMusolf:2006vr,dim6ngtulin,deVries:2012ab,Dekens:2013zca}.

To evolve these operators further down to the energy scale $\Lambda_{{\rm had}}\gtrsim 1$ GeV, also the heavy SM degrees of freedom such as the top quark, the Higgs, $Z$ and $W$ bosons and other heavy quarks have to be systematically integrated out. Due to the focus of this paper on interactions of nucleons and light mesons, the strange quark is also not retained as an explicit degree of freedom. The effective operators (relevant for this work) at the energy scale $\Lambda_{{\rm had}}$ are known as the four-quark left-right operator (FQLR-term), the quark-chromo EDM (qCEDM) and quark EDM (qEDM), both with isospin-conserving as well as isospin-violating components, the gluon-chromo EDM (gCEDM) and the four-quark term (4q-term)\cite{Grzadkowski:2010es,deVries:2012ab,Dekens:2013zca}:
\begin{alignat}{3}
&\mbox{{\rm FQLR-term}} &&:\quad&&i\nu_1V_{ud}\,(\bar{u}_R\gamma_{\mu}d_R\bar{d}_L\gamma^{\mu}u_L-\bar{d}_R\gamma_{\mu}u_R\bar{u}_L\gamma^{\mu}d_L)\,,\label{eq:4qlr}\\
&&&&&i\nu_8V_{ud}\,(\bar{u}_R\gamma_{\mu}\lambda^{a}d_R\bar{d}_L\gamma^{\mu}\lambda^{a}u_L-\bar{d}_R\gamma_{\mu}\lambda^{a}u_R\bar{u}_L\gamma^{\mu}\lambda^{a}d_L)\,,\label{eq:4qlr8}\\
&{\rm qCEDM} &&:&&i\bar{q}\,(\,\tilde{\delta}_{G}^1\mathds{1}+\tilde{\delta}_{G}^3\tau_3)\,\sigma^{\mu\nu}\gamma_5
\lambda^a q\,G^{a}_{\mu\nu}\,,\label{eq:qcedm}\\
&{\rm qEDM} &&: &&i\bar{q}\,(\,\tilde{\delta}_{F}^1\mathds{1}+\tilde{\delta}_{F}^3\tau_3)\,\sigma^{\mu\nu}\gamma_5q\,F_{\mu\nu}\,,\label{eq:qedm}\\
&{\rm gCEDM}&&:&&\beta_G\,f^{abc}\epsilon^{\mu\nu\alpha\beta}\, G^{a}_{\alpha\beta}G^{b}_{\mu\rho}G^{c,\rho}_{\nu}\,,\label{eq:gcedm}\\
&\mbox{{\rm 4q-term}}&&:&&i\mu_1\,(\bar{u}u\,\bar{d}\gamma_5 d+\bar{u}\gamma_5u\,\bar{d}d-\bar{d}\gamma_5u\,\bar{u}d-\bar{d}u\,\bar{u}\gamma_5d)\,,\label{eq:4q}\\
&&&&&i\mu_8\,(\bar{u}\lambda^au\,\bar{d}\gamma_5 \lambda^a d+\bar{u}\gamma_5\lambda^a u \,\bar{d}\lambda^ad-\bar{d}\gamma_5\lambda^a u\,\bar{u}\lambda^ad-\bar{d}\lambda^a u \,\bar{u}\gamma_5\lambda^ad)\,.\label{eq:4q8}
\end{alignat}
Here $V_{ud}\simeq 1$ is an element of the CKM matrix, $F_{\mu\nu}$ the electromagnetic field-strength tensor and $f^{abc}$ are the structure constants of the Lie algebra of the color $SU(3)$ group. All these effective operators are of mass-dimension six and constitute the most general set of effective $P$- and $T$-violating operators of this dimension at the energy scale $\Lambda_{{\rm had}}$ if only up and down quarks are considered and if Lorentz invariance
and especially the  $CPT$-theorem are assumed. The coefficients $\delta^{1,3}_{F,G}$, $\tilde{\delta}_{F,G}^{1,3}$, $\nu_{1,8}$, $\mu_{1,8}$ and $\beta_G$ in Eqs.\,(\ref{eq:4qlr})-(\ref{eq:4q8}) depend on the specific BSM theory considered (see {\it e.g.} \cite{newmodelspaper}). These coefficients are independent of each other apart from $\nu_1$ and $\nu_8$. The coefficient $\nu_8$ is generated in renormalization-group running \cite{Dekens:2013zca} and thus completely determined by $\nu_1$.\footnote{According to Eq.\,(3.35) in Ref.\,\cite{Dekens:2013zca}, the relation between $\nu_1$ and $\nu_8$ is given by $\nu_8\approx 1.3\,\nu_1$.} The qEDM, Eq.\,(\ref{eq:qedm}), and the qCEDM, Eq.\,(\ref{eq:qcedm}), are of dimension six since the coefficients $\delta_{F,G}^{1,3}$ are proportional to the Higgs vacuum expectation value (VEV) \cite{deVries:2012ab,Dekens:2013zca}. The gCEDM and the two color structures of the 4q-term are chiral singlets and thus cannot be distinguished
by their chiral transformation properties. The same is true for the two different color structures of the FQLR-term. We will only consider one representative for each source of $P$ and $T$ violation explicitly in our analysis. Other $P$- and $T$-violating multi-quark terms are of dimension larger than six.

The aim of this paper is to present the underlying aspects of our EDM computation for light nuclei whose results have been presented in \cite{edmoln}. These comprise (a) the derivation of the $P$- and $T$-violating terms in the low-energy effective Lagrangian induced by the QCD $\theta$-term and the effective dimension-six sources and (b) the derivation of scattering and Faddeev equations that are the basis of our numerical computation of nuclear EDMs. The significance of our formalism to derive the $P$- and $T$-violating Lagrangians and of our numerical technique is not limited to the computation of EDMs. The formalism presented to derive the Lagrangians induced by the effective dimension-six sources admits a straightforward extension to other effective BSM sources (of even higher mass dimension) which do not have to be $P$- or $T$-violating. For instance, each dimension-six source of $P$ and $T$ violation has $P$- and $T$-conserving isospin-multiplet partners. Also the terms in the effective low-energy Lagrangians induced by the latter are derived in this paper. In the course of our analysis, we will demonstrate among other findings the following central statement on the basis of group-theoretical considerations: any {\it chiral structure} ({\it i.e.} an effective term modulo source fields and low-energy constants), that may be induced by an arbitrary quark multilinear is already encountered in the standard pion-sector and pion-nucleon $\chi$EFT Lagrangians, which up to fourth order are presented in \cite{GassLeut1,GassLeut2,GasserSvarc,MeissnerFettes3}. An effective quark multilinear only shifts the order at which a particular chiral structure appears and introduces in general new, independent low-energy constants (LECs). This is a particularly convenient result since the set of independent chiral terms that are induced by an arbitrary quark multilinear can in principle be obtained from already existing $\chi$EFT Lagrangians by a duplication of standard terms and an appropriate replacement of LECs and chiral source fields (and a replacement of $P$- and $T$-conserving chiral-singlet structures by their $P$- and $T$-violating counterparts, {\it e.g.} the electromagnetic field-strength tensor by its dual, in some cases). Since effective quark multilinears shift the orders at which particular chiral structures occur, the $\chi$EFT Lagrangians of orders larger than four gain significance. 

The only assumption that enters our derivation of the scattering and Faddeev equations is that {\it additional} nuclear interactions constitute only small perturbations of the standard ($P$- and $T$-conserving) $\chi$EFT two-nucleon and three-nucleon potentials. These equations can thus equally well be used as a basis to compute all kinds of electric and magnetic moments of light nuclei (induced by BSM physics), regardless of whether they are $P$- and/or $T$-violating or not.
 
The QCD $\theta$-term, expressed as a complex phase of the quark mass matrix, and the set of effective dimension-six sources serve as the starting point of the analysis presented in this work. In order to describe the induced hadronic interactions at energies below $\Lambda_{{\rm QCD}}\!\sim\! 200$MeV, the framework of Chiral Perturbation Theory (ChPT) for the two lightest quark flavors and its extension to nucleons, called Chiral Effective Field Theory ($\chi$EFT), is employed. Since the $\theta$-term is a genuine feature of standard QCD, the induced hadronic $P$- and $T$-violating operators can be accounted for within standard $\chi$EFT \cite{GassLeut1,GassLeut2,GasserSvarc,MeissnerFettes3,Mereghetti:2010tp}. In particular, the leading LECs in the standard effective Lagrangians are quantitatively known, which means that the coupling constants of the leading induced $P$- and $T$-violating vertices are explicitly computable as functions of the parameter $\bar{\theta}$ with well-defined {\em hadronic} uncertainties.

The derivation of the effective Lagrangians induced by the effective dimension-six sources is more intricate. The effective dimension-six sources constitute an amendment of the standard QCD Lagrangian. They do not {\em transform} into any standard dimension-four
QCD terms by chiral $SU(2)_L\!\times\! SU(2)_R$ rotations \cite{deVries:2010ah,deVries:2012ab}. This observation translates into a corresponding amendment of the $\chi$EFT Lagrangian by terms with new source fields (including the above mentioned chiral-singlet ones) {\em and} independent, quantitatively unknown LECs. 
Some of the new source fields may even exhibit apparently novel transformation properties under chiral $SU(2)_L\!\times\! SU(2)_R$ rotations. 
However, {\em compositions} of conventional source fields can always  be arranged in the standard $\chi$EFT Lagrangians which --- with respect to chiral rotations --- 
transform in an identical manner as even these source structures.
In this sense all relevant chiral structures can of course
be encountered already in the standard $\chi$EFT Lagrangian at a sufficiently high chiral order. Nevertheless,
the $P$- and $T$-conserving standard QCD source terms can not be rotated  in such a way that the BSM terms 
{\em including} their dedicated sources and LECs are generated.
We call the low-energy effective field theory of QCD generalized to BSM quark multilinears {\it amended} $\chi$EFT. A reliable technique to numerically assess the new LECs in the amended $\chi$EFT Lagrangian is Lattice QCD. Although such Lattice QCD computations are subject of ongoing efforts (see {\it e.g.} \cite{Bhattacharya:2014cla,Shindler:2014oha}), at least for the time being one still has to rely on naive dimensional analysis (NDA) \cite{Manohar:1983md,Weinberg:1989dx} or model calculations (see {\it e.g.} \cite{Pospelov_Ritz,Pitschmann:2014jxa}) to obtain order-of-magnitude estimates. In the case of the FQLR-term, however, the dominant contributions to the leading coupling constants of $P$- and $T$-violating vertices prove to be functions of the coupling constant of the three-pion vertex. Contributions with further unknown LECs are found to be of subleading orders in the standard chiral counting, which allows for an explicit computation of the leading coupling constants induced by the FQLR-term as functions of the three-pion coupling constant (defined in Eq.\,(\ref{eq:impcoup})) only.

The derivation of the chiral effective Lagrangians induced by the $\theta$-term and the effective dimension-six sources has been the subject of earlier publications \cite{Mereghetti:2010tp,deVries:2010ah,deVries:2012ab} within the Weinberg formulation of $SU(2)$ $\chi$EFT (see {\it e.g.} chapter 19 of \cite{WeinbergII}). The main difficulty of this formulation is in the derivation of higher-order terms and in the generalization to three-flavor $SU(3)$ $\chi$EFT. Within the Gasser-Leutwyler formulation of $\chi$EFT \cite{GassLeut1,GassLeut2,GasserSvarc}, the derivation of higher-order terms is significantly less tedious and the extension of the analysis to three quark flavors is rather straightforward once particular quark multilinears with strange-quark content have been specified. The derivation of the induced (amended) $\chi$EFT Lagrangians in the pion and pion-nucleon sector is presented within the Gasser-Leutwyler formulation of $\chi$EFT for the first time in this paper. Our independent derivation leads to final results for the leading coupling constants of $P$- and $T$-violating vertices which are consistent with those of \cite{Mereghetti:2010tp,deVries:2010ah,deVries:2012ab}. 
The $P$- and $T$-violating vertices relevant for our analysis of nuclear EDMs with their respective coefficients are defined 
by\footnote{The last line containing isospin-violating four-nucleon operators is only relevant for the case of the FQLR-term.} (see {\it e.g.} \cite{jbepja,deVries:2011an,edmoln})
\begin{eqnarray}\label{eq:impcoup}
\lefteqn{\mathcal{L}_{\pi}^{(2)}+\mathcal{L}_{\pi}^{(4)}+\mathcal{L}_{\pi N}^{(2)}+\mathcal{L}_{\pi N}^{(4)}+\mathcal{L}_{4N}^{(2)}}\nonumber\\
&=&\quad m_N\Delta_3\pi_3\pi^2+g_0N^{\dagger}\vec{\pi}\cdot\vec{\tau}N+g_1N^{\dagger}\pi_3N\nonumber\\
&&-\,2d_0N^{\dagger}S^{\mu}v^{\nu}NF_{\mu\nu}-2d_1N^{\dagger}\tau_3S^{\mu}v^{\nu}NF_{\mu\nu}\nonumber\\
&&+\,C_1N^{\dagger}N\mathcal{D}_{\mu}(N^{\dagger}S^{\mu}N)+C_2N^{\dagger}\vec{\tau}N\cdot\mathcal{D}_{\mu}(N^{\dagger}S^{\mu}\vec{\tau}N)\nonumber\\
&&+\,C_3N^{\dagger}\tau_3N\mathcal{D}_{\mu}(N^{\dagger}S^{\mu}N)+C_4N^{\dagger}N\mathcal{D}_{\mu}(N^{\dagger}\tau_3S^{\mu}N)+\cdots\,,
\end{eqnarray}
where $v^{\mu}$ is the nucleon four-velocity, $S^{\mu}$ is the nucleon spin, $\mathcal{D}_{\mu}$ is the covariant derivative, \mbox{$N=(p,n)^T$} is the doublet of nucleon fields and $\pi_i$ are the pion fields. 
The notation of \cite{BKM1995} is adopted throughout this paper.
While $\Delta_3$ is the coefficient of an isospin-violating three-pion vertex, $g_0$ and $g_1$ are the coefficients of isospin-conserving and isospin-breaking $P$-and $T$-breaking pion-nucleon interactions, respectively. The coefficients $d_0$ and $d_1$ are the isoscalar and isovector single-nucleon EDMs, which are regarded as parameters in this work. The $C_i$ terms are isospin-conserving ($i=1,2$) and isospin-violating ($i=3,4$) short-range 
$P$- and $T$-violating nucleon-nucleon interactions.  As a main result, expressions for these coefficients induced by the QCD $\theta$-term, the FQLR-term and the qCEDM will be given and we explain the hierarchies of these coefficients for the cases of the qEDM, the 4q-term and the gCEDM. For the QCD $\theta$-term the coefficients $g_0$, $g_1$ and $\Delta_3$ and for the FQLR-term the coefficients $g_0$ and  $g_1$  
are computed explicitly as functions of $\bar{\theta}$ and $\Delta_3$, respectively.

This paper is structured as follows: Section~\ref{sec:chptlag} is concerned with the derivation of the chiral effective Lagrangians induced by the QCD $\theta$-term, FQLR-term, qCEDM, qEDM, gCEDM and 4q-term. Subsection~\ref{sec:standardchpt} contains a brief summary of aspects of standard $\chi$EFT that are used in our analysis, {\it i.e.} fundamental building blocks, constants etc. Our procedure to derive the $P$- and $T$-violating low-energy effective Lagrangians induced by the considered sources is presented in Subsection~\ref{sec:ptvlag}. Group theoretical considerations utilized in this derivation are explained in Appendix~\ref{app:o4rep}. Additional chiral-symmetry breaking terms in general alter the ground state of (amended) QCD and analogously the one of the effective field theory. Our ground-state-selection procedure and the impact of the resulting shifts of the (amended) effective Lagrangians on the hadronic interactions in Eq.\,(\ref{eq:impcoup}) are presented in Subsection~\ref{sec:selection}. All our derivations have been performed in the Gasser-Leutwyler formulation of $\chi$EFT. In order to draw a connection to the Weinberg formulation of $\chi$EFT employed to derive the effective Lagrangians in \cite{Mereghetti:2010tp,deVries:2010ah,deVries:2012ab}, Appendix~\ref{GassWein} provides a translation procedure between both formulations. Subsection~\ref{sec:intsummary} gives a brief summary of the essential aspects of Section~\ref{sec:chptlag}. The primary aim of Section~\ref{sec:edmcomp} is to derive scattering and Faddeev equations that can be used to compute electromagnetic form factors of light nuclei. To illustrate the utility of our approach, the results of the EDM computation for light nuclei published in \cite{edmoln} are shown, which are based on the findings of Section~\ref{sec:chptlag} and Section~\ref{sec:edmcomp} of this paper. The set of $P$- and $T$-violating operators relevant for the EDM computation is shown in Subsection~\ref{sec:ptvff}, while the actual derivation of scattering and Faddeev equations is presented in detail in Subsection~\ref{sec:numana}. The results of the EDM computations in \cite{edmoln} and the different resulting hierarchies among single-nucleon and nuclear EDM contributions for all considered sources of $P$ and $T$ violation listed in Eqs.\,(\ref{eq:4qlr})-(\ref{eq:4q8}) are discussed individually in Subsection~\ref{sec:results}. The main conclusions of this paper are briefly summarized and discussed in Section~\ref{sec:concl}.

\section{\boldmath{$P$}- and \boldmath{$T$}-violating effective low-energy Lagrangians}
\label{sec:chptlag}

The main part of this section is the formal derivation the effective hadronic interactions induced by effective quark multilinears. The reader primarily interested in the nuclear EDM computation may move on directly to the intermediate summary of this section in Subsection \ref{sec:intsummary}.


\subsection{Standard Chiral Effective Field Theory}
\label{sec:standardchpt}
$\chi$EFT is the low-energy effective field theory of QCD. The aspects of standard $\chi$EFT of relevance for the subsequent parts of this section are briefly summarized here. This comprises the definitions of all relevant objects, the standard second-order and fourth-order pion-sector Lagrangians as well as the standard first-order and second-order pion-nucleon Lagrangians (see {\it e.g.} \cite{BKM1995,Scherer:2002tk,Bernard:2006gx,Kubis:2007iy,veroniquereview} for comprehensive summaries). 

If the $SU(2)_L\!\times\!SU(2)_R$ symmetry of massless two-flavor QCD were reflected by a symmetry of the ground state or the particle spectrum, parity doubling would be observable and another triplet of bosons with even parity would exist, since axial rotations couple odd and even parity states. The absence of such degeneracy implies the spontaneous symmetry breakdown 
\mbox{$SU(2)_L\!\times\!SU(2)_R\rightarrow SU(2)_V$}, where $SU(2)_V$ is the diagonal subgroup of $SU(2)_L\!\times\! SU(2)_R$. The three resulting Goldstone bosons are identified with the odd-parity triplet of pions $\pi_{i}$. There is a one-to-one correspondence between the space of Goldstone bosons and the space of left cosets $SU(2)_L\!\times\!SU(2)_R/SU(2)_V$, which leads to the following definition of the matrix $U$ and its transformation property under $SU(2)_L\!\times\!SU(2)_R$ \cite{Leut,LeutProceedings}:
\begin{equation}\label{eq:su2u}
\bigl(\pi_1(x),\pi_2(x),\pi_3(x)\bigr)\mapsto U(x):=\exp\left(\frac{i}{F_{\pi}}\tau_j\pi_{j}(x)\right)\,,\quad U(x)\rightarrow R(x)U(x)L^{\dagger}(x)\,.
\end{equation}
The validity of the effective field theory with pions as degrees of freedom is limited to momenta well below the chiral-symmetry breaking scale $\Lambda_{\chi}\sim4\pi\fpi\sim1.2$ GeV, where $\fpi=92.2$ MeV \cite{PDG} is the pion decay constant.

The $\chi$EFT Lagrangian exhibits the same global symmetries as the underlying Lagrangian of QCD (see {\it e.g.}\cite{WeinbergChPTEarly,WeinbergChPTEarly2,WeinbergChPTEarly3,LiPagelsChPTEarly}). However, Ward identities are a consequence of the invariance of the generating functional under particular \textit{local} transformations. Green functions and Ward identities of particular (composite) operators can be studied by regarding these operators as additional currents in the Lagrangian which couple to external fields \cite{GassLeut1,GassLeut2,Leut}. To this end, the most general non-kinetic quark terms in the QCD Lagrangian can be written as \cite{GassLeut1,GassLeut2}
\begin{equation}\label{eq:currentsqcd}
\mathcal{L}_{{\rm QCD}}=\cdots+\bar{q}\gamma_{\mu}(v_j^{\mu}\tau_{j}+v^{(s),\mu}+a_j^{\mu}\tau_j\gamma_5)q-\bar{q}(s_0+s_j\tau_{j}-i\gamma_5p_0-i\gamma_5p_j\tau_{j})q\,,
\end{equation}
where the functions $v^{\mu}\!=\!\{v^{\mu}_{j},v^{(s),\mu}\}$, $a^{\mu}\!=\!\{a^{\mu}_{j}\}$, $s\!=\!\{s_{j},s_{0}\}$ and $p\!=\!\{p_{j},p_0\}$, $j=1,2,3$, are the source fields for the additional currents.
The external source fields are assigned 
transformation properties under the local $SU(2)_L\!\times\!SU(2)_R\!\times\!U(1)_V$ transformation 
to render the QCD Lagrangian invariant ($a_j^{\mu}=r_j^{\mu}-l_j^{\mu},\, v_j^{\mu}=r_j^{\mu}+l_j^{\mu},\, v_{\mu}^{(s)}=r_{\mu}^{(s)}+l_{\mu}^{(s)}$):
\begin{alignat}{5}
r_i^{\mu}\tau_i&\mapsto\,\,&& R\,r_i^{\mu}\tau_iR^{\dagger}+iR\,\partial^{\mu}R^{\dagger}\,,
&&\phantom{----}l_i^{\mu}\tau_i&&\mapsto\,\,&& L\,l_i^{\mu}\tau_iL^{\dagger}+iL\,\partial^{\mu}L^{\dagger}\,,\label{eq:qcdsfli}\\
r_{\mu}^{(s)}&\mapsto&& R\,r_{\mu}^{(s)}R^{\dagger}+iR\,\partial_{\mu}R^{\dagger}\,,
&&\phantom{----}l_{\mu}^{(s)}&&\mapsto&& L\,l_{\mu}^{(s)}L^{\dagger}+iL\,\partial_{\mu}L^{\dagger}\,,\label{eq:qcdsfl}\\
(s_i\tau_i\!+\!ip_i\tau_i)&\mapsto&& R\,(s_i\tau_i+ip_i\tau_i)L^{\dagger}\,,\quad\,\,
&&(s_i\tau_i\!-\!ip_i\tau_i)&&\mapsto&& R\,(s_i\tau_i-ip_i\tau_i)L^{\dagger}\,,\label{eq:qcdsfpi}\\
(s_0+ip_0)&\mapsto&& R\,(s_0+ip_0)\,L^{\dagger}\,,
&&\phantom{-}(s_0-ip_0)&&\mapsto&& R\,(s_0-ip_0)\,L^{\dagger}\,.\label{eq:qcdsfp}
\end{alignat}
The $P$- and $T$-conserving quark-mass matrix $\mathcal{M}$ and the quark-charge matrix $\mathcal{Q}$ for two flavors given by
\begin{equation}\label{eq:chargemassmatrix}
\mathcal{M}=\bar{m}\,\mathds{1}+\epsilon\bar{m}\,\tau_3\,,\quad\mathcal{Q}=\frac{1}{6}\mathds{1}+\frac{1}{2}\tau_{3}
\end{equation}
are recovered by an appropriate replacement of the source fields $s_{j,0}$, $p_{0,j}$, $v^{\mu}_j$, $v^{(s),\mu}$ and $a^{\mu}_j$. In particular, $s_0$ and $s_3$ have to be replaced by
\begin{equation}\label{eq:s0s3rep}
s_0=\bar{m}\,,\quad s_3=\bar{m}\epsilon\,.
\end{equation}

The effective Lagrangian has to be constructed to yield exactly the same Ward identities as the underlying theory when inserted into the effective generating functional  \cite{GassLeut1,GassLeut2}:
\begin{eqnarray}
W_{\rm QCD}[s,p,a_{\mu},v_{\mu}]&=&W_{\rm eff}[s,p,a_{\mu},v_{\mu}]\nonumber\\
&=&\int \mathcal{D}U(x)\exp\left(i\int d^4x\, \mathcal{L}_{\rm eff}[U,s,p,v^{\mu},a^{\mu}]\right)\,.
\end{eqnarray}
It has to be a hermitian Lorentz scalar and invariant under \textit{local} $SU(2)_L\!\times\! SU(2)_R$ transformations \cite{GassLeut1,GassLeut2},
\begin{equation}
\mathcal{L}_{\rm eff}=\mathcal{L}_{\rm eff}(U,D_{\mu}U,D^2U,\cdots,s,p,a_{\mu},v_{\mu},\cdots)\,,
\end{equation}
where $D_{\mu}$ is the covariant derivative 
\begin{equation}
D_{\mu}U(x)=\partial_{\mu}U-ir_{\mu}(x)U(x)+iU(x)l_{\mu}(x)
\end{equation}
and the ellipses denote higher-order derivative terms. The covariant derivative generates the right- and left-handed field-strength tensors
\begin{equation}
R_{\mu\nu}:=\partial_{\mu}r_{\nu}-\partial_{\nu}r_{\mu}-i[r_{\mu},r_{\nu}]\,,\quad L_{\mu\nu}:=\partial_{\mu}l_{\nu}-\partial_{\nu}l_{\mu}-i[l_{\mu},l_{\nu}]\,,
\end{equation}
which transform under $SU(2)_L\!\times\!SU(2)_R$ by
\begin{equation}
R_{\mu\nu}\mapsto R(x)\,R_{\mu\nu}\,R^{\dagger}(x)\,,\quad L_{\mu\nu}\mapsto L(x)\,L_{\mu\nu} L^{\dagger}(x)\,.
\end{equation}
The source fields of the scalar and pseudo-scalar currents in Eq.\,(\ref{eq:currentsqcd}) can be combined into one object $\chi=2B(s+ip)$  \cite{GassLeut1,GassLeut2} which transforms under $SU(2)_L\!\times\! SU(2)_R$ by
\begin{equation}\label{eq:chidef}
\chi=2B(s+ip)\mapsto R(x)2B(s+ip)L^{\dagger}(x)\,,
\end{equation}
where $B$ is a constant related to the squared mass of the charged pions at leading order by $M_{\pi^{\pm}}^2=B(m_u+m_d)+\cdots$  \cite{GellMannOakesRenner}.

The low-energy effective Lagrangian is an infinite sum over terms that admit an ordering by the number of covariant derivatives and pion masses each term contains. The pion-sector Lagrangians of second and fourth order are given by \cite{GassLeut1} (with $\langle\cdots\rangle$ denoting the isospin trace)
\begin{equation}\label{eq:pilag2}
\mathcal{L}^{(2)}_{\pi}=\frac{F_{\pi}^2}{4}\langle D_{\mu}U(D^{\mu}U)^{\dagger}\rangle+\frac{F_{\pi}^2}{4}\langle \chi U^{\dagger}+U\chi^{\dagger}\rangle\,,
\end{equation}
and \cite{GassLeut1,Scherer:2002tk}
\begin{align}\label{eq:pilag4}
\begin{split}
\mathcal{L}^{(4)}_{\pi}=&\frac{l_1}{4}\langle D_{\mu}U(D^{\mu}U)^{\dagger}\rangle^2+\frac{l_2}{4}\langle D_{\mu}U(D_{\nu}U)^{\dagger}\rangle\langle D^{\mu}U(D^{\nu}U)^{\dagger}\rangle\\
&+\frac{l_3}{16}\langle \chi U^{\dagger}+U\chi^{\dagger}\rangle^2+\frac{l_4}{4}\langle D_{\mu}U(D^{\mu}\chi)^{\dagger}+D_{\mu}\chi(D^{\mu} U)^{\dagger}\rangle\\
&+l_5\left[\langle R_{\mu\nu}UL^{\mu\nu}U^{\dagger}\rangle-\frac{1}{2}\langle L_{\mu\nu}L^{\mu\nu}+R_{\mu\nu}R^{\mu\nu}\rangle \right]\\
&+i\frac{l_6}{2}\langle R_{\mu\nu}D^{\mu}U(D^{\nu}U)^{\dagger}+L_{\mu\nu}(D^{\mu}U)^{\dagger}D^{\nu}U\rangle\\
&-\frac{l_7}{16}\langle\chi U^{\dagger}-U\chi^{\dagger}\rangle^2\\
&+\frac{h_1+h_3}{4}\langle\chi\chi^{\dagger}\rangle+\frac{h_1-h_3}{16}\bigg[\langle\chi U^{\dagger}+U\chi^{\dagger}\rangle^2\\
&+\langle\chi U^{\dagger}-U\chi^{\dagger}\rangle^2-2\langle\chi U^{\dagger}\chi U^{\dagger}+U\chi^{\dagger}U\chi^{\dagger}\rangle\bigg]\\
&-2h_2\langle L_{\mu\nu}L^{\mu\nu}+R_{\mu\nu}R^{\mu\nu}\rangle\,,
\end{split}
\end{align}
respectively. The LECs $l_i$ are not constrained by any symmetry considerations and encode also physics of 
hadron resonances which are not explicit degrees of freedom of the effective field theory. The LECs have to be determined from experimental data by relating them to measurable quantities or from Lattice QCD. The LEC $l_7$, for instance, is related to the square of the strong mass difference of charged and neutral pions \cite{GassLeut1,GassLeut2},
\begin{equation}\label{eq:dmpi1}
(\delta M_{\pi}^2)^{\rm str}=(M_{\pi^+}^2-M_{\pi^0}^2)^{\rm str}=(m_u-m_d)^2\frac{2B^2}{F_{\pi}^2}l_7(1+\mathcal{O}(\mathcal{M}))\,,
\end{equation}
which is driven by the $\eta$-meson due to $\eta\!-\!\pi$-mixing \cite{GassLeut2}. For a detailed discussion of the LECs we refer to references \cite{GassLeut1,GassLeut2}.
The so-called high-energy constants $h_i$ in Eq.\,(\ref{eq:pilag4}) do not appear in physical low-energy processes.

The additional degrees of freedom in the pion-nucleon sector in the extreme non-relativistic limit (see {\it e.g.} \cite{Georgi:1985kw,GasserSvarc,Jenkins:1990jv,Bernard:1992qa,BKM1995,MeissnerFettes3}) are combined into the heavy component of the proton-neutron $SU(2)$ doublet
\begin{equation}
N(x)=\frac{1}{2}(\mathds{1}+\slashed{v})e^{im_Nv_{\mu}x^{\mu}}\binom{p(x)}{n(x)}\,,
\end{equation}
where $v^{\mu}$ is the four-velocity with $v^2=1$. In this limit, any product of Dirac matrices can be expressed as a combination of the velocity $v^{\mu}$ and the Pauli-Lubanski spin operator \mbox{$S^{\mu}=i\gamma^5\sigma^{\mu\nu}v_{\nu}/2$}, which equal $v=(1,0,0,0)$ and $S=(0,\vec{\sigma})/2$  in the nucleon rest-frame. The heavy nucleon field $N$ and the $SU(2)$ matrix $u$ defined by $u^2=U$ transform under $SU(2)_L\!\times\!SU(2)_R$ by
\begin{equation}
N\mapsto KN\,,\quad u\mapsto RuK^{\dagger}=KuL^{\dagger}\,.
\end{equation}
$K=K(R,L,U)\in SU(2)$ is called the compensator field. The set of fundamental building blocks in the pion-nucleon sector comprises -- apart from $N$ and $u$ -- the covariant derivative $\mathcal{D}_{\mu}$, the connection $\Gamma_{\mu}$ and the structures $u_{\mu}$, $\chi_{\pm}$ and $f^{\mu\nu}_{\pm}$:
\begin{eqnarray}
\mathcal{D}_{\mu}N&=&(\partial_{\mu}+\Gamma_{\mu}-iv_{\mu}^{(s)})N\,,\\
\Gamma_{\mu}&=&\frac{1}{2}[u^{\dagger}(\partial_{\mu}-ir_{\mu})u+u(\partial_{\mu}-il_{\mu})u^{\dagger}]\,,\\
u_{\mu}&=&i[u^{\dagger}(\partial_{\mu}-ir_{\mu})u-u(\partial_{\mu}-il_{\mu})u^{\dagger}]\,,\label{eq:pinumu}\\
\chi_{\pm}&=&u^{\dagger}\chi u^{\dagger}\pm u\chi^{\dagger}u\,,\label{eq:chi}\\
f_{\pm}^{\mu\nu}&=&uR^{\mu\nu}u^{\dagger}\pm u^{\dagger}L^{\mu\nu}u\,.\label{eq:fmunupm}
\end{eqnarray}
The pion-nucleon Lagrangian in the extreme non-relativistic limit is a power series in $\mpi/m_N$, where $\mpi\!=138.01$~MeV is the average pion mass and $m_N\!=\!938.92$~MeV is the average nucleon mass \cite{PDG}. Up to the second order it is given by \cite{Bernard:1992qa,Bernard:1996gq,MeissnerFettes1,MeissnerFettes3}\footnote{The hats over the LECs in the heavy baryon ChPT Lagrangians of \cite{MeissnerFettes1,MeissnerFettes3} are omitted throughout this paper.}
\begin{eqnarray}
\mathcal{L}_{\pi N}^{(1)}&=&N^{\dagger}\left(iv\cdot\mathcal{D}+g_AS\cdot u\right)N\,,\\
\mathcal{L}_{\pi N}^{(2)}&=&\frac{1}{2m_N}N^{\dagger}\left((v\cdot\mathcal{D})^2-\mathcal{D}^2-ig_A\{S\cdot\mathcal{D},v\cdot u\}\right)\nonumber\\
&&+N^{\dagger}\bigg(c_1\langle\chi_+\rangle-\frac{c_2}{2}\langle(v\cdot u)^2\rangle+\frac{c_3}{2}\langle u\cdot u\rangle+\frac{c_4}{2}[S^{\mu},S^{\nu}][u_{\mu},u_{\nu}]\nonumber\\
&&+c_5\hat{\chi}_+-i\frac{c_6}{4m_N}[S_{\mu},S_{\nu}]f_+^{\mu\nu}-i\frac{c_7}{4m_N}[S_{\mu},S_{\nu}]\langle f_+^{\mu\nu}\rangle\bigg) N\,,\label{eq:pin2}
\end{eqnarray}
where here the hat denotes the traceless component of a general chiral structure $A$ ({\it e.g.} $\chi_{\pm}$),
\begin{equation}
\hat{A}:=A-\frac{1}{2}\langle A\rangle\,,
\end{equation}
and the quantities $c_i$ are LECs of standard $\chi$EFT. The axial coupling $g_A=1.269$ \cite{PDG} is related to the pion-nucleon coupling constant $g_{\pi NN}\approx m_Ng_A/\fpi$ at leading order. The LEC $c_5$ in Eq.\,(\ref{eq:pin2}), for instance, is related to the quark-mass-induced contribution to the neutron-proton mass difference:
\begin{equation}\label{eq:dmnp}
\delta m^{\rm str}_{np}:=(m_n-m_p)^{\rm str}=4B_0(m_u-m_d)\,c_5+\cdots=(2.44\pm0.18)\,\rm{MeV}\,,
\end{equation}
where the 
ellipses denote higher-order corrections. The value for the strong neutron-proton mass splitting $\delta m^{\rm str}_{np}$ used here is the weighted average of the Lattice-QCD values compiled in Ref. \cite{Walker-Loud:2014iea} and of Ref. \cite{Borsanyi:2014jba}. The LEC $c_1$ can be related to the $\pi N$-sigma term. 
Reference~\cite{BaruHanhartc1-1} provides a compilation of various extractions of $c_1$ \cite{c1-2,c1-3,c1-4} and gives a value of
\begin{equation}\label{eq:c1pred}
c_1=(-1.0\pm0.3)\,\rm{GeV}^{-1}\,.
\end{equation}

The power-counting scheme by Weinberg \cite{WeinbergPowerCounting,weinbergpiNN} amounts to computing the naive engineering dimension, or chiral dimension $D$, of the relevant chiral operators or diagrams. In the one-nucleon sector this
chiral dimension of a diagram  is given by (see {\it e.g.} \cite{Bernard:2006gx,Kubis:2007iy,veroniquereview})
\begin{equation}
D=2L+1+\sum_kN_k^{\pi\pi}(k-2)+\sum_{k'}N_{k'}^{\pi N}(k'-1)\,,
\label{eq:power_counting}
\end{equation}
where $L$ denotes the number of loops, $N_k^{\pi\pi}$ the number of pion vertices of order $k$ and $N_{k'}^{\pi N}$ the number of pion-nucleon-nucleon vertices of order $k'$ in a considered diagram.


\subsection{Derivation of the \boldmath{$P$}- and \boldmath{$T$}-violating effective Lagrangians}
\label{sec:ptvlag}
This section is concerned with the derivation of $P$- and $T$-violating terms in the $\chi$EFT Lagrangian that are induced by the QCD $\theta$-term, Eq.\,(\ref{eq:lagtheta}), and the effective dimension-six sources, Eqs.\,(\ref{eq:4qlr})\,-\,(\ref{eq:4q8}). The transition from (amended) QCD to (amended) $\chi$EFT involves the introduction of (new) source fields. Characteristic transformation properties under chiral \mbox{$SU(2)_L\!\times\! SU(2)_R$} rotations are assigned to each source field that render the combination of a particular source field and its associated quark term invariant. Therefore, this implies that an investigation of  the \mbox{$SU(2)_L\!\times\!SU(2)_R$} transformation 
properties of quark terms is required to establish the connection between QCD and $\chi$EFT. A $P$- and $T$-violating quark term induces in general an infinite set of $P$- and $T$-violating terms of different orders in the (amended) $\chi$EFT Lagrangian. This ordering of terms in the (amended) $\chi$EFT Lagrangian can be different for each source of $P$ and $T$ violation. The aim of the remaining sections of this chapter is to identify the hierarchies of coupling constants of the leading $P$- and $T$-violating vertices for each considered source of $P$ and $T$ violation.

\subsubsection{Chiral transformation properties of quark bilinears and quark quadrilinears}
As noted in \cite{GassLeut1} and demonstrated in Appendix \ref{app:o4rep}, quark multilinears in the two-flavor case transform as states of $O(4)$ isospin representations and can be decomposed into quark multilinears which transform as basis states of irreducible representations of $O(4)$. According to appendix~\ref{app:o4rep}, an irreducible $O(4)$ representation can be labelled by a pair of half-integers or integers $(j_1,j_2)$ with $j_1+j_2\in\mathds{N}$. In the case of $j_1\!=\!j_2\!=\!j$, there are two different irreducible $O(4)$ representations which are labelled by the superscript $\pm$: $(j,j)^{\pm}$. 
 The list of quark bilinears (two-quark terms) and the irreducible representations of $O(4)$ (including their dimensions) to which they belong is given by
\begin{alignat}{4}
&\text{dim}=1\quad &&(0,0)^+ &&: \quad&&\bar{q}\gamma^{\mu}q\,,\label{eq:qb00p}\\
&\text{dim}=1 &&(0,0)^- &&: &&i\bar{q}\gamma^{\mu}\gamma_5q\,,\label{eq:qb00m}\\
&\text{dim}=4 &&(1/2,1/2)^+&&: &&(i\bar{q}\gamma_5\tau_iq,\bar{q}q), (i\bar{q}\sigma^{\mu\nu}\gamma_5\tau_iq,\bar{q}\sigma^{\mu\nu}q) \,,\label{eq:qb12p}\\
&\text{dim}=4 &&(1/2,1/2)^-&&: &&(\bar{q}\tau_iq,i\bar{q}\gamma_5q), (\bar{q}\sigma^{\mu\nu}\tau_iq,i\bar{q}\sigma^{\mu\nu}\gamma_5q) \,,\label{eq:qb12m}\\
&\text{dim}=6 &&(1,0)\oplus(0,1)&&: &&(\bar{q}\gamma^{\mu}\tau_iq,\bar{q}\gamma^{\mu}\gamma_5\tau_iq) \label{eq:qb1001}
\end{alignat}
with $i=1,2,3$.
The Dirac vectors and tensors in this list are understood to combine with other fields to form Lorentz invariant objects.

Quark quadrilinears (four-quark terms) transform as states of \textit{symmetric tensor products} of the irreducible representations of $O(4)$ in Eqs.\,(\ref{eq:qb00p})-(\ref{eq:qb1001}). An arbitrary quark quadrilinear decomposes in general into a sum of quark quadrilinears 
in which each quark quadrilinear transforms as a basis state of a particular irreducible representations of~$O(4)$.
The list of Lorentz invariant quark quadrilinears which transform as basis states of particular irreducible representation of $O(4)$ is derived in Appendix \ref{app:o4rep} (see Eq.\,(\ref{eq:o4-qq1})-(\ref{eq:o4-qq7})) and is given by ($i,j,k,l=1,2,3$, summation over $i,j$ implied, $k,l$ are fixed)
\begin{alignat}{6}
&(0,0)^-\quad&&1\,&&\slashed{P}\quad&&\text{-state}\quad&&:\quad&&\bar{q}\gamma_{\mu}\tau_iq\bar{q}\gamma^{\mu}\gamma_5\tau_iq\,,\label{eq:qq1}\\
&(1,1)^+&&3&&\slashed{P}\slashed{T}&&\text{-states}&&:&&\epsilon^{kij}\bar{q}\gamma_{\mu}\tau_iq\bar{q}\gamma^{\mu}\gamma_5\tau_jq\,,\label{eq:4qlrtrans}\\
&(2,0)\oplus(0,2)\quad&&5&&\slashed{P}&&\text{-states}&&:&&(\delta_{ik}\delta_{jl}\!+\!\delta_{jk}\delta_{il}\!-\!2\delta_{ij}\delta_{kl})\bar{q}\gamma_{\mu}\tau_iq\bar{q}\gamma^{\mu}\gamma_5\tau_jq\,,\\
&(0,0)^-&&1&&\slashed{P}\slashed{T}&&\text{-state}&&:&&\bar{q}q\bar{q}i\gamma_5q-\bar{q}\tau_iq\bar{q}i\gamma_5\tau_iq\,,\\
&(1,1)^{+}&&3&&\slashed{P}\slashed{T}&&\text{-states}&&:&&\bar{q}q\bar{q}i\gamma_5\tau_kq\pm\bar{q}\tau_kq\bar{q}i\gamma_5q \,,\label{eq:qq11p2}\\
&(1,1)^-&&6&&\slashed{P}\slashed{T}&&\text{-states}&&:&&\bar{q}\tau_kq\bar{q}i\gamma_5\tau_lq\pm\bar{q}\tau_lq\bar{q}i\gamma_5\tau_kq\quad k\neq l\,,\nonumber\\
&&&&&&&&&&&\bar{q}\tau_{k}q\bar{q}i\gamma_5\tau_{k}q+\bar{q}q\bar{q}i\gamma_5q\,.\label{eq:qq8}
\end{alignat}
The first column contains the $(j_1,j_2)$ labels of the irreducible representation of $O(4)$ of dimensions $(2j_1+1)(2j_2+1)$. The second column shows the number of $P$-violating (and $T$-violating) quark quadrilinears in a particular irreducible representation which are listed in the third column. Note that the two different relative signs in Eq.\,(\ref{eq:qq11p2}) (and also in Eq.\,(\ref{eq:qq8})) define two different sets of tensors which both transform as the same basis states of the $(1,1)^+$ (or the $(1,1)^-$) representation (as demonstrated in Appendix \ref{app:o4rep}). This list of quark quadrilinears reveals Fierz identities between two quark quadrilinears:  in order for a Fierz identity between two quark bilinears or two quark quadrilinears to exist, they have to transform as the same basis state of the same irreducible representation as identical tensors. This is the case for the quark quadrilinears in Eq.\,(\ref{eq:4qlrtrans}) and the quark quadrilinears in Eq.\,(\ref{eq:qq11p2}) with relative minus signs.

Further Lorentz invariant quark quadrilinears can emerge when external fields such as the photon field are taken into consideration. They are of higher dimension, therefore irrelevant for the analysis in this paper and are thus not discussed here. This holds also for quark quadrilinears involving internal photon loops which arise from integrating out the photon field.  In this way, effective tree-level nuclear operators without external photon lines are generated, but with a loop factor of $\alpha_{{\rm em}}/(4\pi)$ which leads to a further suppression beyond the 
one resulting from the pertinent inverse powers of the characteristic mass scale of BSM physics.

Only the following two of the eight quark quadrilinears in Eqs.\,(\ref{eq:qq1})-(\ref{eq:qq8}) are of relevance for our analysis: the 4q-term Eq.\,(\ref{eq:4q}) can be re-expressed in terms of quark flavor doublets by
\begin{eqnarray}\label{eq:4qdem}
&&i(\bar{u}u\bar{d}\gamma_5d+\bar{u}\gamma_5u\bar{d}d-\bar{d}u\bar{u}\gamma_5d-\bar{d}\gamma_5u\bar{u}d)\nonumber\\
=&&i(\bar{q}\gamma_5q\bar{q}q-\bar{q}\gamma_5\tau_3q\bar{q}\tau_3q-\bar{q}\gamma_5\tau_2q\bar{q}\tau_2q-\bar{q}\gamma_5\tau_1q\bar{q}\tau_1q)/2\,,
\end{eqnarray}
which demonstrates that it transforms as a basis state of the $(0,0)^-$ representation of $O(4)$. The FQLR-term Eq.\,(\ref{eq:4qlr}) can equally be rewritten in terms of quark-flavor doublets by 
\begin{eqnarray}\label{eq:4qlrdem}
&&i\bar{u}_R\gamma_{\mu}d_R\bar{d}_L\gamma^{\mu}u_L-i\bar{d}_R\gamma_{\mu}u_R\bar{u}_L\gamma^{\mu}d_L\nonumber\\
=&&-\epsilon^{3ij}\bar{q}_R\gamma^{\mu}\tau_iq_R\bar{q}_L\gamma_{\mu}\tau_jq_L/2\nonumber\\
=&&\epsilon^{3ij}\bar{q}\gamma^{\mu}\tau_iq\bar{q}\gamma_{\mu}\gamma_5\tau_jq/4\,,
\end{eqnarray}
and thus exhibits the transformation properties of a basis state of the $(1,1)^+$ representation of $O(4)$. Although, for completeness, the above list (\ref{eq:qq1})-(\ref{eq:qq8}) contains further $P$- and $T$-violating quark quadrilinears, Eq.\,(\ref{eq:4qdem}) and Eq.\,(\ref{eq:4qlrdem}) specify the only 
ones that contribute to dimension-six BSM terms  and are not additionally  suppressed 
as described in {\it e.g.} \cite{deVries:2012ab}.

It has been explained in Section \ref{sec:standardchpt} that the connection between standard QCD and $\chi$EFT is drawn by the introduction of source fields for each quark term in the standard QCD Lagrangian. We extend this method to QCD amended by effective dimension-six terms, which requires the introduction of further source fields. In order to ensure that the effective field theory obeys the same chiral Ward identities as the underlying theory, the source fields are assigned transformation properties under {\it local}  $SU(2)_L\!\times\! SU(2)_R$ transformations and the group $\mathds{Z}_2$ of 
the parity transformation which render the Lagrangian of the underlying theory -- QCD or amended QCD -- locally invariant. The effective Lagrangian is then obtained by compiling for each source of $P$ and $T$ violation the set of all possible combinations of these source fields with the fundamental building blocks of standard $\chi$EFT. These source fields are defined and the resulting terms in the (amended) $\chi$EFT Lagrangians are subsequently discussed in this section.

\subsubsection{The \boldmath{$\chi$}EFT Lagrangian from the QCD \boldmath{$\theta$}-term}\label{sec:chiefteta}
The $\theta$-term is rotated into a complex phase of the quark-mass matrix by an axial $U_A(1)$ transformation via the $U_A(1)$ anomaly. The resulting $P$- and $T$-violating quark bilinears are given by Eq.\,(\ref{eq:lagtheta}),
\begin{equation}
i\bar{q}\gamma_5q\quad\text{and}\quad i\bar{q}\tau_3\gamma_5q\,,
\end{equation}
which transform as basis states of the $(1/2,1/2)^-$ and $(1/2,1/2)^+$ irreducible representations of $O(4)$ according to 
Eqs.\,(\ref{eq:qb12m}) and (\ref{eq:qb12p}), respectively. They are thus connected by \mbox{$SU(2)_L\!\times\! SU(2)_R$} transformations to either the isospin-violating component or the isospin-conserving component of the quark-mass matrix. This implies that their corresponding source fields are given by the well-known source fields $p_0\mathds{1}$ and $p_3\tau_3$ of standard $\chi$EFT combined in the building block $\chi$. Their assigned transformation properties under local $SU(2)_L\!\times\! SU(2)_R$ transformations are identical to their global transformation properties and naturally reflect those of their corresponding quark bilinears, {\it i.e.} the source fields transform as the same basis states of the same irreducible representations \cite{GassLeut1}: 
\begin{alignat}{3}
&(1/2,1/2)^- &&: &&(s_1\tau_1,s_2\tau_2,s_3\tau_3,ip_0)\,,\label{eq:thetamult1}\\
&(1/2,1/2)^+\quad &&:\quad&&(ip_1\tau_1,ip_2\tau_2,ip_3\tau_3,s_0)\,.\label{eq:thetamult2}
\end{alignat}
Therefore, the $\theta$-term induced effective Lagrangian is just the standard $SU(2)$ $\chi$EFT Lagrangian as provided by \cite{GassLeut1,GassLeut2} for the pion sector and by {\it e.g.} \cite{MeissnerFettes1,Bernard:1992qa,MeissnerFettes3} for the pion-nucleon sector. The terms which are induced by the $\theta$-term are the $p_0$ and $p_3$ components of all terms with insertions of the building block $\chi$.

Before discussing the $P$- and $T$-violating terms in the $\chi$EFT Lagrangian induced by the \mbox{$\theta$-term} in detail, a few general remarks which apply to all sources of $P$ and $T$ violation are required. The decomposition of the fundamental building block $\chi_+$ into a traceless component ($\hat{\chi}_+$) and the trace itself ($\langle\chi_+\rangle$) corresponds to the decomposition of the eight dimensional representation of source fields combined in $\chi$ into the two irreducible representations $(1/2,1/2)^-$ of Eq.\,(\ref{eq:thetamult1}) and $(1/2,1/2)^+$ of Eq.\,(\ref{eq:thetamult2}), respectively. The analog is true for the building block $f_{\pm}^{\mu\nu}$, which contains the source fields $v^{\mu}_i\tau_i$, $a^{\mu}_i\tau_i$ and $v^{(s),\mu}$ and constitutes a reducible seven-dimensional representation. The decomposition of this building block into the components $\hat{f}_+^{\mu\nu}$ and $\langle f_+^{\mu\nu}\rangle$ corresponds to the decomposition into the irreducible representations $(1,0)\oplus(0,1)$ and $(0,0)^+$ (see Eq.\,(\ref{eq:qb00p}) and Eq.\,(\ref{eq:qb1001})), respectively.

Combinations of these fundamental building blocks correspond to tensor products of the above mentioned irreducible representations of $O(4)$. Consider a building block containing a \mbox{$P$- and $T$-}violating source field denoted by $\tilde{A}_{\slashed{P}\slashed{T}}$ for which another $P$- and $T$-conserving counterpart $A_{PT}$ exists that transforms identically under $SU(2)_L\!\times\! SU(2)_R$ transformations. Let this building block be combined with a
$P$- and $T$-conserving building block $B_{PT}$ that also has a \mbox{$P$- and $T$-}violating counterpart $\tilde{B}_{\slashed{P}\slashed{T}}$. In the language of group theory, the combination $\tilde{A}_{\slashed{P}\slashed{T}}B_{PT}$ transforms identically to the combination of building blocks $A_{PT}\tilde{B}_{\slashed{P}\slashed{T}}$, in which the former building block is replaced by the building block associated with its $P$- and $T$-conserving partner source field and in which the latter building block is replaced by its $P$- and $T$-violating counterpart. This is the justification for the definition of the building block $\chi_-$ in Eq.\,(\ref{eq:chi}), in which a particular source field is combined with the chiral structure of its partner source field. It is apparent that this building block can only occur in combination with other building blocks:  there is no term such as $\fpi^2\langle i\chi_-\rangle/4$ in the pion-sector Lagrangian of Eq.\,(\ref{eq:pilag2}), for instance, whereas the term $-l_7\langle\chi_-\rangle^2/16$ exists in Eq.\,(\ref{eq:pilag4}) (this is just a reflection of the fact that the representation $(1/2,1/2)^+\!\otimes\! (1/2,1/2)^+$ is identical to the representation $(1/2,1/2)^-\!\otimes\!(1/2,1/2)^-$). This reasoning extends to the pion-nucleon sector where the property of $P$ and/or $T$ violation can also be absorbed by products of Dirac matrices which act on nucleons fields and are contracted with other quantities. 

The terms of the $\chi$EFT Lagrangian that are induced by the $\theta$-term are contained in the expressions in the standard $\chi$EFT Lagrangians that include insertions of the building blocks 
$\langle\chi_{\pm}\rangle$ and $\hat{\chi}_{\pm}$.
Thus according to Eq.\,(\ref{eq:s0s3rep}) and Eq.\,(\ref{eq:lagtheta}), the sources
$s_0$, $p_0$, $s_{i}$ and $p_{i}$, \mbox{$i=1,2,3$}, of Eq.\,(\ref{eq:currentsqcd}) have to be replaced   simultaneously by
\begin{equation}\label{eq:thetafundrepl}
s_0\rightarrow\bar{m}\,,\quad s_{1,2}\rightarrow 0\,,\quad s_3\rightarrow\bar{m}\epsilon\,,\quad p_0\rightarrow\frac{\bar{\theta}}{2}\bar{m}\,,\quad p_{1,2}\rightarrow 0\,,\quad p_3\rightarrow\frac{\bar{\theta}}{2}\bar{m}\epsilon\,.
\end{equation}
However, it will be demonstrated in the next Subsection \ref{sec:selection} that the presence of chiral-symmetry breaking $P$- and $T$-violating terms in the QCD Lagrangian alters the ground state of the theory and requires a redefinition of the quark fields by an axial $SU(2)_L\!\times\! SU(2)_R$ rotation. As a result of this redefinition, all terms with a factor of $p_3$ vanish. The $P$- and $T$-violating Lagrangians discussed in this subsection are thus the \textit{generic} Lagrangians in $\chi$EFT induced by the $\theta$-term, {\it i.e.} the $\chi$EFT Lagrangians before the correct ground state of the theory has been selected. Furthermore, it will be shown in the next section that the ground-state-selection procedure ensures parameterization-invariant leading-order terms in the pion-sector Lagrangian. In order to obtain the induced terms in the generic $P$- and $T$-violating Lagrangian in an arbitrary parameterization, the matrix $U$ can be replaced by the generalized $U$ matrix (see \cite{Hanhart:2007mu} and references therein) with a parameterization function $g$,
\begin{equation}\label{eq:ugenpara1}
U=\exp\left(i\frac{\vec{\pi}\cdot\vec{\tau}}{\fpi}g\left(\frac{\pi^2}{\fpi^2}\right)\right)\,,\quad g\left(\frac{\pi^2}{\fpi^2}\right)=1+\left[
\alpha_{\rm p}+\frac{1}{6}\right]\frac{\pi^2}{\fpi^2}+\cdots\,,
\end{equation}
where $\alpha_{\rm p}$ is a real constant defining the parameterization ($\alpha_{\rm p}\!=\!0$ corresponds to the so called $\sigma$-parameterization of $SU(2)$ $\chi$EFT \cite{BKM1995}. The parameterization used in this paper is the {\it exponential} parameterization corresponding to $\alpha_{\rm p}\!=\!-1/6$). This replacement is equivalent to the following replacement of terms at leading order in powers of pion fields, where $A(\pi_i)$ is a generic homogeneous polynomial in $\pi_i$:
\begin{equation}
A(\pi_i)\left(1-\frac{1}{6}\frac{\pi^2}{\fpi^2}\right)\rightarrow A(\pi_i)\left(1+\alpha_{\rm p}\frac{\pi^2}{\fpi^2}\right)\,.
\end{equation}

The terms induced by the $\theta$-term in the leading-order pion-sector Lagrangian of \cite{GassLeut1} are given by Eq.\,(\ref{eq:pilag2}),
\begin{eqnarray}\label{eq:thetapi2}
\mathcal{L}_{\pi}^{(2)}&=&\frac{\fpi^2}{4}\langle\chi_+\rangle+\cdots\nonumber\\
&=&(2Bp_3)\fpi\pi_3\left(1-\frac{\pi^2}{6\fpi^2}\right)+\cdots\,,
\end{eqnarray}
where the ellipses denote -- throughout this section -- terms which are either of higher order in the pion-field expansion or $P$- and $T$-conserving.

The fourth-order pion-sector Lagrangian of \cite{GassLeut1} leads to the following generic $P$- and $T$-violating terms induced by the $\theta$-term (see Eq.\,(\ref{eq:pilag4})):
\begin{eqnarray}\label{eq:thetapi4}
\mathcal{L}_{\pi}^{(4)}&=&\frac{l_3}{16}\langle\chi_+\rangle^2-\frac{l_7}{16}\langle\chi_-\rangle^2+\cdots\nonumber\\
&=&2l_3(2Bs_0)(2Bp_3)\frac{\pi_3}{\fpi}\left(1-\frac{2\pi^2}{3\fpi^2}\right)-2l_7(2Bs_3)(2Bp_0)\frac{\pi_3}{\fpi}\left(1-\frac{2\pi^2}{3\fpi^2}\right)\nonumber\\
&&+\cdots\,.
\end{eqnarray}
The generic terms in the second-order pion-nucleon sector Lagrangian Eq.\,(\ref{eq:pin2}) read (see also \cite{MeissnerFettes3})
\begin{eqnarray}\label{eq:thetapin2}
\mathcal{L}_{\pi N}^{(2)}&=&c_1\langle\chi_+\rangle N^{\dagger}N+\cdots+c_5N^{\dagger}\hat{\chi}_+N+\cdots\nonumber\\
&=&4c_1(2Bp_3)\frac{\pi_3}{\fpi}\left(1-\frac{\pi^2}{6\fpi^2}\right)N^{\dagger}N+2c_5(2Bp_0)N^{\dagger}\frac{\vec{\pi}\cdot\vec{\tau}}{\fpi}\left(1-\frac{\pi^2}{6\fpi^2}\right)N\nonumber\\
&&+\cdots\,.
\end{eqnarray}
The generic fourth-order pion-nucleon Lagrangian of \cite{MeissnerFettes3} contains identical structures at leading order in the pion-field expansion:\footnote{The alternative  $\langle \chi_-\rangle^2$ is linearly dependent on the two displayed sources,
$\langle \chi_+\rangle^2$ and $\langle \hat{\chi}_+ \hat{\chi}_+\rangle$.}
\begin{eqnarray}\label{eq:thetapin41}
\mathcal{L}_{\pi N}^{(4)}\!\!\!&=&\!\!\!e_{38}\langle\chi_+\rangle^2N^{\dagger}N+e_{40}\langle\hat{\chi}_+\hat{\chi}_+\rangle N^{\dagger}N+\cdots\nonumber\\
&=&e_{38}32(2Bs_0)(2Bp_3)\frac{\pi_3}{\fpi}\left(1-\frac{2\pi^2}{3\fpi^2}\right)N^{\dagger}N+e_{40}16(2Bs_3)(2Bp_0)\frac{\pi_3}{\fpi}\left(1-\frac{2\pi^2}{3\fpi^2}\right)N^{\dagger}N\nonumber\\
&&+\cdots\,.
\end{eqnarray}
These terms constitute corrections to the leading $P$- and $T$-violating $\pi NN$ vertices of $\mathcal{L}_{\pi N}^{(2)}$ shown in Eq.\,(\ref{eq:thetapin2}), but prove to be negligible as demonstrated in the next subsection. The leading-order $P$- and $T$-violating $\gamma NN$ vertices also emerge from the fourth-order pion-nucleon Lagrangian $\mathcal{L}_{\pi N}^{(4)}$ of \cite{MeissnerFettes3} and read
\begin{eqnarray}\label{eq:thetapin4}
\mathcal{L}_{\pi N}^{(4)}&=&e_{110}i\langle \chi_-\rangle \langle f_+^{\mu\nu}\rangle N^{\dagger}S_{\mu}v_{\nu}N+e_{111}i\langle \chi_-\rangle N^{\dagger}\hat{f}_+^{\mu\nu}S_{\mu}v_{\nu}N\nonumber\\
&&+e_{112}i\langle f_+^{\mu\nu}\rangle N^{\dagger}\hat{\chi}_-S_{\mu}v_{\nu}N+e_{113}i\langle f_+^{\mu\nu}\hat{\chi}_-\rangle N^{\dagger}S_{\mu}v_{\nu}N\nonumber\\
&&+e_{109}iN^{\dagger}[\hat{f}_-^{\mu\nu},\hat{\chi}_+]S_{\mu}v_{\nu}N+\cdots\nonumber\\
&=&e_{110}8e(2Bp_0)N^{\dagger}S_{\mu}v_{\nu}NF^{\mu\nu}+e_{111}4e(2Bp_0)N^{\dagger}\tau_3S_{\mu}v_{\nu}NF^{\mu\nu}\nonumber\\
&&+e_{112}2e(2Bp_3)N^{\dagger}\tau_3S_{\mu}v_{\nu}NF^{\mu\nu}+e_{113}4e(2Bp_3)N^{\dagger}S_{\mu}v_{\nu}NF^{\mu\nu}\nonumber\\
&&+\cdots\,.
\end{eqnarray}
There are further terms in \cite{MeissnerFettes3} which contribute to magnetic quadrupole moments \cite{Mereghetti:2013bta} but are irrelevant for the computation of light-nuclei EDMs:
\begin{eqnarray}\label{eq:thetapin42}
\mathcal{L}_{\pi N}^{(4)}&=&e_{105}i\langle \chi_+\rangle \langle f_+^{\mu\nu}\rangle N^{\dagger}[S_{\mu},S_{\nu}]N+e_{106}i\langle\chi_+\rangle N^{\dagger}\hat{f}_+^{\mu\nu}[S_{\mu},S_{\nu}]N\nonumber\\
&&+e_{107}i\langle f_+^{\mu\nu}\rangle N^{\dagger}\hat{\chi}_+[S_{\mu},S_{\nu}]N+e_{108}i\langle \hat{f}_+^{\mu\nu}\hat{\chi}_+\rangle N^{\dagger}[S_{\mu},S_{\nu}]N\nonumber\\
&&+e_{114}iN^{\dagger}[\hat{f}_-^{\mu\nu},\hat{\chi}_-][S_{\mu},S_{\nu}]N+\cdots\,.
\end{eqnarray}
The set of fourth-order terms displayed here is the maximal set of independent terms. The \mbox{$P$- and $T$-}violating terms above extracted from \cite{MeissnerFettes3} have also been found by \cite{Mereghetti:2010tp} in the Weinberg formulation of $SU(2)$ $\chi$EFT.

There are numerous four-nucleon terms induced by the $\theta$-term which lead to the following leading-order generic four-nucleon Lagrangian induced by the $\theta$-term:
\begin{eqnarray}\label{eq:theta4n}
\mathcal{L}_{4N}^{(2)}&=&\quad \!\bar{C}_1\langle\chi_+\rangle N^{\dagger}NN^{\dagger}N+\bar{C}_2\langle\chi_+\rangle N^{\dagger}S_{\mu}NN^{\dagger}S^{\mu}N\nonumber\\
&&+\bar{C}_3N^{\dagger}\hat{\chi}_+NN^{\dagger}N+\bar{C}_4N^{\dagger}\hat{\chi}_+S_{\mu}NN^{\dagger}S^{\mu}N\nonumber\\
&&+\bar{C}_5i\langle\chi_-\rangle N^{\dagger}N\mathcal{D}_{\mu}(N^{\dagger}S^{\mu}N)+\bar{C}_6i\langle\chi_-\rangle N^{\dagger}\vec{\tau}N\cdot\mathcal{D}_{\mu}(N^{\dagger}S^{\mu}\vec{\tau}N)\nonumber\\
&&+\bar{C}_7iN^{\dagger}\hat{\chi}_-N\mathcal{D}_{\mu}(N^{\dagger}S^{\mu}N)+\bar{C}_8iN^{\dagger}\{\hat{\chi}_-,\vec{\tau}\}N\cdot\mathcal{D}_{\mu}(N^{\dagger}S^{\mu}\vec{\tau}N)/2\nonumber\\
&&+\cdots\nonumber\\
&=&\quad \! \bar{C}_14(2Bp_3)\pi_3 N^{\dagger}NN^{\dagger}N/\fpi+\bar{C}_24(2Bp_3)\pi_3 N^{\dagger}S_{\mu}NN^{\dagger}S^{\mu}N/\fpi\nonumber\\
&&+\bar{C}_32(2Bp_0)N^{\dagger}\vec{\pi}\cdot\vec{\tau}NN^{\dagger}N/\fpi+\bar{C}_42(2Bp_0)N^{\dagger}\vec{\pi}\cdot\vec{\tau}S_{\mu}NN^{\dagger}S^{\mu}N/\fpi\nonumber\\
&&-\bar{C}_54(2Bp_0)N^{\dagger}N\mathcal{D}_{\mu}(N^{\dagger}S^{\mu}N)-\bar{C}_64(2Bp_0)N^{\dagger}\vec{\tau}N\cdot\mathcal{D}_{\mu}(N^{\dagger}\vec{\tau}S^{\mu}N)\nonumber\\
&&-\bar{C}_72(2Bp_3)N^{\dagger}\tau_3N\mathcal{D}_{\mu}(N^{\dagger}S^{\mu}N)-\bar{C}_82(2Bp_3)N^{\dagger}N\cdot\mathcal{D}_{\mu}(N^{\dagger}S^{\mu}\tau_3N)\nonumber\\
&&+\cdots\,.
\end{eqnarray}
The generic set of $P$- and $T$-violating terms induced by $\theta$-term presented here is in agreement with the one derived in the Weinberg formulation of $SU(2)$ $\chi$EFT in \cite{Mereghetti:2010tp,Maekawa:2011vs} at all relevant orders.

\subsubsection{Amended \boldmath{$\chi$}EFT Lagrangian from the qCEDM}\label{sec:chptqcedm}
As pointed out in \cite{deVries:2010ah,deVries:2012ab}, the isospin-conserving and the isospin-violating components of the qCEDM, Eq.\,(\ref{eq:qcedm}), transform identically to the corresponding components of the $\theta$-term in Eq.\,(\ref{eq:lagtheta}) as basis states of the $(1/2,1/2)^-$ and $(1/2,1/2)^+$ irreducible representations of $O(4)$ (see Eq.\,(\ref{eq:qb12p}) and Eq.\,(\ref{eq:qb12m})):
\begin{align}
(1/2,1/2)^-:\quad&(\bar{q}\vec{\tau}\sigma^{\mu\nu}G_{\mu\nu}^{a}\lambda^a q,i\bar{q}\sigma^{\mu\nu}\gamma_5G_{\mu\nu}^{a}\lambda^{a}q)\,,\\
(1/2,1/2)^+:\quad&(i\bar{q}\vec{\tau}\sigma^{\mu\nu}\gamma_5G_{\mu\nu}^{a}\lambda^a q,\bar{q}\sigma^{\mu\nu}G_{\mu\nu}^{a}\lambda^{a}q)\,.
\end{align}
The components of the qCEDM constitute additional, separate isospin multiplets which are not connected by $SU(2)_L\!\times\!SU(2)_R$ rotations to the corresponding isospin multiplets of the standard QCD Lagrangian containing the components of the $\theta$-term in Eq.\,(\ref{eq:lagtheta}) (see Eq.\,(\ref{eq:qb12p}) and Eq.\,(\ref{eq:qb12m})).

In the Gasser-Leutwyler formulation of $\chi$EFT, this observation requires the introduction of an additional set of source fields, which can be combined in the new object $\tilde{\chi}$ in analogy to the definition of $\chi$ in standard $\chi$EFT:
\begin{eqnarray}
\chi&=&2B(s_0+s_1\tau_1+s_2\tau_2+s_3\tau_3+ip_0+ip_1\tau_1+ip_2\tau_2+ip_3\tau_3)\,,\\
\tilde{\chi}&=&2C(\tilde{s}_0+\tilde{s}_1\tau_1+\tilde{s}_2\tau_2+\tilde{s}_3\tau_3+i\tilde{p}_0+i\tilde{p}_1\tau_1+i\tilde{p}_2\tau_2+i\tilde{p}_3\tau_3)\,.\label{eq:chitilde}
\end{eqnarray}
While the quantity $B$ of Eq.\,(\ref{eq:chidef}) in standard 
$\chi$EFT is proportional to the scalar quark condensate, $\tilde{\chi}$ contains an additional corresponding quantity denoted by $C$ which is defined by
\begin{eqnarray}
\langle 0|\bar{u}u|0\rangle &=&-\fpi^2B+\cdots\,,\\
\langle 0|\bar{u}\sigma^{\mu\nu}G^{a}_{\mu\nu}\lambda^{a}u|0\rangle&=&-F_{\pi}^2C+\cdots\,.
\end{eqnarray}
This leads to the definition of further fundamental blocks $\tilde{\chi}_{\pm}$, in analogy to $\chi_{\pm}$, by
\begin{eqnarray}
\chi_{\pm}&=&u^{\dagger}\chi u^{\dagger}\pm u\chi^{\dagger}u\,,\\
\tilde{\chi}_{\pm}&=&u^{\dagger}\tilde{\chi} u^{\dagger}\pm u\tilde{\chi}^{\dagger}u\,.
\end{eqnarray}
For the qCEDM, $\tilde{p}_{0}, \tilde{p}_i$, $i=1,2,3$, in Eq.\,(\ref{eq:chitilde}) 
have to be subsequently replaced according to Eq.\,(\ref{eq:qcedm}) by
\begin{equation}
\tilde{p}_0\rightarrow-\tilde{\delta}_G^0\,,\quad \tilde{p}_{1,2} \to 0\,,\quad
\tilde{p}_3\rightarrow- \tilde{\delta}_G^3\,.
\end{equation}
Furthermore, \mbox{$s_{1,2}=\tilde{s}_{1,2} =0$} and 
\mbox{$p_{0,1,2,3}=0$} in the absence of the $\theta$-term, while $s_0$ and $s_3$ are
specified in Eq.\,(\ref{eq:s0s3rep}) and $\tilde{s}_0$ and $\tilde{s}_3$ are tiny but in general
non-zero constants: \mbox{$0\leq |\tilde{s}_{0,3}| \ll |s_{0,3}|$}.

The existence of new source fields has serious implications for the generic $P$- and $T$-violating Lagrangian induced by the qCEDM. Since the two additional isospin multiplets give rise to new, independent LECs as emphasized in \cite{deVries:2010ah,deVries:2012ab}, all $P$- and $T$-conserving observables (such as $\delta m_{np}^{\rm str}$ and $(\delta M_{\pi}^2)^{\rm str}$) originally related only to source fields in $\chi$ are now also related to the source fields in $\tilde{\chi}$. The new LECs encode BSM physics which is expected to yield only minor modifications to the SM at the energy scale $\Lambda_{{\rm had}}$ and below. The contributions from the new LECs to $P$- and $T$-conserving observables can then safely be considered insignificant compared to the ones from the standard LECs. Therefore, it is in general impossible to infer the values of these new LECs from measurements of $P$- and $T$-conserving observables \cite{deVries:2010ah,deVries:2012ab}.

The leading $P$- and $T$-violating terms induced by the qCEDM in the modified pion-sector Lagrangians $\mathcal{L}_{\pi N}^{(2)}$ and $\mathcal{L}_{\pi N}^{(4)}$ are contained in the structures
\begin{equation}\label{eq:dim6pion2}
\mathcal{L}^{(2)}_{\pi}=\frac{\fpi^2}{4}\langle\chi_++\tilde{\chi}_+\rangle+\cdots\,,
\end{equation}
and
\begin{eqnarray}\label{eq:dim6pion4}
\mathcal{L}^{(4)}_{\pi}&=&\frac{l_3}{16}\langle \chi_+\rangle^2+\frac{\tilde{l}_{3}}{16}\langle\tilde{\chi}_+\rangle^2+\frac{l_3'}{16}\langle \chi_+\rangle\langle \tilde{\chi}_+\rangle\nonumber\\
&&-\frac{l_7}{16}\langle \chi_-\rangle^2-\frac{\tilde{l}_{7}}{16}\langle\tilde{\chi}_-\rangle^2-\frac{l_7'}{16}\langle \chi_-\rangle\langle\tilde{\chi}_-\rangle+\cdots\,.
\end{eqnarray}
In the absence of the $\theta$-term, only those terms in $\mathcal{L}_{\pi N}^{(4)}$ proportional to the LECs $\tilde{l}_3$, $l_3'$, $\tilde{l}_7$ and $l_7'$ contain $P$- and $T$-violating components which are induced by the qCEDM. Due to the insignificance of the \mbox{$P$- and $T$-}conserving components of $\langle\tilde{\chi}_{\pm}\rangle$ compared to those of $\langle\chi_{\pm}\rangle$, the \mbox{$P$- and $T$-}violating terms from the structures proportional to the LECs $l_3'$ and $l_7'$ clearly dominate. 

All these statements are transferable to the pion-nucleon sector: the leading $P$- and $T$-violating terms induced by the qCEDM in the pion-nucleon Lagrangian are contained in
\begin{eqnarray}\label{eq:dim6nucpion2}
\mathcal{L}_{\pi N}^{(2)}&=&c_1\langle \chi_+\rangle N^{\dagger}N+\tilde{c}_1\langle \tilde{\chi}_+\rangle N^{\dagger}N\nonumber\\
&&+c_5N^{\dagger}\hat{\chi}_+N+\tilde{c}_5N^{\dagger}\hat{\tilde{\chi}}_+N+\cdots\,.
\end{eqnarray}
If the qCEDM is the sole source of $P$ and $T$ violation, only the structures proportional to the LECs $\tilde{c}_1$ and $\tilde{c}_5$ contain $P$- and $T$-violating terms. By the same procedure of duplicating structures with insertions of $\chi_{\pm}$, higher-order $P$- and $T$-violating $\pi NN$-, $\gamma NN$- and $4N$-terms are obtained from $\mathcal{L}_{\pi N}^{(3)}$, $\mathcal{L}_{\pi N}^{(4)}$ and $\mathcal{L}_{4N}$ as described above for the $\theta$-term case. All $P$- and $T$-violating terms listed above for the $\theta$-term case are also induced by the qCEDM when the source fields and LECs are replaced by the ones for the qCEDM. The findings regarding the generic set of $P$- and $T$-violating terms induced by the qCEDM presented here are in agreement with those of \cite{deVries:2010ah,deVries:2012ab} derived in the Weinberg formulation of $SU(2)$ $\chi$EFT at all relevant orders.

\subsubsection{Amended \boldmath{$\chi$}EFT Lagrangian from the FQLR-term}
The FQLR-term  defined in Eq.\,(\ref{eq:4qlr}) transforms as a basis state of the $(1,1)^+$ irreducible representation of $O(4)$ according to Eq.\,(\ref{eq:4qlrtrans}). Expressing the positive parity basis states of the $(1,1)^+$ representation in Eq.\,(\ref{eq:o4-11p2}) in terms of symmetric tensor products of quark bilinears, one obtains the new and separate isospin multiplet 
\begin{equation}\label{eq:4qlrisomult}
(1,1)^+\,:\quad(\bar{q}\tau^{i}\gamma^{\mu}q\,\bar{q}\tau^{j}\gamma_{\mu}q-\bar{q}\tau^{i}\gamma^{\mu}\gamma_5q\,\bar{q}\tau^{j}\gamma_{\mu}\gamma_5q, \epsilon^{klm}\,\bar{q}\tau_l\gamma^{\mu}q\,\bar{q}\tau_m\gamma_{\mu}\gamma_5q)\,,
\end{equation}
where $i,j,k=1,2,3$ and the summation only over the indices $l,m$ in the three $P$- and $T$-violating terms is implied. The isospin multiplet Eq.\,(\ref{eq:4qlrisomult}) consists of six $P$- and $T$-conserving (left part) and three $P$- and $T$-violating (right part, see Eq.\,(\ref{eq:qq11p2})) quark quadrilinears.

Due to the absence of source fields in standard $\chi$EFT which transform as basis states of the $(1,1)^+$, a genuinely new set of source fields has to be defined (by the symmetric tensor-product-form of Eq.\,(\ref{eq:o4-11p1}) and Eq.\,(\ref{eq:o4-11p2})):
\begin{equation}\label{eq:4qlrsf1}
q_{ij}\big((\tau_i)_R(\tau_j)_L+(\tau_i)_L(\tau_j)_R\big)+r_k\epsilon^{klm}(\tau_l)_R(\tau_m)_L
\end{equation}
with $i,j,k,l,m=1,2,3$ and summation only over the indices $l,m$ in the second term. The symmetric tensor $q_{ij}$ and the vector $r_k$ are the $(1,1)^+$ counterparts of the quantities $s_0,s_i$ and $p_0,p_i$ associated with the $(1/2,1/2)^{\pm}$  representations in standard $\chi$EFT. Note that the implied product of $\tau$-matrices in this formula is the symmetric tensor product of matrices (see Eq.\,(\ref{eq:o4-symten})), whose constituent $\tau$-matrices transform under $SU(2)_L\!\times\! SU(2)_R$ as indicated by the subscripts $L$ and $R$ (see Appendix \ref{app:o4rep} for details):
\begin{eqnarray}
&&q_{ij}\big((\tau_i)_R(\tau_j)_L+(\tau_i)_R(\tau_j)_L\big)+r_k\epsilon^{klm}(\tau_l)_R(\tau_m)_L\nonumber\\
&\mapsto& q_{ij}\big((R\tau_iR^{\dagger})_R(L\tau_jL^{\dagger})_L+(L\tau_iL^{\dagger})_L(R\tau_jR^{\dagger})_R\big)+r_k\epsilon^{klm}(R\tau_lR^{\dagger})_R(L\tau_mL^{\dagger})_L\,.
\end{eqnarray}

According to Eq.\,(\ref{eq:o4-11p5}) and Eq.\,(\ref{eq:o4-11p6}) in Appendix \ref{app:o4rep}, there is another basis of the $(1,1)^+$ representation which exhibits the chiral structure associated with the product $p_3s_0$ of familiar source fields,
\begin{eqnarray}\label{eq:4qlrsf2}
&&\frac{1}{2}ir_k\left([(\tau_k)_4-(\tau_k)_4^{\dagger}][(\mathds{1})_4+(\mathds{1})_4^{\dagger}]-(\tau_k\leftrightarrow \mathds{1})\right)\nonumber\\
&\mapsto&\frac{1}{2}ir_k\left([(L\tau_kR^{\dagger})_4-(L\tau_kR^{\dagger})_4^{\dagger}][(L\mathds{1}R^{\dagger})_4+(L\mathds{1}R^{\dagger})_4^{\dagger}]-(\tau_k\leftrightarrow\mathds{1})\right)\,,
\end{eqnarray}
where the subscript 4 in $(t)_4$, $t=\mathds{1},\tau_1,\tau_2,\tau_3$, and the dagger indicate the above defined 
\mbox{$SU(2)_L\!\times\!SU(2)_R$} transformation behavior of $\tau_k$ and $\mathds{1}$. The three source fields in Eq.\,(\ref{eq:4qlrsf2}) correspond to the $P$- and $T$-violating basis states of the  $(1,1)^+$ representation in decompositions of the symmetric tensor products of the irreducible representations $(1/2,1/2)^{\pm}\!\otimes\! (1/2,1/2)^{\pm}$ (see Eq.\,(\ref{eq:o4-prod12pp})). As previously mentioned, the two expressions of the source fields associated with the $P$- and $T$-violating basis states of the $(1,1)^+$ representation in Eq.\,(\ref{eq:4qlrsf1}) and Eq.\,(\ref{eq:4qlrsf2}) are Fierz equivalent.

The FQLR-term then induces all terms in the amended $\chi$EFT Lagrangian with insertions of the new building block $\eta_+$, which is defined for the two expressions of the source fields Eq.\,(\ref{eq:4qlrsf1}) and Eq.\,(\ref{eq:4qlrsf2}) by ($u=\sqrt{U}$, summation over $l,m$ implied)
\begin{eqnarray}
\eta_{+}&:=&(2Dr_{k})\epsilon^{klm}(u^{\dagger}\tau_lu)(u\tau_mu^{\dagger})+\cdots\,,\label{eq:4qlrsfdef1}\\
&:=&i(Dr_k)\big((u^{\dagger}\tau_ku^{\dagger}-u\tau_ku)(u\mathds{1}u+u^{\dagger}\mathds{1}u^{\dagger})-(\tau_k\leftrightarrow \mathds{1})\big)+\cdots\nonumber\\
&=&i(2Dr_k)\big((u^{\dagger}\tau_ku^{\dagger})(u\mathds{1}u)-(u\tau_ku)(u^{\dagger}\mathds{1}u^{\dagger})\big)+\cdots\,,\label{eq:4qlrsfdef2}
\end{eqnarray}
where the ellipses denote terms proportional to the $P$- and $T$-conserving source fields $q_{ij}$ and the product between the brackets ({\it i.e.} $(\cdots)(\cdots)$) is understood to be the symmetric tensor product of matrices. The quantity $D$ is defined by
\begin{equation}
\langle 0| \bar{u}\gamma^{\mu}u\bar{u}\gamma_{\mu}u-\bar{u}\gamma^{\mu}\gamma_5u\bar{u}\gamma_{\mu}\gamma_5u|0\rangle=D+\cdots\,,
\end{equation}
where the ellipses denote higher-order terms in the pion-field expansion. The $P$- and $T$-conserving component of the building block $\eta_+$ obtained from Eq.\,(\ref{eq:4qlrsf1}),
\begin{equation}\label{eq:4qlrbbpe1}
\eta_+=(2Dq_{ij})((u\tau_iu^{\dagger})(u^{\dagger}\tau_ju)+(u^{\dagger}\tau_iu)(u\tau_ju^{\dagger}))+\cdots\,,
\end{equation}
has been utilized here. In analogy to the definition of the building block $i\chi_{-}$ as a $P$- and $T$-violating counterpart of $\chi_+$ in standard $\chi$EFT, the building block $\eta_-$ (which corresponds to the $(1,1)^-$ representation) can be defined. The three $P$- and $T$-conserving terms in this building block are (see Eq.\,(\ref{eq:o4-11m5}) and Eq.\,(\ref{eq:o4-11m6}))
\begin{equation}
\eta_{-}=(2Dr_k)\big((u^{\dagger}\tau_ku^{\dagger})(u\mathds{1}u)+(u\tau_ku)(u^{\dagger}\mathds{1}u^{\dagger})\big)+\cdots\,.
\end{equation}

The terms induced by the FQLR-term in the effective Lagrangian are obtained by compiling the list of all chiral structures with insertions of $\eta_{\pm}$ and by the subsequent replacement (see Eq.\,(\ref{eq:4qlr}) 
\begin{equation}
r_3\rightarrow \frac{\nu_1V_{ud}}{2},\quad r_1=r_2=q_{ij}=0\,,
\end{equation}
where
here $\nu_1$ is  understood as an effective coupling which also includes  the $\nu_8$ contribution.

In the pion sector, the first expression of $\eta_+$ defined by Eq.\,(\ref{eq:4qlrsfdef1}) yields at leading order the 
\mbox{$P$- and $T$-}violating terms\footnote{Note that the ground-state-selection procedure discussed in the next subsection ensures parameterization-invariant $3\pi$ vertices. The generic $3\pi$ vertex in Eq.\,(\ref{eq:4qlrsfdef1}) is not the same for all parameterizations.}
\begin{equation}\label{eq:4qlr-pi1}
(2Dr_k)\epsilon^{klm}\langle \tau_lU\tau_mU^{\dagger}\rangle=(2Dr_k)8\frac{\pi_k}{\fpi}\left(\mathds{1}-\frac{2}{3}\frac{\pi^2}{\fpi^2}\right)+\cdots\,,
\end{equation}
whereas the second expression of $\eta_+$ in Eq.\,(\ref{eq:4qlrsfdef2}) gives a more familiar structure with identical $P$- and $T$-violating terms:
\begin{eqnarray}\label{eq:4qlr-pi2}
i(2Dr_k)(\langle\tau_kU^{\dagger}\rangle\langle U\rangle-\langle\tau_kU\rangle\langle U^{\dagger}\rangle)&=&\frac{i(2Dr_k)}{2}\langle \tau_kU^{\dagger}-\tau_kU\rangle\langle U+U^{\dagger}\rangle\nonumber\\
&=&(2Dr_k)8\frac{\pi_k}{\fpi}\left(\mathds{1}-\frac{2}{3}\frac{\pi^2}{\fpi^2}\right)+\cdots\,.
\end{eqnarray}
The chiral structure in Eq.\,(\ref{eq:4qlr-pi2}) 
resembles the $P$- and $T$-violating component of the familiar chiral structure $\langle\chi_{+}\rangle^2$ in $\mathcal{L}_{\pi}^{(4)}$ of Eq.\,(\ref{eq:pilag4}) with the product of conventional source fields $(2Bs_0)(2Bp_3)$ replaced by $(2Dr_3)$. In this sense, an insertion of $\eta_+$ is equivalent to two insertions of {\it e.g.} $\chi_+$ with the replacement $(2Bp_3)(2Bs_0)\rightarrow (2Dr_3)$.
The leading term in the pion sector is thus given by
\begin{equation}\label{eq:4qlrlotad}
\langle\chi_{+}\rangle^2\leadsto  (2Dr_k)8\frac{\pi_3}{\fpi}\left(\mathds{1}-\frac{2}{3}\frac{\pi^2}{\fpi^2}\right)+\cdots\,.
\end{equation}
The next pion-tadpole term arises from a chiral structure with a simultaneous insertion of $\eta_+$ and $\chi_+$ combined with the new LEC $l_{4qLR}$:
\begin{equation}\label{eq:lagpi44qlr}
\frac{l_{4qLR}}{16}(2Dr_3)(2Bs_0)\epsilon^{3lm}\langle\tau_lU\tau_mU^{\dagger}\rangle\langle U^{\dagger}+U\rangle=l_{4qLR}8Dr_3Bs_0\frac{\pi_3}{\fpi}+\cdots\,.
\end{equation}

The leading-order $P$- and $T$-violating $\pi NN$ vertices induced by the FQLR-term can be extracted from the fourth-order Lagrangian $\mathcal{L}^{(4)}_{\pi N}$ by the above described replacement of the source fields $(2Bs_0)(2Bp_3)$ and the LEC\footnote{Note that the $3\pi NN$ vertex in Eq.\,(\ref{eq:4qlrpin}) is not in a parameterization-invariant form. To obtain this vertex in another parameterization, Eq.\,(\ref{eq:4qlrpin}) has to be replaced by
\begin{equation}
c_{4qLR}(2Dr_k)8\frac{\pi_3}{\fpi}\left(\mathds{1}-\frac{2}{3}\frac{\pi^2}{\fpi^2}\right)N^{\dagger}N\rightarrow c_{4qLR}(2Dr_k)8\frac{\pi_3}{\fpi}\left(\mathds{1}-\frac{1}{2}\frac{\pi^2}{\fpi}+\alpha_{\rm p}\frac{\pi^2}{\fpi^2}\right)N^{\dagger}N\,.
\end{equation}

}:
\begin{equation}\label{eq:4qlrpin}
\langle\chi_+\rangle^2N^{\dagger}N\leadsto c_{4qLR}(2Dr_k)8\frac{\pi_3}{\fpi}\left(\mathds{1}-\frac{2}{3}\frac{\pi^2}{\fpi^2}\right)N^{\dagger}N+\cdots\,.
\end{equation}
The constant $c_{4qLR}$ here is the new and independent LEC induced by the FQLR-term.

The FQLR-term also induces all isospin-violating terms of the Lagrangian $\mathcal{L}_{4N}$ in Eq.\,(\ref{eq:theta4n}) at leading order in the expansion of pion fields, {\it i.e.} those proportional to the LECs $\bar{C}_1$, $\bar{C}_2$, $\bar{C}_7$, 
and $\bar{C}_8$. The corresponding terms induced by the FQLR-term are obtained by insertions of $\eta_{\pm}$. The remaining isospin-conserving terms in Eq.\,(\ref{eq:theta4n}) at leading order in the pion-field expansion are obtained by simultaneous insertions of $\eta_{\pm}$ and $\chi_{\pm}$ and are therefore suppressed. The $\gamma NN$ vertices in Eq.\,(\ref{eq:thetapin4}) at leading order in the pion-field expansion are also induced by the FQLR-term. The isospin-conserving term in Eq.\,(\ref{eq:thetapin4}) is in this case generated by a simultaneous insertion of $\eta_-$ and $\hat{f}_+^{\mu\nu}$, whereas the isospin-violating term arises from an insertion of $\eta_-$ and $\langle f_+^{\mu\nu}\rangle$. The generic set of $P$- and $T$-violating terms induced by the FQLR-term derived here is in agreement with the results of \cite{deVries:2010ah,deVries:2012ab} derived in the Weinberg formulation of $SU(2)$ $\chi$EFT.

\subsubsection{Amended \boldmath{$\chi$}EFT Lagrangians from the qEDM}
The isospin-conserving and isospin-violating components of the qEDM given in Eq.\,(\ref{eq:qedm}) transform identically to the corresponding components of the qCEDM as basis states of the $(1/2,1/2)^-$ and $(1/2,1/2)^+$ irreducible representations of $O(4)$. In contrast to the qCEDM, the qEDM has an explicit insertion of the photon field. This requires the new source field to be identified with the photon field and to transform identically to the source fields $\tilde{p}_0$ and $\tilde{p}_3\tau_3$ in $\tilde{\chi}$. The set of all \mbox{$P$- and $T$-}violating terms in the amended $\chi$EFT Lagrangian is therefore obtained from all possible insertions of the building blocks $\tilde{\chi}_{\pm}$ and $F^{\mu\nu}$ by replacing the qCEDM source fields $(2C\tilde{p}_0)$ and $(2C\tilde{p}_3)$ by the newly defined qEDM counterparts $v_0$ and $w_3$. All resulting terms contain at least one photon field, which has to be integrated out in order to generate \mbox{$P$- and $T$-}violating pion-, $\pi NN$- and $4N$ vertices at the price of picking up a loop factor of $\alpha_{em}/(4\pi)$. Therefore, all qEDM contributions involving  such vertices are heavily suppressed by a factor of at least $\alpha_{em}/(4\pi)$ with respect to qEDM contributions involving only one {\em external} photon --- {\it cf.} the $d_0$ and $d_1$ terms in Eq.\,(\ref{eq:impcoup}). 
The leading $P$- and $T$-violating $\gamma NN$-terms are obtained from Eq.\,(\ref{eq:thetapin4}) and Eq.\,(\ref{eq:thetapin42}). The terms relevant for the computation of light-nuclei EDMs are given by
\begin{eqnarray}
i\langle\tilde{\chi}_-\rangle N^{\dagger}S_{\mu}v_{\nu}NF^{\mu\nu}&\leadsto&4v_0\left(1-\frac{\pi^2}{2\fpi^2}\right)N^{\dagger}S_{\mu}v_{\nu}NF^{\mu\nu}+\cdots\,,\\
iN^{\dagger}\hat{\chi}_-S_{\mu}v_{\nu}NF^{\mu\nu}&\leadsto&2w_3N^{\dagger}\left(\tau_3-\frac{\vec{\pi}\cdot\vec{\tau}\pi_3}{2\fpi^2}\right)S_{\mu}v_{\nu}NF^{\mu\nu}+\cdots\,.
\end{eqnarray}
These terms constitute the leading isoscalar and isovector contributions to the single-nucleon EDMs induced by the qEDM. The full set of induced terms that can be easily obtained by the procedure explained here is in agreement with the results derived in the Weinberg formulation of $SU(2)$ $\chi$EFT which are published in \cite{deVries:2010ah,deVries:2012ab}. 

\subsubsection{Amended \boldmath{$\chi$}EFT Lagrangians from the  4q-term and gCEDM}
The 4q-term and the gCEDM are chiral singlets and thus transform as the basis state of the $(0,0)^-$ irreducible representation of $O(4)$. They are not connected to other terms in standard QCD by $SU(2)_L\!\times\!SU(2)_R$ or $U(1)_A$ transformations as the $\theta$-term. In order to derive all induced terms in the amended $\chi$EFT Lagrangian, a new $P$- and $T$-violating source field has to be introduced that transforms as an $SU(2)_L\!\times\!SU(2)_R$ singlet. This leads to the definition of the new fundamental building block $\varsigma^-$ and its $P$- and $T$-conserving partner chiral-singlet building block $\varsigma^+$ (analogous to the definition of $i\chi_-$ as a partner building block for $\chi_+$). Since $\varsigma^-$ is a $P$- and $T$-violating source field, there do not exist any non-vanishing chiral structures with one insertion of $\varsigma^-$ (as a $P$-violating counterpart to the standard source field $v_{\mu}^{(s)}$ would also not generate non-vanishing terms). The complete list of $P$- and $T$-violating terms in the amended $\chi$EFT Lagrangian induced by the 4q-term and the gCEDM can thus in principle be obtained by all possible combinations of $\varsigma^+$ with the fundamental building blocks of standard $\chi$EFT (and setting $p_0=p_3=0$ if $\bar{\theta}=0$). This procedure can be illustrated by the following example: let $A_{PT}$ and $B_{PT}$ be two conventional fundamental building blocks with $P$- and $T$-violating partner building blocks  $\tilde{A}_{\slashed{P}\slashed{T}}$ and $\tilde{B}_{\slashed{P}\slashed{T}}$. Chiral structures induced by the 4q-term and the gCEDM are then of the form:
\begin{equation}
\varsigma^+\tilde{A}_{\slashed{P}\slashed{T}}B_{PT},\quad \varsigma^+A_{PT}\tilde{B}_{\slashed{P}\slashed{T}}\,.
\end{equation}
Some leading chiral structures induced by the 4q-term and the gCEDM obtained in this manner are given by
\begin{alignat}{3}\label{eq:gcedmlag}
&\mathcal{L}_{\pi}&&:&&i\varsigma^+\langle\chi_-\rangle,\,i\varsigma^+\langle\chi_-\rangle\langle\chi_+\rangle,\,\cdots\,,\nonumber\\
&\mathcal{L}_{\pi N}\,&&:\quad&&i\varsigma^+\langle\chi_-\rangle N^{\dagger}N,\,i\varsigma^+N^{\dagger}\hat{\chi}_-N,\,i\varsigma^+N^{\dagger}[S\cdot u,v\cdot u]N,\nonumber\\
&&&&&\varsigma^+\langle f_+^{\mu\nu}\rangle N^{\dagger}S_{\mu}v_{\nu}N,\,\varsigma^+N^{\dagger}\hat{f}_+^{\mu\nu}S_{\mu}v_{\nu}N,\,\cdots\,,\nonumber\\
&\mathcal{L}_{4N}&&:&&\varsigma^+ N^{\dagger}N\mathcal{D}_{\mu}(N^{\dagger}S^{\mu}N),\,\varsigma^+N^{\dagger}\vec{\tau}N\cdot\mathcal{D}_{\mu}(N^{\dagger}\vec{\tau}S^{\mu}N),\,\cdots\,.
\end{alignat}
Induced vertices involving pions emerge from terms with insertions of $\chi_{\pm}$ (or multiple derivatives (see the third term of $\mathcal{L}_{\pi N}$ in Eq.\,(\ref{eq:gcedmlag}))) and are thus suppressed by $\mpi^2/m_N^2$.
This observation is just a reflection of Goldstone's theorem. The leading $P$- and $T$-violating $\gamma NN$ and $4N$ vertices emerge from the third and fourth line of Eq.\,(\ref{eq:gcedmlag}), respectively.
These results are in agreement with the findings of \cite{deVries:2010ah,deVries:2012ab} derived in the Weinberg formulation of $SU(2)$ $\chi$EFT.


\subsection{Selection of the ground state}\label{sec:selection}
If the ChPT action functional is invariant under chiral $SU(2)_L\!\times\!SU(2)_R$ transformations, the $SU(2)_V$ subgroup to which $SU(2)_L\!\times\! SU(2)_R$ breaks down is not unique. The presence of terms in the (amended) QCD Lagrangian or equivalently in $\mathcal{L}_{\pi}$ which explicitly violate the \mbox{$SU(2)_L\!\times\! SU(2)_R$} symmetry impose a constraint on the selection of the $SU(2)_V$ subgroup such that it is then in general well defined \cite{Dashen:1970et,Baluni}. The definition of this subgroup also implies the definition of the ground state of (amended) ChPT around which the effective field theory is expanded. The ground-state selection for the QCD $\theta$-term and the effective dimension-six sources within the Weinberg formulation of $\chi$EFT has recently been presented in \cite{Mereghetti:2010tp,deVries:2012ab}. 

This section provides a thorough derivation of the ground-state-selection procedure and its impact on the effective Lagrangians in the Gasser-Leutwyler formulation of $SU(2)$ ChPT. This section is organized as follows: before computing the ground states of (amended) $\chi$EFT for all considered sources of $P$ and $T$ violation, the general selection procedure of the ground state in standard QCD and standard $\chi$EFT in the Gasser-Leutwyler formulation is discussed. The set of $P$- and $T$-violating vertices relevant for our analysis is given by Eq.\,(\ref{eq:impcoup}). As the main result of this subsection, the coupling constants in Eq.\,(\ref{eq:impcoup}) are either calculated explicitly or estimated by the means of NDA and the relative ordering by their absolute values is identified for each source of $P$ and $T$ violation. By a detailed study, all other vertices which are not displayed in Eq.\,(\ref{eq:impcoup}) prove to yield negligible contributions to the EDMs of light nuclei and are not discussed in this subsection.

\subsubsection{Selection of the ground state in standard QCD and \boldmath{$\chi$}EFT}
The correct $SU(2)_V$ subgroup of standard QCD is identified by minimizing the QCD potential
\begin{equation}\label{eq:qcdgspot}
V=\int d^4x\,(s_0\bar{q}q+s_3\bar{q}\tau_3q-ip_0\bar{q}\gamma_5q-ip_3\bar{q}\gamma_5\tau_3q)\,.
\end{equation}
Since explicit chiral-symmetry breaking constitutes at most a small perturbation, the minimum can be identified by an infinitesimal variation of the quark fields defined by
\begin{equation}
q\mapsto\exp(i\tau_3\,\delta\alpha_V^3+i\gamma_5\tau_3\,\delta\alpha_A^3)q\,,
\end{equation}
{\it i.e.} by the multiplication of $q$ by a diagonal charge-conserving matrix (this corresponds to the procedure presented for $SU(3)$ $\chi$EFT in \cite{Borasoy:2000pq}). This variation yields the ground-state condition
\begin{equation}\label{eq:qcdgs}
\delta V=2\int d^4x \,\bar{q}(s_0i\gamma_5\tau_3+s_3i\gamma_5+p_0\tau_3+p_3)q\,\delta\alpha_A^3=0\,.
\end{equation}
The quark fields can be redefined by
\begin{equation}\label{eq:vacrot}
q\mapsto\exp(i\gamma_5\tau_3\,\beta/2)q\,,
\end{equation}
in order to obey the ground-state condition Eq.\,(\ref{eq:qcdgs}). However, the ground-state condition is only fulfilled if $p_3/s_0=p_0/s_3$ holds, which is in general not true. As argued in \cite{Mereghetti:2010tp}, the assumption that the ground state has to be $P$ and $T$ conserving as well as isospin conserving requires Eq.\,(\ref{eq:qcdgs}) to be evaluated at 
\begin{equation}
\bar{q}\tau_3q=\bar{q}i\gamma_5q=\bar{q}i\gamma_5\tau_3q=0\,,
\end{equation}
which reduces the ground-state condition in Eq.\,(\ref{eq:qcdgs}) to the requirement that the coefficient of $i\bar{q}\gamma_5\tau_3q$ in Eq.\,(\ref{eq:qcdgspot}),  $p_3$, has to vanish. This condition is fulfilled if the angle $\beta$ in the transformation of Eq.\,(\ref{eq:vacrot}) applied to the quark fields in Eq.\,(\ref{eq:qcdgspot}) is chosen to be
\begin{equation}\label{eq:qcdgs-beta}
\beta=\arctan\left(\frac{p_3}{s_0}\right)\,.
\end{equation}


The ground-state-selection procedure can equivalently be carried out in $\chi$EFT. The ground state is identified by minimizing the leading-order potential in the pion-sector Lagrangian of $\chi$EFT \cite{GassLeut2} which is given by
\begin{equation}\label{eq:chpt-pot}
V=-\int d^4 x \frac{\fpi^2}{4}\langle \chi U^{\dagger}+U\chi^{\dagger}\rangle
\end{equation}   
with
\begin{equation}
\chi=2B(s_0\mathds{1}+s_3\tau_3+ip_0\mathds{1}+ip_3\tau_3)\,,
\end{equation} 
as usual. The minimum of this functional is identified by a variation of $U=u^2$: since for each pair $g, g'\in SU(2)$ there is a $\tilde{g}\in SU(2)$ such that $g'=\tilde{g} g$, a variation of the field $U(x)\in SU(2)$ amounts to the left-multiplication by an element $G(x)\in SU(2)$:
\begin{equation}
U(x)\mapsto G(x)U(x)=\exp(i\vec{\tau}\cdot\vec{\alpha}(x))U(x)\,.
\end{equation}  
The minimization of $V$ leads to
\begin{equation}
\delta{V}=i\frac{\fpi^2}{4}\int d^4x \langle -\chi U^{\dagger}\tau_i+\tau_iU\chi^{\dagger}\rangle\alpha_i\,=\,0\,,
\end{equation}
which gives the three ground-state conditions
\begin{equation}\label{eq:chptgs}
\langle \tau_iU\chi^{\dagger}-\chi U^{\dagger}\tau_i\rangle=0\,,\quad i=1,2,3\,.
\end{equation}
For $p_0\!=\!p_3\!=\!0$, these conditions reduce to the set of equations
\begin{alignat}{3}
i=1,2,3:\qquad& &s_0\langle\tau_i(U-U^{\dagger})\rangle+s_3\langle \tau_3\tau_iU-\tau_i\tau_3U^{\dagger}\rangle&=0\,,\nonumber\\
&\Leftrightarrow\quad &s_0\langle\tau_i(U-U^{\dagger})\rangle+is_3\epsilon^{3ij}\langle \tau_j(U+U^{\dagger})\rangle&=0\,,\nonumber\\
&\Leftrightarrow\quad &s_0\langle\tau_i(U-U^{\dagger})\rangle&=0\,,
\end{alignat}
which has the unique minimum solution $U\!=\!\mathds{1}$ (up to a phase). The case
$p_0\neq0$ does not alter the ground state since the corresponding terms in the ground-state conditions of Eq.\,(\ref{eq:chptgs}),
\begin{equation}
-ip_0\langle\tau_i(U+U^{\dagger})\rangle=0\,,
\end{equation}
vanish trivially for $i\!=\!1,2,3$ (there is no term proportional to $p_0$ in $\mathcal{L}_{\pi}^{(2)}$ of Eq.\,(\ref{eq:pilag2})). The situation is entirely different for $p_3\neq 0$, which alters the ground-state conditions in Eq.\,(\ref{eq:chptgs}) to
\begin{alignat}{2}
i=1,2&:\quad s_0\langle\tau_i(U-U^{\dagger})\rangle+p_3\epsilon^{3ij}\langle\tau_j(U-U^{\dagger})\rangle&=0\,,\label{eq:chptgs-p31}\\
i=3&:\quad s_0\langle\tau_3(U-U^{\dagger})\rangle-ip_3\langle U+U^{\dagger}\rangle&=0\,.\label{eq:chptgs-p32}
\end{alignat}
Equation\,(\ref{eq:chptgs-p31}) requires the $U$ matrix to be of the form 
\begin{equation}
U=\exp(i\tau_3\beta)=\cos(\beta)+i\tau_3\sin(\beta)\,.
 \label{eq:Ubeta}
\end{equation}
The insertion of this expression for $U$ into Eq.\,(\ref{eq:chptgs-p32}) yields the familiar ground-state condition of Eq.\,(\ref{eq:qcdgs-beta}):
\begin{equation}\label{eq:betachpt}
\beta=\arctan\left(\frac{p_3}{s_0}\right)\,.
\end{equation}
This demonstrates that the $U$ matrix has to be transformed in the presence of a non-vanishing $P$-, $T$-, and isospin-violating source field $p_3$ by an axial rotation 
\mbox{$A\!=\!R\!=\!L^{\dagger}\!=\!\exp(i\tau_3\beta/2)$}:
\begin{equation}
U\mapsto AUA\,.
\end{equation}
The ground state itself is then given by $U_0=A^2\approx\exp(i\,\tau_3\,p_3/s_0)$. 

The ground-state conditions Eq.\,(\ref{eq:chptgs-p32}) in the presence of a non-vanishing source field $p_3$ is obvious due to the following group theoretical argument: as demonstrated in the previous subsection and in particular in Appendix \ref{app:o4rep}, the source fields transform inversely to their associated quark bilinears as basis states of the $(1/2,1/2)^{\pm}$ irreducible representations of $O(4)$: 
\begin{alignat}{3}
&(1/2,1/2)^+\quad &&:\quad&&(ip_1\tau_1,ip_2\tau_2,ip_3\tau_3,s_0)\,,\\
&(1/2,1/2)^- &&: &&(s_1\tau_1,s_2\tau_2,s_3\tau_3,ip_0)\,.
\end{alignat}
Any $SU(2)_L\!\times\!SU(2)_R$ transformation of the quark fields or the matrix $U$, therefore, is equivalent to the inverse $SU(2)_L\!\times\!SU(2)_R$ transformation of the source fields:
\begin{alignat}{5}
&(1/2,1/2)^+&&:\quad&&(ip_1\tau_1,ip_2\tau_2,ip_3\tau_3,s_0)&&\mapsto&&(ip_1'\tau_1,ip_2'\tau_2,ip_3'\tau_3,s_0')\,,\\
&(1/2,1/2)^- &&: &&(s_1\tau_1,s_2\tau_2,s_3\tau_3,ip_0)&&\mapsto\,&&(s_1'\tau_1,s_2'\tau_2',s_3'\tau_3,ip'_0)\,.
\end{alignat}
The axial transformation of the source fields can then be chosen to yield $p_3'=0$ and the ground state is again simply given by $U_0'=\mathds{1}$. If this axial transformation is undone, the actual ground state for $p_3\neq 0$ is on the path defined by the one-parameter subgroup $\exp(i\tau_3\beta)$.
The infinitesimal axial $SU(2)_L\!\times\! SU(2)_R$ transformation of the quark bilinears or the $U$ matrix to remove $p_3$ corresponds the following inverse axial $SU(2)_L\!\times\! SU(2)_R$ transformation of the source fields $s_0$, $s_3$, $p_0$ and $p_3$:
\begin{alignat}{2}
s_0'&=s_0+\beta p_3+\cdots&&=s_0+p_3^2/s_0+\cdots\,,\label{eq:corrot1}\\
s_3'&=s_3+\beta p_0+\cdots&&=s_3+p_3p_0/s_0+\cdots\,,\label{eq:corrot2}\\
p_0'&=p_0-\beta s_3+\cdots&&=p_0-p_3s_3/s_0+\cdots\,,\label{eq:corrot3}\\
p_3'&=p_3-\beta s_0+\cdots&&=p_3-p_3+\cdots=0+\cdots\,.\label{eq:corrot4}
\end{alignat}
As a result of the axial transformation, the original terms proportional to $p_3$ are absent from the QCD Lagrangian and equivalently from the entire effective Lagrangian.
 
The ground-state-selection procedure ensures the absence of leading-order pion tadpole terms. However, a further pion tadpole term emerges in the pion sector Lagrangian $\mathcal{L}_{\pi}^{(4)}$ of Eq.\,(\ref{eq:pilag4}),
\begin{equation}\label{eq:thetapitad2}
-\frac{l_7}{16}\langle \chi U^{\dagger}-U\chi^{\dagger}\rangle^2=\cdots-2l_7(2Bp_0)(2Bs_3)\frac{\pi_3}{\fpi}\left(1-\frac{2}{3}\frac{\pi^2}{\fpi^2}\right)+\cdots\,,
\end{equation}
which can be removed by another axial rotation $A'=\exp(i\tau_3\beta'/2)$, where the angle $\beta'$ is given by \mbox{$\beta'=-4Bl_7s_3p_0/(\fpi^2s_0)+\cdots$}. Pion tadpoles that occur at subleading orders cannot be rotated away within one specific order, since such a rotation would reintroduce the tadpole terms previously removed from lower orders, in this case the leading order. The axial rotation has to be chosen such that no tadpoles occur at all orders up to and including the one from which the tadpole is to be removed. In the above case, this entails that the axial rotation has to generate terms in the leading-order pion-sector Lagrangian $\mathcal{L}_{\pi}^{(2)}$ (Eq.\,(\ref{eq:pilag2})) which cancel the tadpole term in $\mathcal{L}_{\pi}^{(4)}$ (Eq.\,(\ref{eq:pilag4})). This procedure may be iterated up to any desired order. 
Thus, in an alternative interpretation,  
the angle $\beta$ defining the rotation (\ref{eq:Ubeta}) of the ground state can 
be understood to admit a chiral expansion of its own.

Before concluding the general discussion about the selection of the ground state, a few further issues have to be mentioned. $\mathcal{L}^{(2)}_{\pi}$ in Eq.\,(\ref{eq:chpt-pot}) does not contain the source fields $v_j^{\mu}$, $a_j^{\mu}$ and $v^{(s),\mu}$, which do thus not affect the selection of the ground state at this order.  In the sole presence of the electromagnetic field $A_{\mu}$, all  terms apart from the $l_3$ and the $l_7$ term in $\mathcal{L}_{\pi}^{(4)}$ of Eq.\,(\ref{eq:pilag4}) are invariant under the axial rotation $A$: the $l_1$ and the $l_2$ term are invariant since $[D_{\mu},A]=0$. The $l_4$ term vanishes due to
\begin{equation}
D_{\mu}\chi=ie[\mathcal{Q},\chi]=0\,,
\end{equation}
where $\mathcal{Q}$ is the $SU(2)$ quark-charge matrix of Eq.\,(\ref{eq:chargemassmatrix}) and $e$ is the
elementary charge. The $l_5$ and the $l_6$ term are invariant under $A$ since
\begin{equation}
[R_{\mu\nu},A]=[L_{\mu\nu},A]=0\,,
\end{equation}
and the $h_i$ terms are designed to be chiral singlets. Therefore, the presence of the electromagnetic field $A_{\mu}$ does not affect the selection of the ground state. 

The selection of the ground state also ensures parameterization-invariant leading-order terms in the pion sector. This statement can be illustrated by the following example: let the $U$ matrix be the one defined in Eq.\,(\ref{eq:ugenpara1}). 
Eq.\,(\ref{eq:thetapitad2}) then reads 
\begin{equation}\label{eq:parainv1}
-\frac{l_7}{16}\left(\chi U^{\dagger}-U\chi^{\dagger}\right)^2=-2l_7(2Bp_0)(2Bs_3)\frac{\pi_3}{\fpi}\left(1-\frac{1}{2}\frac{\pi^2}{\fpi^2}+\alpha_{\rm p}\frac{\pi^2}{\fpi^2}\right)+\cdots\,.
\end{equation}
The second axial rotation $A'=\exp(i\tau_3\beta'/2)$ to remove the subleading tadpole term causes a shift of $\mathcal{L}_{\pi}^{(2)}$ of Eq.\,(\ref{eq:pilag2}) which cancels the pion-tadpole term as well as the parameterization-dependent component of the $3\pi$ vertex in Eq.\,(\ref{eq:parainv1}):
\begin{equation}
\frac{\fpi^2}{4}\langle\chi U^{\dagger}+\chi^{\dagger}U\rangle\rightarrow -\beta'(2Bs_0)\fpi\pi_3\left(1+\alpha_{\rm p}\frac{\pi^2}{\fpi^2}\right)+\cdots\,.
\end{equation}

\subsubsection{Selection for the \boldmath{$\theta$}-term and the hierarchy of coupling constants}
We will now focus on the selection of the ground state in the presence of the $\theta$-term. We start from the expression of the $\theta$-term as a complex phase of the quark-mass matrix given by Eq.\,(\ref{eq:lagtheta}). The correct ground state is obtained by subjecting the matrix $U$ to an axial rotation $A\!=\!\exp(i\tau_3\beta/2)$ with (see Eq.\,(\ref{eq:qcdgs-beta}))
\begin{equation}
\beta=\arctan\left(\frac{p_3}{s_0}\right)=\frac{\bar{\theta}}{2}\epsilon+\mathcal{O}(\bar{\theta}^3)\,,
\end{equation}
where Eq.\,(\ref{eq:s0s3rep}) and Eq.\,(\ref{eq:thetafundrepl}) have been utilized. According to Eq.\,(\ref{eq:corrot3}), this implies a simultaneous shift of the parameter $p_0$ by
\begin{equation}\label{eq:thetap0fin}
p_0=\frac{\bar{\theta}}{2}\bar{m}\rightarrow p_0'= \frac{\bar{\theta}}{2}\bar{m}\left(1-\epsilon^2\right)=\bar{\theta}\frac{m_um_d}{m_u+m_d}\equiv\bar{\theta}m^{\ast}\,,
\end{equation}
where $m^{\ast}$ is the reduced quark mass. The $\theta$-term has thus been rotated into the isospin-conserving component of the quark-mass matrix \cite{Baluni,Mereghetti:2010tp}:
\begin{equation}
\bar{\theta}m^{\ast}\bar{q}i\gamma_5q\,.
\end{equation}

This redefinition of the quark fields also alters the generic set of $P$- and $T$-violating terms in the $\chi$EFT Lagrangian of Subsection \ref{sec:chiefteta} by setting $p_0=\bar{\theta}m^{\ast}$ and $p_3=0$ in Eqs.\,(\ref{eq:thetapi2})-(\ref{eq:theta4n}). Furthermore, the subleading pion-tadpole term in Eq.\,(\ref{eq:thetapitad2}) has to be canceled by a small shift of the leading-order $P$- and $T$-conserving term proportional to $s_0$ in $ \fpi^2 \langle\chi_+\rangle/4$ of Eq.\,(\ref{eq:thetapi2}) that is induced by a second axial rotation of the ground state defined by the angle $\beta'$ (see Eq.\,(\ref{eq:corrot4}) and by utilizing Eq.\,(\ref{eq:s0s3rep})):
\begin{equation}
\beta'(\bar{\theta})=-\frac{4Bl_7s_3p_0}{\fpi^2 s_0}+\cdots=-l_7 (1-\epsilon^2)\epsilon\frac{\mpi^2}{\fpi^2}\btheta+\mathcal{O}(\bar{\theta}^2) \,.
\end{equation}
This shift of $\fpi^2\langle\chi_+\rangle/4$ removes the tadpole term and modifies the term proportional to $\pi_3\pi^2$ in $-l_7\langle\chi_-\rangle^2/16$ (see Eq.\,(\ref{eq:thetapi4}) and Eq.\,(\ref{eq:thetapitad2})) \cite{jbepja}:
\begin{equation}\label{eq:3pithetamod}
\frac{\fpi^2}{4}\langle\chi_+\rangle\rightarrow -\beta'(\bar{\theta})(2Bs_0)\fpi\pi_3\left(1-\frac{\pi^2}{6\fpi^2}\right)+\cdots\,.
\end{equation}
The leading contribution to the $3\pi$ coupling constant $\Delta_3$ in Eq.\,(\ref{eq:impcoup}) induced by the $\theta$-term, $\Delta^{\theta}_3$, is obtained by adding the second term on the right-hand side of Eq.\,(\ref{eq:3pithetamod}) to the $3\pi$ term in Eq.\,(\ref{eq:thetapi4}) \cite{edmoln}:
 \begin{equation}
 \Delta^{\theta}_3=\frac{1}{m_N}\left(\frac{\beta'(\bar{\theta})(2Bs_0)}{6\fpi}+\frac{4l_7(2Bs_3)(2Bp_0)}{3\fpi^3}\right)+\cdots
 =\frac{(\delta\mpi^2)^{\rm str}(1-\epsilon^2)}{4\fpi m_N\epsilon}\,\bar{\theta}+\cdots\,.
 \end{equation}
The relation \cite{GassLeut_eta}
\begin{equation}\label{eq:strpimass}
     \Mpistr \simeq\frac{B}{4}\frac{(m_u{-}m_d)^2}{m_s{-}(m_u{+}m_d)/2}\simeq\frac{\epsilon^2}{4}\frac{\mpi^4}{M_K^2-\mpi^2}
\end{equation}
yields \cite{edmoln}
\begin{equation}
\Delta_3^{\theta}=\frac{ \epsilon (1-\epsilon^2)}{16 F_\pi m_N} \frac{M_\pi^4}{M_K^2 - M_\pi^2} \,\bar\theta+\cdots= ( -0.37\pm 0.09) \cdot 10^{-3}\, \bar\theta\,,
\end{equation}
with the averaged kaon mass $M_K\!=\!494.98$ MeV \cite{PDG}. The prediction for the quark-mass ratio of \cite{Aoki:2013ldr}, $m_u/m_d=0.46\pm0.03$, has been used here to compute $\epsilon$. 

The redefinition of the ground state generates new structures in the pion-nucleon Lagrangian $\mathcal{L}_{\pi N}^{(2)}$ of Eq.\,(\ref{eq:thetapin2}) as pointed out in \cite{Mereghetti:2010tp,jbepja}:
\begin{eqnarray}
    c_1\langle \chi_+\rangle N^\dagger N  &\rightarrow& -4\beta'(\btheta)c_1 \mpi^2 \frac{\pi_3}{\fpi}\left(1-\frac{\pi^2}{6\,\fpi^2}\right) N^\dagger N+\cdots\,,  \label{eq:thetapinc1bp} \\
   c_5 N^\dagger \hat{\chi}_+N  &\rightarrow& -2\beta'(\btheta)c_5 \epsilon \mpi^2  N^\dagger \!\left(\frac{\vec{\pi}\cdot\vec{\tau}}{\fpi} -\frac{(1{-}\epsilon^2)\bar\theta\,\tau_3}{2\epsilon}\!\right) N+\cdots\,. \label{eq:thetapinc5bp}
\label{eq: new structures}
\end{eqnarray}
As stated before, the ellipses denote omitted $P$- and $T$-conserving as well as higher-order $P$- and $T$-violating 
terms.
The terms proportional to the LECs
$c_2$, $c_3$, $c_4$, $c_6$ and $c_7$ in the pion-nucleon Lagrangian of Eq.\,(\ref{eq:pin2}) are invariant under the axial rotation $A$ when the electromagnetic field is the sole external current.
The second term on the right-hand side of Eq.\,(\ref{eq:thetapinc5bp}) is $\mathcal{O}(\bar{\theta}^2)$ and can be disregarded. The leading contributions to the coupling constant $\gzero$ are then obtained by inserting Eq.\,(\ref{eq:thetap0fin}) into Eq.\,(\ref{eq:thetapin2}) and adding Eq.\,(\ref{eq:thetapinc5bp}) \cite{Mereghetti:2010tp,jbepja}:
\begin{equation}\label{eq:thetag0full}
c_5\left(4Bm^{\ast}\bar{\theta}-2\beta'(\btheta) \epsilon \mpi^2\right)N^{\dagger}\frac{\vec{\pi}\cdot\vec{\tau}}{\fpi}N+\cdots\,.
\end{equation}
These terms are proportional to the LEC $c_5$ and thus related to the to the quark-mass-induced part
of the proton--neutron mass difference $ \Mstr$ according to Eq.\,(\ref{eq:dmnp}). The first term in Eq.\,(\ref{eq:thetag0full}) is the dominating contribution and equals \cite{jbepja,edmoln}
\begin{equation}
    \gzero=  \frac{\Mstr (1 -\epsilon^2)}{4 F_\pi \epsilon}\, 
     \btheta+\cdots =( -0.0155 \pm 0.0019)\,\bar\theta \,. 
     \label{gzero_form}
\end{equation}
This expression agrees with \cite{Mereghetti:2010tp}. The second term in Eq.\,(\ref{eq:thetag0full}) constitutes a small correction to $\gzero$  which is given by \cite{jbepja}
\begin{equation}
        \delta\gzero
        = \frac{\Mstr\,(1-\epsilon^2)}{4\fpi\,\epsilon}\,\bar{\theta}\,\frac{ \Mpistr }{\mpi^2}+\cdots
        = \gzero\,\frac{ \Mpistr }{\mpi^2}\,,
\end{equation}
reproducing the corresponding term in Eq.\,(113) in \cite{Mereghetti:2010tp}.

The dominating contribution to $\gone$ of Eq.\,(\ref{eq:impcoup}) is given by Eq.\,(\ref{eq:thetapinc1bp}) and equals at leading order \cite{jbepja,edmoln,Mereghetti:2010tp}
\begin{equation}
  \gone(c_1)=\frac{2\,c_1\,\Mpistr\,(1-\epsilon^2)}{\fpi\,\epsilon}\btheta=8c_1m_N\Delta_3^{\theta}+\cdots\,.
  \label{eq: gone first}
 \end{equation}
Inserting Eq.\,(\ref{eq:strpimass}) and Eq.\,(\ref{eq:c1pred}) into Eq. (\ref{eq: gone first}) yields \cite{jbepja,edmoln}
 \begin{equation}\label{eq:gone3}
\gone(c_1)= \frac{c_1(1-\epsilon^2)\epsilon}{2\fpi} \,\frac{\mpi^4}{M_K^2-\mpi^2}\,\btheta+\cdots
 = (0.0028\pm 0.0011) \,\bar \theta \, ,
\end{equation}
where the uncertainty is dominated by the one of $c_1$. Eq.\,(\ref{eq:gone3}) exactly agrees with the corresponding expression derived from $\eta$--$\pio$ mixing
in \cite{Lebedev_Olive} for a negligible strange-quark content.

In addition to Eq.\,(\ref{eq:thetapinc1bp}), there is an another independent operator structure that contributes to $\gone$ given by Eq.\,(\ref{eq:thetapin41}), which has also been emphasized in Ref. \cite{Mereghetti:2010tp}:
\begin{equation}
e_{40}\langle\hat{\chi}_+\hat{\chi}_+\rangle N^{\dagger}N=e_{40}16(2Bs_3)(2Bp_0)\frac{\pi_3}{\fpi}\left(1-\frac{2\pi^2}{3\fpi^2}\right)N^{\dagger}N + \cdots \ . 
\label{chimi2}
\end{equation} 
We denote this contribution to $\gone$ by $\gone(e_{40})$\footnote{This $\gone$ contributions is denoted by $\tilde{g}_1^{\theta}$ in Ref. \cite{edmoln}.}. Unfortunately, this operator structure in Eq.\,(\ref{chimi2}) contributes
to $P$- and $T$-conserving observables at such a high order that it cannot be constrained
from a study of, say, $\pi N$ scattering. The value of $e_{40}$ has therefore to be estimated differently. The term in Eq.\,(\ref{chimi2}) could have been replaced in Eq.\,(\ref{eq:thetapin41}) by the term $N^{\dagger}\langle\chi_-\rangle^2 N$, which has the same $P$- and $T$-violating structure but does not explicitly appear in Eq.\,(\ref{eq:thetapin41}) since it is not independent of $N^{\dagger}\langle\hat{\chi}_+\hat{\chi}_+\rangle N$. Therefore, the assessment of the contribution of the $e_{40}$ term to $\gone$ can equally be performed by considering $N^{\dagger}\langle\chi_-\rangle^2 N$. The assessment of the size of this term has been done by resonance saturation in \cite{jbepja}, yielding a result for $\gone$ of \cite{edmoln}
\begin{equation}
\gone =\gone(c_1)+\gone(e_{40})+\cdots= (0.0034\pm 0.0015)\, \bar{\theta} \,.
\label{eq: gone second}
\end{equation}
 
In order to summarize our results, we list the numerical results and NDA estimates of leading tree-level contributions to the coupling constants $\gzero$, $\gone$, $\Delta_3^{\theta}$ and $d_{0,1}^{\theta}$, $C_{1,2}^{\theta}$ as defined in Eq.\,(\ref{eq:impcoup}), respectively:
\begin{eqnarray}
\gzero&=&\frac{\delta m_{np}^{\rm str}(1-\epsilon^2)}{4\fpi\epsilon}\bar{\theta}+\cdots=-(0.0155\pm0.0019)\,\bar{\theta}\,,\label{eq:sumg0theta}\\
\gone&=&\frac{2c_1(\delta M_{\pi}^2)^{\rm str}(1-\epsilon^2)}{\fpi \epsilon}\bar{\theta}+\gone(e_{40})+\cdots=(0.0034\pm 0.0015)\,\bar{\theta}\,,\label{eq:sumg1theta}\\
\Delta^{\theta}_3&=&\frac{(\delta\mpi^{2})^{\rm str}(1-\epsilon^2)}{4\fpi m_N\epsilon}\bar{\theta}+\cdots=-(0.00037\pm0.00009)\,\bar{\theta}\,,\label{eq:delta3final}\\
d_0^{\theta}&=&4e_{110}\, e\,  \mpi^2(1-\epsilon^2)\bar{\theta}+\cdots=\mathcal{O}\left(\bar{\theta}e\frac{\mpi^2}{m_N^3}\right)\,,\\
d_1^{\theta}&=&2e_{111}\, e\, \mpi^2(1-\epsilon^2)\bar{\theta}+\cdots=\mathcal{O}\left(\bar{\theta}e\frac{\mpi^2}{m_N^3}\right)\,,\\
C_{1,2}^{\theta}&=&\mathcal{O}\left(\frac{\gzero}{\fpi m_N^2}\right)\sim 2\,\bar{\theta}\cdot 10^{-3}\,{\rm fm}^3\,.\label{eq:sumctheta}
\end{eqnarray}
Here and in the following the elementary charge $e$ is defined to be negative, $e<0$. The NDA estimates for $C_{1,2}^{\theta}$ have been derived in \cite{edmoln} by comparing the $C_{1,2}^{\theta}$ vertices to the \mbox{$\gzero$-induced} two-pion-exchange diagrams. They are related to the isospin-violating pion production in nucleon-nucleon collisions \cite{CSBpn2dpi} and can in principle be deduced from a refined analysis. The above {\it NDA estimates} are in agreement with those in \cite{Mereghetti:2010tp,Maekawa:2011vs}.

\subsubsection{Selection for the qCEDM and the hierarchy of coupling constants}
In the presence of effective dimension-six sources, a few differences to the above discussed selection of the ground state for the $\theta$-term occur which are due to the existence of additional LECs. Since the numerical values of these LECs are in general unknown, quantitative assessments of the pion-nucleon coupling constants are now impossible within the framework of $\chi$EFT alone. Until Lattice QCD might be able to provide numerical values for the new LECs, any estimates of the relevant coupling constants in Eq.\,(\ref{eq:impcoup}) induced by the effective dimension-six sources rely on NDA. As demonstrated in Section \ref{sec:chptqcedm}, the source fields of the qCEDM are $\tilde{p}_0\neq p_0$ and $\tilde{p}_3\neq p_3$. Due to the existence of new separate isospin multiplets, a simultaneous removal of $p_3$ and $\tilde{p}_3$ by an axial rotation $A=\exp(i\tau_3\beta/2)$ is impossible and only the leading pion tadpole is removed from the amended $\chi$EFT Lagrangian. Assuming a vanishing $\theta$-term, one has $p_0=p_3=0$ and of 
course $s_0$ and $s_3$ as in Eq.\,(\ref{eq:s0s3rep}) and $s_1=s_2=p_1=p_2=0$.

The minimization of the potential in Eq.\,(\ref{eq:chpt-pot}) with the generalized mass term of Eq.\,(\ref{eq:dim6pion2}) leads to the ground-state conditions in the presence of the qCEDM, which are obeyed if $U_0=\exp(i\tau_3\beta)$ with
\begin{equation}
\beta=\arctan\left(\frac{C\tilde{p}_3}{B s_0+C\tilde{s}_0}\right)=\frac{C\tilde{p}_3}{B s_0+C\tilde{s}_0}+\cdots\,.
\end{equation}
A chiral rotation $A=\exp(i\tau_3\beta/2)$ of the quark fields or equivalently of the matrix $U=u^2$ in the effective Lagrangian results in corrections of all other terms in the amended QCD and the amended $\chi$EFT Lagrangians. Utilizing Eqs.\,(\ref{eq:corrot1})-(\ref{eq:corrot4}), $\chi$ and $\tilde{\chi}$ are shifted by
\begin{eqnarray}
\tilde{\chi}&\mapsto& \tilde{\chi}'=2C\big(\tilde{s}_0\mathds{1}+\tilde{s}_3\tau_3+i(\tilde{p}_0-\beta \tilde{s}_3)\mathds{1}+i(\tilde{p}_3-\beta \tilde{s}_0)\tau_3\big)+\cdots\,,\label{eq:tilchip}\\
\chi&\mapsto&\chi'=2B\big(s_0\mathds{1}+s_3\tau_3-i\beta s_3\mathds{1}-i\beta s_0\tau_3)+\cdots\,,\label{eq:chip}
\end{eqnarray}
where the ellipses denote terms which are proportional to higher powers of $\tilde{p}_{0,3}$. 

The axial rotation $A$ profoundly affects the pion-sector Lagrangian $\mathcal{L}_{\pi}^{(4)}$ Eq.\,(\ref{eq:dim6pion4}) in the presence of the qCEDM source fields.
Utilizing Eq.\,(\ref{eq:tilchip}) and Eq.\,(\ref{eq:chip}), the next-to-leading-order pion-tadpole term after this axial rotation reads
\begin{eqnarray}
\mathcal{L}^{(4)}_{\pi}&=&\big[l_7(2B\beta s_3)(2Bs_3)-\tilde{l}_7(2C(\tilde{p}_0-\beta\tilde{s}_3))(2C\tilde{s}_3)\nonumber\\
&&-l_7'(2C(\tilde{p}_0-\beta\tilde{s}_3))(2Bs_3)+l_7'(2B\beta s_3)(2C\tilde{s}_3)\nonumber\\
&&+\tilde{l}_3(2C\tilde{s}_0)(2C(\tilde{p}_3-\beta\tilde{s}_0)-l_3(2Bs_0)(2B\beta s_0)\nonumber\\
&&+l_3'(2C(\tilde{p}_3-\beta\tilde{s}_0))(2Bs_0)-l_3'(2B\beta s_0)(2C\tilde{s}_0)\big]\nonumber\\
&&\times 2\frac{\pi_3}{\fpi}\left(1-\frac{2\pi^2}{3\fpi^2}\right)+\cdots\,.
\end{eqnarray}
Since $|\tilde{s}_{0}|\ll |s_0|$ and $|\tilde{s}_3|\ll |s_3|$, the tadpole term essentially reduces to
\begin{eqnarray}\label{eq:lagpi4qcedm}
\mathcal{L}_{\pi}^{(4)}&=&\big[l_7(2B\beta s_3)(2Bs_3)-l_7'(2C\tilde{p}_0)(2Bs_3)\nonumber\\
&&-l_3(2Bs_0)(2B\beta s_0)+l_3'(2C\tilde{p}_3)(2Bs_0)\big]\nonumber\\
&&\times 2\frac{\pi_3}{\fpi}\left(1-\frac{2\pi^2}{3\fpi^2}\right)+\cdots\,.
\end{eqnarray}
This next-to-leading-order tadpole term is removed by subjecting the effective Lagrangian to another axial rotation $A'=\exp(i\tau_3\beta'/2)$ with 
\begin{equation}
\beta'=4\frac{l_7B^2\beta s_3^2-l_7'BC\tilde{p}_0s_3-l_3B^2\beta s_0^2+l_3'BC\tilde{p}_3s_0}{\fpi^2(Bs_0+C\tilde{s}_0)}\,,
\end{equation}
which causes a shift of the leading-order mass terms $\fpi^2\langle\chi_+\rangle/4$ and $\fpi^2\langle\tilde{\chi}_+\rangle/4$ in Eq.\,(\ref{eq:dim6pion2}) in analogy to the $\theta$-term case:
\begin{equation}\label{eq:3pimodqcedm}
\frac{\fpi^2}{4}(\langle\chi_+\rangle+\langle\tilde{\chi}_+\rangle)\rightarrow-\fpi\beta'[(2Bs_0)+(2C\tilde{s}_0)]\pi_3\left(1-\frac{\pi^2}{6\fpi^2}\right)+\cdots\,.
\end{equation}
This second axial rotation $A'$ corresponds to a second axial rotation of the source fields in \mbox{$\chi$ and $\tilde{\chi}$:} 
\begin{eqnarray}
\chi&\mapsto&\chi'=2B(s_0\mathds{1}+s_3\tau_3-i(\beta+\beta')s_3\mathds{1}-i(\beta+\beta')s_0\tau_3)+\cdots\,,\label{eq:tilchip2}\\
\tilde{\chi}&\mapsto&\tilde{\chi}'=2C(\tilde{s}_0\mathds{1}+\tilde{s}_3\tau_3+i(\tilde{p}_0-(\beta+\beta')\tilde{s}_3)\mathds{1}+i(\tilde{p}_3-(\beta+\beta')\tilde{s_0})\tau_3)+\cdots\,.
\label{eq:chip2}
\end{eqnarray}

The coupling constant $\Delta_3$ of the $P$- and $T$-violating $3\pi$ vertex in Eq.(\ref{eq:impcoup}) is then obtained by adding the three-pion term on the right-hand side of Eq.\,(\ref{eq:3pimodqcedm}) to the $3\pi$-term in Eq.\,(\ref{eq:lagpi4qcedm}):
\begin{equation}
\Delta_3=-\frac{\beta'(Bs_0+C\tilde{s}_0)}{\fpi m_N}+\cdots\,.
\end{equation}
By inserting Eq.\,(\ref{eq:tilchip2}) and Eq.\,(\ref{eq:chip2}) into Eq.\,(\ref{eq:dim6nucpion2}), the leading $P$- and $T$-violating $\pi NN$ coupling constants $g_0$ and $g_1$ of Eq.\,(\ref{eq:impcoup}) are obtained in analogy to the $\theta$-term case  ($\hat{\beta}:=\beta+\beta'$):
\begin{equation}
g_0=\frac{4\tilde{c}_5C\tilde{p}_0-4Bc_5s_3\hat{\beta}}{\fpi}+\cdots\,,\quad g_1=\frac{8\tilde{c}_1C\tilde{p}_3-8Bc_1s_0\hat{\beta}}{\fpi} +\cdots\,.
\end{equation}
Similar expression for the coupling constants $d_{0,1}$ and $C_{1,2}^{0,3}$ in Eq.\,(\ref{eq:impcoup}) involve further new LECs (corresponding to {\it e.g.} $e_{110}$ and $e_{111}$ in Eq.\,(\ref{eq:thetapin4}) and to $C_5$-$C_8$ in Eq.\,(\ref{eq:theta4n})) which are not discussed here since they are not required for NDA estimates.

All coupling constants of Eq.\,(\ref{eq:impcoup}) induced by the qCEDM are proportional to the quark mass as in the case of the $\theta$-term since the qCEDM emerges from a coupling to the Higgs field at higher energies \cite{deVries:2012ab,Dekens:2013zca}. Since the qCEDM has an isospin-conserving as well as an isospin-violating component, the only difference to the hierarchy of coupling constants for the $\theta$-term is that $g_0$ and $g_1$ are now induced at the same order and are thus expected to be numerically comparable. The same is true for the three pairs of coupling constants $d_{0,1}$, $C_{1,2}$ and $C_{3,4}$, respectively. Since the $P$- and $T$-violating $3\pi$ vertex arises from an additional insertion of the quark mass ({\it i.e.} and additional insertion of the building block $\chi_{\pm}$), the coupling constant $\Delta_3$ is suppressed by a factor of $\mpi^2/m_N^2$ with respect to $g_0$ and $g_1$. The NDA estimates of the leading tree-level contributions to the coupling constants in Eq.\,(\ref{eq:impcoup}) are then given by
\begin{alignat}{2}
g_0&=g_1=\mathcal{O}\left(c_{\rm g}\frac{\mpi^2}{\fpi m_N}\right)\,,&\Delta_3&=\mathcal{O}\left(c_{\rm g}\frac{\mpi^4}{\fpi m_N^3}\right)\,,\nonumber\\
d_0&=d_1=\mathcal{O}\left(c_{\rm g}\,e\frac{\mpi^2}{m_N^3}\right)\,,\quad
&C_{1,2}&=\mathcal{O}\left(c_{\rm g}\frac{\mpi^2}{\fpi^2 m_N^3}\right)\,,\quad C_{3,4}=\mathcal{O}\left(c_{\rm g}\frac{\mpi^2}{\fpi^2 m_N^3}\right)\,,
\end{alignat}
where $c_{\rm g}$ is a generic dimensionless constant parametrizing BSM physics. It is suppressed by the scale $\Lambda_{{\rm had}}\gtrsim 1$ GeV over the mass scale characteristic for BSM physics. The NDA estimates are in agreement with those in \cite{deVries:2010ah,deVries:2012ab}.

\subsubsection{Selection for the FQLR-term and the hierarchy of coupling constants}
In the case of the FQLR-term, the only non-vanishing $P$- and $T$-violating source field is the one given by Eq.\,(\ref{eq:4qlrsf1}), {\it i.e.} $p_0=p_3=0$ and $r_3\neq0$. The most convenient form of the leading-order ChPT potential $V$ is obtained by utilizing Eq.\,(\ref{eq:4qlrsfdef1}) and Eq.\,(\ref{eq:4qlrbbpe1}):
\begin{alignat}{3}
V&=-\int d^4x&&\bigg(&&\frac{\fpi^2}{4}\langle 2B(s_0\mathds{1}+s_3\tau_3)^{\dagger} U\rangle+\frac{\fpi^2}{4}\langle2B(s_0\mathds{1}+s_3\tau_3) U^{\dagger}\rangle\nonumber\\
&&&\,+&&\,(2Dq_{ij})\langle U\tau_iU^{\dagger}\tau_j+U^{\dagger}\tau_iU\tau_j\rangle+(2Dr_3)\epsilon^{3lm}\langle\tau_lU\tau_mU^{\dagger}\rangle\bigg)\,.
\end{alignat}
By employing the variation of $U$,
\begin{equation}
U(x)\mapsto G(x)U(x)=\exp(i\vec{\tau}\cdot\vec{\alpha}(x))U(x)\,,
\end{equation}  
with $\alpha_1=\alpha_2=0$ and $\alpha_3\neq0$,
the ground-state condition for a non-vanishing FQLR-term is obtained (neglecting terms proportional to $|2Dq_{ij}|\ll |2Bs_0|,|2Bs_3|$):
\begin{equation}
\frac{i\fpi^2Bs_0}{2}\langle\tau_3U-\tau_3U^{\dagger}\rangle+4Dr_3\langle\tau_1U\tau_1U^{\dagger}+\tau_2U\tau_2U^{\dagger}\rangle=0\,.
\end{equation}
This ground-state condition requires the ground state to be $U_0=A^2=\exp(i\tau_3\beta)$ with
\begin{equation}\label{eq:4qlrbeta}
\beta=\frac{8(2Dr_3)}{(2Bs_0)\fpi^2}+\cdots\,.
\end{equation}
The numerator as well as the denominator contain only leading-order terms, and the corresponding rotation will generate leading-order contributions to the relevant coupling constants. This is not true for the shift in the angle $\beta$ when the fourth-order pion-sector Lagrangian is taken into account. The higher-order tadpole term Eq.\,(\ref{eq:lagpi44qlr}) is removed by a second axial rotation $A'=\exp(i\tau_3\beta')$ with $\beta'$ defined by
\begin{equation}
\beta'=\frac{4l_{4qLR}Dr_3}{\fpi^2}+\cdots\,.
\end{equation}
The matrix $U$ in the amended $\chi$EFT Lagrangian has therefore to be subjected to an axial rotation $A'A=\exp(i\tau_3\hat{\beta}/2)$ with $\hat{\beta}\!=\!\beta\!+\!\beta'$ in order to adjust the amended $\chi$EFT Lagrangian to the altered ground state.

The leading-order contribution to the coupling constant of the $P$- and $T$-violating $3\pi$ vertex in Eq.\,(\ref{eq:impcoup}), $\Delta_3$, is given by the sum of the $3\pi$ term in Eq.\,(\ref{eq:4qlrlotad}) and the shift of $\fpi^2\langle\chi_+\rangle/4$ caused by the axial rotation $A$:
\begin{equation}\label{eq:delta4qlr}
\Delta_3=\frac{-32Dr_3+Bs_0\beta\fpi^2}{3\fpi^3m_N}+\cdots=-\frac{8Dr_3}{\fpi^3m_N}+\cdots\,.
\end{equation}
The shift from other $P$- and $T$-conserving terms proportional to $|2Dq_{ij}|\ll|2Bs_0|$ are heavily suppressed and have been neglected here.
The second axial rotation $A'$ generates further contributions to $\Delta_3$, which are two orders suppressed with respect to the one displayed in Eq.\,(\ref{eq:lagpi44qlr}) due to the smallness of $\beta'$. The axial rotations also generate further $P$- and $T$-violating terms in the pion-nucleon Lagrangian $\mathcal{L}_{\pi N}^{(2)}$ from $P$- and $T$-conserving terms proportional to the LECs $c_1$ and $c_5$:
\begin{eqnarray}
c_1\langle\chi_+\rangle N^{\dagger}N&\rightarrow& -4\hat{\beta} c_1\mpi^2\frac{\pi_3}{\fpi}\left(1-\frac{\pi^2}{6\fpi^2}\right)N^{\dagger}N+\cdots\,,\label{eq:4qlrc1}\\
c_5N^{\dagger}\hat{\chi}_+N&\rightarrow& -\frac{2\hat{\beta} c_5\epsilon \mpi^2}{\fpi}N^{\dagger}\vec{\pi}\cdot\vec{\tau}N+\cdots\,.\label{eq:4qlrc5}
\end{eqnarray}
The term on the right-hand side of Eq.\,(\ref{eq:4qlrc1}) constitutes a correction to the dominating contribution to the coupling constant $g_1$ of the $P$- and $T$-violating $\pi_3 NN$ vertex in Eq.\,(\ref{eq:4qlrpin}), and the term on the right-hand side of Eq.\,(\ref{eq:4qlrc5}) is the leading contribution to the coupling constant $g_0$:
\begin{equation}\label{eq:4qlrg1}
g_1=\frac{(16c_{4qLR}Dr_3-4\hat{\beta} c_1\mpi^2)}{\fpi}+\cdots\,,\quad g_0= -\frac{2\hat{\beta} c_5\epsilon \mpi^2}{\fpi}+\cdots\,.
\end{equation}
The dominating contributions to $g_1$ is generated by the first axial rotation. When inserting $\beta$ of Eq.\,(\ref{eq:4qlrbeta}) for $\hat\beta$ in Eq.\,(\ref{eq:4qlrg1}), this contribution turns out to be enhanced by a factor of roughly $8$, such that the contributions proportional to $c_{4qlr}$ can be disregarded. The coupling constants $g_{0,1}$ at leading order then equal \cite{deVries:2012ab,edmoln}
\begin{equation}\label{eq:4qlrg0g1}
g_1=8c_1m_N\Delta_3=(-7.5\pm2.3)\,\Delta_3\,,\quad g_0=\delta m_{np}^{\rm str}m_N\Delta_3/\mpi^2=(0.12\pm0.02)\,\Delta_3\,.
\end{equation}
Disregarded contributions are accounted for by the uncertainties of $g_0$ and $g_1$ along the lines of NDA.

Therefore, the following hierarchy of (leading tree-level contributions to the) coupling constants emerges for the FQLR-term \cite{deVries:2012ab,edmoln}: 
\begin{equation}\label{eq:4qlrsum}
-g_1\sim\frac{m_N}{\mpi}\Delta_3\,>\,\Delta_3\,>\,g_0\sim\frac{\mpi}{m_N}\Delta_3\,,\,\,\,\,\,\, |C_{1,2}|\sim\frac{|g_0|}{\fpi m_N^2}\,<\,|C_{3,4}|\sim\frac{|g_1|}{\fpi m_N^2}\,,\,\,\,\,\,|d_{0,1}|\sim\frac{e\fpi}{m_N^2}\Delta_3\,.
\end{equation}\,
The $\gamma NN$ coupling constants are generated by insertions of $\eta_{-}$ and $f_+^{\mu\nu}$. $C_{3,4}$ vertices are obtained by insertions of the building block $\eta_-$, whereas the coupling constants $C_{1,2}$ are generated by the first axial rotation $A$ in the same manner as the coupling constant $g_0$.

\subsubsection{Selection for the qEDM and the hierarchies of coupling constants} 
The selection of the correct ground states for the qEDM  is performed in analogy to those of the previously considered sources of $P$ and $T$ violation. The effects of the ground-state-selection procedure on the coupling constants of leading $P$- and $T$-violating vertices induced by the qEDM are twofold: the already heavily suppressed coupling constants of vertices without photon fields (by a loop factor of $\alpha_{em}/(4\pi)$) receive corrections of the same order. The analog is true for the leading $\gamma NN$ coupling constants $d_{0,1}$. Since the qEDM has both an isospin-conserving and an isospin-violating component, the effect of the ground state selection procedure is predominantly a mixing of the coupling constants of isospin-conserving vertices and of their isospin-violating counterparts. On the basis of NDA, the hierarchy among coupling constants of the leading $P$- and $T$-violating vertices is then the following: $d_0$ and $d_1$ are the leading coupling constants. $g_0$, $g_1$, $\Delta_3$ and the $4N$ coupling constants $C_{1,2,3,4}$ are suppressed by at least a factor of $\alpha_{em}/(4\pi)$ with respect to $d_{0,1}$. Since the qEDM arises from a coupling to the Higgs field at higher energies, the NDA estimates of all coupling constants of vertices induced by the qEDM also include a factor of $\mpi^2/m_N^2$. These findings are in agreement with the ones in \cite{deVries:2012ab}.

\subsubsection{Selection for the 4q-term and gCEDM and the hierarchies of couplings} 
Again in analogy to the previously considered cases  of $P$ and $T$ violation,
the adjustment to the altered ground state in the presence of the
chiral {\em singlet} sources, gCEDM and the 4q-term, removes the leading pion-tadpole term and the leading $3\pi$ vertex and also causes a mixing of terms of equal order. The $C_{1,2}$ and $d_{0,1}$ coupling constants dominate, while the coupling constants of vertices involving pions, $g_0$, $g_1$ and $\Delta_{3}$, are suppressed by a factor $\mpi^2/m_N^2$ as they are protected by Goldstone's theorem. The NDA estimates of the coupling constants of isospin-violating vertices $C_{3,4}$, $g_1$ and $\Delta_3$ contain an additional factor of $\epsilon$. 
Also these findings are in agreement with the ones in \cite{deVries:2012ab}.

\subsection{Intermediate summary and discussion}\label{sec:intsummary}
In this section, we presented a scheme to derive the effective low-energy Lagrangians induced by arbitrary quark multilinears within the Gasser-Leutwyler formulation of $\chi$EFT explicitly for the considered sources of $P$ and $T$ violation. While Subsection \ref{sec:standardchpt} contained a brief overview of aspects of standard $\chi$EFT required for our analysis, Subsection \ref{sec:ptvlag} was concerned with the derivation of the leading chiral interactions induced by the QCD $\theta$-term, FQLR-term, qCEDM, qEDM, gCEDM and 4q-term (as well as their $P$- and $T$-conserving isospin-multiplet partners). Whereas the QCD $\theta$-term allows for a treatment entirely within standard $\chi$EFT, the effective dimension-six sources induce independent effective terms and require an amendment of standard $\chi$EFT. 

Furthermore, the presence of additional terms which break chiral-symmetry  alters the ground state of (amended) QCD and of (amended) $\chi$EFT, {\it i.e.} the $SU(2)_V$ subgroup to which the chiral \mbox{$SU(2)_L\!\times\!SU(2)_R$} symmetry breaks down. The effective Lagrangians have to be adjusted to the altered ground states for each source of $P$ and $T$ violation separately. These shifts of the generic (amended) $\chi$EFT Lagrangians were elaborated in Subsection \ref{sec:selection}. We briefly summarize the final results for the coupling constants of the leading $P$- and $T$-violating operators defined in Eq.\,(\ref{eq:impcoup}) that are relevant for one important application of our findings, the computation of single-nucleon and light-nuclei EDMs as presented in \cite{edmoln}. The leading hadronic coupling constants induced by the QCD $\theta$-term and the FQLR-term are expressible as functions of only one respective parameter. These functions for the $\theta$-term case are given by Eqs.\,(\ref{eq:sumg0theta})-(\ref{eq:delta3final}) and Eq.\,(\ref{eq:sumctheta}),
\begin{alignat}{2}
\Delta_3^{\theta}&=(-0.37\pm0.09)\cdot10^{-3}\cdot\bar{\theta}\,,\quad&\gzero&=(-15.5\pm 1.9)\cdot 10^{-3}\cdot \bar{\theta}\,,
\nonumber \\
\gone&=(3.4\pm 1.5)\cdot 10^{-3}\cdot\bar{\theta}\,,&|C_{1,2}^{\theta}|&\sim 2\cdot 10^{-3}\,{\rm fm}^3\cdot\bar{\theta}\,,
\end{alignat}
and for the FQLR-term case by Eq.\,(\ref{eq:4qlrg1}),
\begin{equation}
g_0=(0.12\pm0.02)\cdot\Delta_3\,,\quad g_1=(-7.5\pm2.3)\cdot\Delta_3\,,\quad|C_{3,4}|\sim 1\,{\rm fm}^3\cdot\Delta_3\,.
\end{equation}
In the $\theta$-term case, the uncertainty of $\Delta_3^{\theta}$ is driven by the one of the quark-mass-induced pion-mass splitting $(\delta M_{\pi}^2)^{\rm str}$. While $\gone$ also depends on $(\delta M_{\pi}^2)^{\rm str}$, its uncertainty has to be largely attributed to the ones of current estimates of the LEC $c_1$ as well as to the unknown LEC $e_{40}$ in the standard pion-nucleon Lagrangian \cite{MeissnerFettes3}. Even refined knowledge of $c_1$ would thus not significantly decrease the uncertainty of $\gone$. The already relatively small uncertainty of $\gzero$ has two sources, the error estimates of the current predictions of the neutron-proton mass difference, $\delta m_{np}^{\rm str}$, and \mbox{$\epsilon=(m_u-m_d)/(m_u+m_d)$}. The error estimates of the predictions of $c_1$, $\delta m_{np}^{\rm str}$ and $\epsilon$  also determine the uncertainties of $g_{0}$ and $g_{1}$ in the same manner in the case of the FQLR-term.

The coupling constants of leading hadronic operators induced by the other effective dimension-six sources, however, depend on several new and quantitatively unknown LECs which only Lattice QCD might be able to accurately compute. Eq.\,(\ref{eq:impcoup}) encompasses the complete set of hadronic interaction that are relevant to the EDM computation in Section \ref{sec:edmcomp} for each of the $P$- and $T$-violating dimension-four and dimension-six quark terms.   However, the hierarchies of coupling constants is in general different for each of these sources of $P$ and $T$ violation. The final results of our independent derivation of the induced hadronic interactions are consistent with those derived earlier within the Weinberg formulation of $\chi$EFT \cite{Mereghetti:2010tp,deVries:2010ah,deVries:2012ab}. 

In more general terms, this section provides a general and systematic investigation of new, independent quark multilinears and their induced hadronic operators, which can be easily extended to further quark multilinears not discussed here. The chiral structures ({\it i.e.} effective terms module their LECs and source fields) induced by arbitrary non-chiral-singlet quark multilinears are in principle already encountered in the standard $\chi$EFT Lagrangians. The reason for this is that each quark multilinear transforms in general as the sum of basis states of irreducible isospin $O(4)$ representations. Quark multilinears of higher mass dimensions transform as basis states of irreducible representations that are contained in the symmetric tensor products of lower-dimension irreducible representations. The chiral structures in the standard $\chi$EFT Lagrangians reflect the transformation properties of the standard quark multilinears, which transform as the basis states of all lowest-dimensional irreducible $O(4)$ representations. The higher-order terms in the standard $\chi$EFT Lagrangians constitute tensor products of these lowest-dimension irreducible representations. Hence $\chi$EFT already contains all chiral structures that transform as basis states of higher-dimensional irreducible $O(4)$ representations. The terms in the amended $\chi$EFT Lagrangians induced by arbitrary non-chiral-singlet effective quark multilinears can thus  be obtained from the standard $\chi$EFT Lagrangians by an appropriate replacement of LECs and the chiral source fields defined in Eq.\,(\ref{eq:currentsqcd}) as explained in this section explicitly for the sources of $P$ and $T$ violation in Eqs.\,(\ref{eq:lagtheta})-(\ref{eq:4q8}). The new LECs and source fields are in general independent of the conventional ones. This entails that the orders at which particular chiral structures occur are in general different for each effective quark multilinear, {\it e.g.} of lower order in the FQLR-term case, increasing the relevance of higher-order Lagrangian terms in standard $\chi$EFT. The procedure to obtain the amended $\chi$EFT Lagrangians induced by effective {\em chiral-singlet} quark multilinears involves one additional step, the replacement of one $P$- and $T$-conserving structure in a duplicated term by its $P$- and $T$-violating counterpart. Therefore, the significance of the work in \cite{GassLeut1,GassLeut2,GasserSvarc,MeissnerFettes3,Mereghetti:2010tp} is, in general, well beyond {\it standard} QCD and {\it standard} $\chi$EFT.

\section{The EDMs of the deuteron, helion and triton}\label{sec:edmcomp}
The isoscalar and isovector single-nucleon EDMs, {\it i.e.} $d_0$ and $d_1$ in Eq.\,(\ref{eq:impcoup}), receive leading tree-level as well as one-loop contributions for most sources of $P$ and $T$ violation \cite{Ottnad:2009jw,Mereghetti:2010kp,deVries:2010ah,Guo:2012vf,Seng:2014pba}. Eq.\,(\ref{eq:thetapin4}) and its counterparts for the effective dimension-six sources contain these leading tree-level contributions. Since the LECs of the terms in Eq.\,(\ref{eq:thetapin4}) are quantitatively unknown even in the case of the QCD $\theta$-term, the single-nucleon EDMs can not be computed within the framework of $\chi$EFT alone. Supplementary input from {\it e.g.} Lattice QCD is required to numerically assess these LECs \cite{Guo:2012vf,Akan:2014yha}. However, the loop contributions induced by the pion-nucleon vertices in Eq.\,(\ref{eq:impcoup}) determine the momentum dependence of the $P$- and $T$-violating photon-nucleon form factors \cite{Mereghetti:2010kp,deVries:2010ah} (except for the qEDM), such that some of the coefficients of these vertices can be extracted from higher-order electromagnetic moments. 

Especially tree-level nuclear few-body contributions might dominate the EDMs of light nuclei, as pointed out in \cite{Khriplovich:1999qr}, which indicates that $\chi$EFT computations of the EDMs of these systems become feasible for some sources of $P$ and $T$ violation.

The focus of this section is to explain the technique that underlies our computation of the EDM contributions of the deuteron, helion and triton. The results of this computation up-to and including next-to-leading order (\NLO) have already been presented in \cite{edmoln}, in which the nuclear power-counting scheme of \cite{Liebig:2010ki,jbepja,dissertation} was employed. The structure of this section is the following: the set of $P$- and $T$-violating operators relevant up to \NLO\ as listed in \cite{edmoln} is briefly discussed. The scattering equations in the two-nucleon case (Faddeev equations in the three-nucleon case) that have to be solved in order to obtain the EDM contributions are derived subsequently. The results of our computation shown in \cite{edmoln} -- which utilize our findings of sections \ref{sec:ptvlag} and \ref{sec:selection} -- are briefly summarized afterwards.

\subsection{The \boldmath{$P$}- and \boldmath{$T$}-violating form factor}\label{sec:ptvff}
The EDMs of the deuteron, helion and triton, denoted by $d_A$ with $A\!=\!{}^2{\rm H},\he,\hy$, are extracted from their respective $P$- and $T$-violating photon-nucleus form factors $F_3^{A}$. In the Breit-frame, the out-going four-momentum of the photon can be chosen to equal $q^{\mu}=(0,0,0,q)$. The form factors $F_3^{A}$ then depend only on the z-component $q$, and the EDM of the nucleus $A$ is defined by  
\begin{equation}
d_{A}=\lim_{q^2\to 0}\frac{F_3^{A}(q^2)}{2\,m_{A}}\,,\quad -iq\frac{F^{A}_3(q^2)}{2\,m_{A}}=\langle \psi_A;J,J_z=J,P'|\tilde{J}_{\slashed{P}\slashed{T}}^0(q)|\psi_A;J,J_z=J,P\rangle \,.
\end{equation}
Here $m_A$ is the mass of the considered nucleus 
and $\tilde{J}_{\slashed{P}\slashed{T}}^{\mu}(q)$ is the total $P$- and $T$-violating transition current. $J$ denotes the total angular momentum of the nucleus and $J_z$ its $z$-component. The total incoming (outgoing) momentum $P$ ($P'$) 
of the nucleus in the Breit-frame is given by \mbox{$P\!=\!(0,\vec{q}/2)$} (\mbox{$P'=(0,-\vec{q}/2)$}). 

A two-nucleon ($NN$) or three-nucleon ($3N$) operator is called {\em irreducible}
if one of the following two criteria holds~\cite{weinbergpiNN}:  (i) either it does not admit a decomposition into a product of operators containing the free $P$- and $T$-conserving $NN$ or $3N$ propagator $G_0$ or (ii)  the moduli of all   {\em energy denominators} of the free $NN$ or $3N$ propagators, as determined in time-dependent perturbation theory for cases where (i) fails, are of the order $M_\pi$ 
or larger.\footnote{In fact, the EDM contributions resulting from operators according to
the irreducibility criterion (ii)  are chirally suppressed by at least 
two orders~\cite{edmoln}, because loop diagrams are 
necessarily involved --- {\it cf.} Eq.\.(\ref{eq:power_counting}).
Moreover, in the deuteron case  these  EDM contributions 
either cancel against their counter parts involving crossed $NN$  terms 
or contribute only at next-to-next-to-next-to-leading order (\NthreeLO)~\cite{jbepja}.}
Since the full $P$- and $T$-conserving $NN$ or $3N$ propagator $G$ can be written as a series in which each term contains at least one factor of $G_0$, any product of operators involving $G$ is also reducible by default. Let $J_{PT}^{\mu}$ and $J_{\slashed{P}\slashed{T}}^{\mu}$ denote the irreducible $P$- and $T$-conserving and $P$- and $T$-violating $NN$ or $3N$ transition-current operators, respectively. The total $P$- and $T$-violating transition-current operator -- which is a sum of reducible {\it and} irreducible operators -- can be written as
\begin{equation}\label{eq:tcurrpt}
\tilde{J}_{\slashed{P}\slashed{T}}^{\mu}=J_{\slashed{P}\slashed{T}}^{\mu}+V_{\slashed{P}\slashed{T}}\,G\,J_{PT}^{\mu}+J^{\mu}_{PT}\,G\,V_{\slashed{P}\slashed{T}}+\cdots\,.
\end{equation}
Here $V_{\slashed{P}\slashed{T}}$ is the irreducible $NN$ or $3N$ $P$- and $T$-violating potential operator and the ellipses denote terms with more than one irreducible $P$- and $T$-violating operator. Since $P$ and $T$ violation constitutes a small perturbative effect, only products with one power of $V_{\slashed{P}\slashed{T}}$ need to be considered.

The nucleons in a nucleus $A$ can be labelled by an index $i=1,2(,3)$. An operator acting on one, two or three particular nucleons can be assigned one, two or three indices, respectively. There are only a few irreducible $P$- and $T$-violating operators relevant for this work. The irreducible transition-current operators up to \NLO\  are given by \cite{BKM1995,edmoln}
\begin{equation}
J_{PT}^{\mu}=-\sum_{i=1}^{N}\frac{e}{2}\left(\mathds{1}_{(i)}+\tau^3_{(i)}\right)v^{\mu}+\cdots\,,\quad J_{\slashed{P}\slashed{T}}^{\mu}=-\sum_{i=1}^{N}i\left(d_0\mathds{1}_{(i)}+d_1\tau^3_{(i)}\right)\,\vec{\sigma}_{(i)}\cdot\vec{q}\,v^{\mu}+\cdots\,,
\end{equation}
where $e<0$ is the elementary charge as in Appendix~A of Ref.\,\cite{BKM1995}, $N=1,2,3$ is the number of nucleons in a nucleus $A$ and the ellipses stand for higher-order operators in the nuclear power counting.
As explained in \cite{deVries:2011an,jbepja}, other irreducible \mbox{$P$- and $T$-conserving} and $P$- and $T$-violating transition-current operators only contribute at orders beyond \NLO.

\begin{figure}[t!]  
	\centering
 \includegraphics[width=1.0\textwidth]{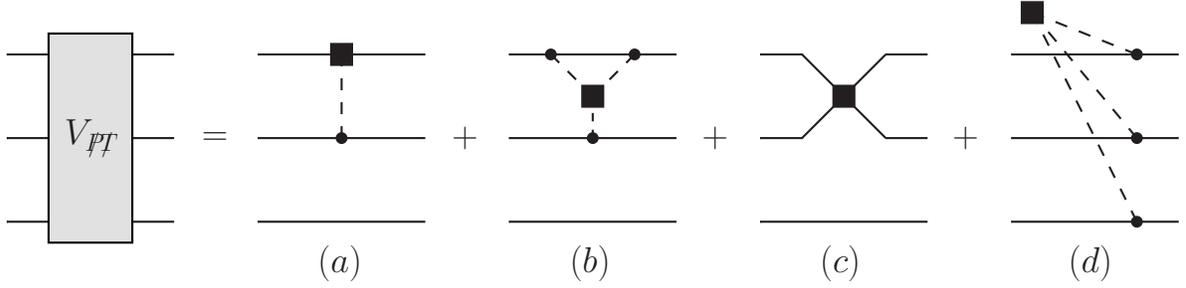}
   \caption{Contributions to the irreducible $P$- and $T$-violating potential operator $V_{\slashed{P}\slashed{T}}$. Only one representative for each diagram class is shown. Continuous lines denote nucleons, while dashed lines stand for pions. A $P$- and $T$-violating vertex is depicted by a black box. A $P$- and $T$-conserving vertex is pictured as a black dot. Only the irreducible $NN$ potential operators (a), (b) and (c) are relevant for the computation of the deuteron EDM.
    \label{fig:ptvpot}}
\end{figure}
Let $V_{\slashed{P}\slashed{T},ij}^{NN}$ denote the completely symmetrized irreducible $P$- and $T$-violating $NN$ potential operator in the (sub)system of nucleons $i$ and $j$, $i < j$, while $V_{\slashed{P}\slashed{T}}^{3N}$ is the completely symmetrized irreducible $P$- and $T$-violating $3N$ potential operator. The irreducible $P$- and $T$-violating potential operators in the $NN$ and $3N$ systems are given by
\begin{equation}
NN\,\text{system}:\,V_{\slashed{P}\slashed{T}}=\sum_{i,j=1,\,i<j}^{N=2}V_{\slashed{P}\slashed{T},ij}^{NN}=V_{\slashed{P}\slashed{T},12}^{NN}\,,\quad 3N\,\text{system}:\,V_{\slashed{P}\slashed{T}}=\sum_{i,j=1,\,i<j}^{N=3}V_{\slashed{P}\slashed{T},ij}^{NN}+V_{\slashed{P}\slashed{T}}^{3N}\,.
\end{equation}
The leading contributions to $V_{\slashed{P}\slashed{T},ij}^{NN}$ and $V_{\slashed{P}\slashed{T}}^{3N}$ induced by vertices in Eq.\,(\ref{eq:impcoup}) are the one-pion-exchange diagram depicted in Fig.\,\ref{fig:ptvpot}~(a), the $P$- and $T$-violating contact interaction pictured in Fig.\,\ref{fig:ptvpot}~(c) and the diagram involving the $P$- and $T$-violating $3\pi$ exchange shown in Fig.\,\ref{fig:ptvpot}~(d), respectively. As pointed out in \cite{deVries:2012ab}, the $g_1$ $\pi NN$ vertex is corrected by the $\pi$-enhanced triangular one-loop diagram depicted in Fig.\,\ref{fig:ptvpot}(b). This correction is denoted by $\Delta_3\, f_{g_1}(|\vec{k}|)$, where the function $f_{g_1}$ is defined by \cite{deVries:2012ab,edmoln}
\begin{equation}
f_{g_1}(|\vec{k}|) \equiv \, -\frac{15}{32}\frac{g_A^2\mpi m_N}{\pi\fpi^2}\left[1+ \left(\frac{1+2\vec{k}^{\,2}/(4\mpi^2)}{3 |\vec{k}\,|/(2\mpi)}\arctan\left(\frac{|\vec{k}\,|}{2\mpi}\right)-\frac{1}{3}\right)\right]\,.
\end{equation}
If the nucleon of the $P$- and $T$-violating $\pi NN$ vertex
is labelled by the index $i$, the vector $\vec{k}_i=\vec{p}_i-\vec{p}\,'\!\!_i$ is the difference of the incoming momentum $\vec{p}_i$ and outgoing momentum $\vec{p}\,'\!\!_i$ of the nucleon. 

Utilizing the definitions $\vec{\sigma}_{(ij)}^{\pm}:=\vec{\sigma}_{(i)}\pm\vec{\sigma}_{(j)}$ and $\tau_{(ij)}^{\pm}:=\tau^3_{(i)}\pm\tau^3_{(j)}$ with $i\!\neq\! j$, the operators $V_{\slashed{P}\slashed{T},ij}^{NN}$ and  $V_{\slashed{P}\slashed{T}}^{3N}$ read \cite{edmoln}:
\begin{eqnarray}\label{eq:ptvpotnn}
V_{\slashed{P}\slashed{T},ij}^{NN}(\vec{k}_{i})=&&i\frac{g_A}{2\fpi}\frac{\vec{k}_{i}}{\vec{k}_{i}^{\,2}+\mpi^2}\cdot\,g_0\,\vec{\sigma}_{(ij)}^{-}\vec{\tau}_{(i)}\cdot\vec{\tau}_{(j)}\nonumber\\
&+&i\frac{g_A}{4\fpi}\frac{\vec{k}_{i}}{\vec{k}_{i}^{\,2}+\mpi^2}\cdot 
\left[g_1+\Delta_3\, \,f_{g_1}(|\vec k_i|) \right](\vec{\sigma}_{(ij)}^{+}\tau_{(ij)}^{-}+\vec{\sigma}_{(ij)}^{-}\tau_{(ij)}^{+})\nonumber\\
&+&\frac{i}{2}\frac{\beta^2\mpi^2\vec{k}_{i}}{\vec{k}_{i}^{\,2}+\beta^2\mpi^2}\cdot 
\Bigl [C_1\,\vec{\sigma}_{(ij)}^{-}+C_2\,\vec{\sigma}_{(ij)}^{-}\vec{\tau}_{(i)}\cdot\vec{\tau}_{(j)}\Bigr]+\cdots\,,
\end{eqnarray}
and \cite{deVries:2012ab,edmoln}
\begin{eqnarray}\label{eq:ptvpot3n}
V_{\slashed{P}\slashed{T}}^{3N}(\vec{k}_1,\vec{k}_2)=&&-i\frac{\Delta_3\,m_Ng_A}{4\fpi^3}(\delta^{ab}\delta^{c3}+\delta^{ac}\delta^{b3}+\delta^{bc}\delta^{a3})\,\tau_{(1)}^{a}\tau_{(2)}^{b}\tau_{(3)}^{c}\nonumber\\
&&\times\frac{(\vec{\sigma}_{(1)}\cdot\vec{k}_{1})(\vec{\sigma}_{(2)}\cdot\vec{k}_{2})(\vec{\sigma}_{(3)}\cdot\vec{k}_{3})}{[\vec{k}_1^{\,2}+\mpi^2][\vec{k}_2^{\,2}+\mpi^2][\vec{k}_3^{\,2}+\mpi^2]}+\cdots\,.
\end{eqnarray}
The $\beta$-dependent term in Eq.\,(\ref{eq:ptvpotnn}) is a  cutoff function introduced during the numerical computation. This cutoff is subsequently removed by taking the limit $\beta\to \infty$.\footnote{Details of the analysis of the short-range EDM contributions are given in \cite{edmoln}. The results that will be  presented below have been computed for $\beta=49$.} It only serves as a tool to compare our EDM results based on $\chi$EFT potentials with those obtained from $P$- and $T$-conserving phenomenological potentials. The ellipses again denote higher-order operators such as two-pion-exchange diagrams, which contribute at orders that higher than those taken into consideration here (see \cite{deVries:2011an,jbepja} for details).

\subsection{Derivation of scattering and Faddeev equations}\label{sec:numana}

Let $|\psi_{A}\rangle$ denote the state of a considered nucleus $A$ and $\tilde{J}^{\mu}_{\slashed{P}\slashed{T}}(q)$ denote the reducible complete $P$- and $T$-violating transition-current operator of Eq.\,(\ref{eq:tcurrpt}). The matrix element to compute the photon-nucleus form factor $F_3^{A}(q^2)$ of a nucleus $A$ is given by
\begin{equation}\label{eq:tcurrptme}
\langle\psi_A|\tilde{J}_{\slashed{P}\slashed{T}}^{\mu}|\psi_A\rangle=\langle\psi_A |(J_{\slashed{P}\slashed{T}}^{\mu}+V_{\slashed{P}\slashed{T}}\,G\,J_{PT}^{\mu}+J^{\mu}_{PT}\,G\,V_{\slashed{P}\slashed{T}}+\cdots)|\psi_A\rangle\,.
\end{equation}
The first matrix element on the right-hand side of Eq.\,(\ref{eq:tcurrptme}) is straightforward to calculate. In order to compute the second and the third matrix element, consider the $P$- and $T$-conserving irreducible transition current $J_{PT}^{\mu}(q)$ acting on the nucleus state $|\psi_{A}\rangle$. $J^{\mu}_{PT}(q)|\psi_A\rangle$ constitutes the initial ket state of particles here, {\it i.e.} the starting seed in the iteration procedure to be described below.

The full $P$- and $T$-conserving propagator $G$ can be rewritten for the $NN$ system in terms of the $NN$ T-matrix $t$ as $G=G_0+G_0\,t\,G_0$ with $G_0=(E-H_0+i\epsilon)^{-1}$. The second term on the right-hand side of eq\,(\ref{eq:tcurrptme}) then becomes for the deuteron case
\begin{equation}\label{eq:h2num1}
\langle\psi_{{}^2\!{\rm H}}|V_{\slashed{P}\slashed{T}}\,G\,J^{\mu}_{PT}(q)|\psi_{{}^2\!{\rm H}}\rangle=\langle\psi_{{}^2\!{\rm H}}|\left(V_{\slashed{P}\slashed{T}}\,G_0\,J^{\mu}_{PT}(q)+V_{\slashed{P}\slashed{T}}\,G_0\,t\,G_0\,J^{\mu}_{PT}(q)\right)|\psi_{{}^2\!{\rm H}}\rangle\,.
\end{equation}
The third matrix element on the right-hand side of Eq.\,(\ref{eq:tcurrptme}) yields an identical contribution to the form factor $F_3^{A}$.

The computation of the second matrix element in Eq.\,(\ref{eq:tcurrptme}) for the $3N$ system of the helion (and analogously for the triton) is more intricate and leads to Faddeev equations. Let $V$ denote the irreducible $P$- and $T$-conserving potential operator. Inserting the resolvent identity for $G$, {\it i.e.}
\begin{equation}\label{eq:residty}
G=G_0+G_0\,V\,G\,,
\end{equation}
into the second matrix element on the right-hand side of Eq.\,(\ref{eq:tcurrptme}) yields
\begin{eqnarray}\label{eq:numedmhe3}
\langle \psi_{\he}|V_{\slashed{P}\slashed{T}}\,G\,J_{PT}^{\mu}(q)|\psi_{\he}\rangle=\langle\psi_{\he}|\left(V_{\slashed{P}\slashed{T}}G_0+V_{\slashed{P}\slashed{T}}\,G_0\,V\,G\right)J^{\mu}_{PT}(q)|\psi_{\he}\rangle\,.
\end{eqnarray}
The irreducible $P$- and $T$-conserving potential $V$ for the $3N$ system comprises $NN$ and $3N$ interactions. Let $V^{ij}_{2N}$ denote the completely symmetrized $P$- and $T$-conserving $NN$-interaction of nucleons $i$ and $j$ for $i,j=1,2,3$ and $i < j$. The $3N$ interaction of nucleons can be decomposed into three parts with each of them being symmetric under an exchange of two nucleons,
\begin{equation}
V=V^{12}_{2N}+V^{(3)}_{3N}+V^{23}_{2N}+V^{(1)}_{3N}+V^{13}_{2N}+V^{(2)}_{3N}\,.
\end{equation}
The $3N$ potential operators $V^{(i)}_{3N}$ are defined by
\begin{equation}
P_{23}\,V^{(1)}_{3N}\,P_{23}=V^{(1)}_{3N}\,,\quad P_{13}\,V^{(2)}_{3N}\,P_{13}=V^{(2)}_{3N}\,,\quad P_{12}\,V^{(3)}_{3N}\,P_{12}=V^{(3)}_{3N}\,.
\end{equation}
The nucleon transposition operator $P_{ij}$ with $i\neq j$ transposes nucleon $i$ and nucleon $j$ in the \mbox{$3N$ system}. The potential $V$ can then be re-expressed solely in terms of $V^{(12)}_{2N}$, $V^{(3)}_{3N}$ and these transposition operators:
\begin{equation}\label{eq:vdecomp}
V=V_{2N}^{12}+V_{3N}^{(3)}+P_{12}P_{23}\,(V_{2N}^{12}+V_{3N}^{(3)})\,P_{23}P_{12}+P_{13}P_{23}\,(V_{2N}^{12}+V_{3N}^{(3)})\,P_{23}P_{13}\,.
\end{equation}

The permutation operators commute with the full $3N$ propagator $G$ and do not alter the initial ket state \mbox{$J^{\mu}_{PT}(q)|\psi_{\he}\rangle$}:
\begin{equation}
P_{23}P_{12}\,G\,J^{\mu}_{PT}(q)|\psi_{{}^3{\rm He}}\rangle=P_{23}P_{13}\,G\,J^{\mu}_{PT}(q)|\psi_{{}^3{\rm He}}\rangle=G\,J^{\mu}_{PT}(q)|\psi_{{}^3{\rm He}}\rangle\,.
\end{equation}
By inserting Eq.\,(\ref{eq:vdecomp}) into Eq.\,(\ref{eq:numedmhe3}), the second term on the right-hand side  becomes (modulo $V_{\slashed{P}\slashed{T}}G_0$)
\begin{eqnarray}
V\,G\,J^{\mu}_{PT}(q)|\psi_{{}^3{\rm H}}\rangle&=&(\mathds{1}+P_{12}P_{23}+P_{13}P_{23})(V_{2N}^{12}+V_{3N}^{(3)})GJ^{\mu}_{PT}(q)|\psi_{{}^3{\rm He}}\rangle\nonumber\\
&\equiv&(\mathds{1}+P)(V_{2N}^{12}+V_{3N}^{(3)})GJ^{\mu}_{PT}(q)|\psi_{{}^3{\rm He}}\rangle\,,
\end{eqnarray}
where $P$ is defined by \mbox{$P=P_{12}P_{23}+P_{13}P_{23}$}, such that \mbox{$\mathds{1}+P$} is the nucleon anti-symmetrization operator. This allows us to define the Faddeev component $|U^{(3)}\rangle$ by
\begin{equation}
V\,G\,J^{\mu}_{PT}(q)\,|\psi_{{}^3{\rm He}}\rangle\equiv(\mathds{1}+P)\,|U^{(3)}\rangle\,.
\end{equation}
A concise introduction into Faddeev equations can be found in \cite{Gloeckle:1995jg}. The Faddeev component $|U^{(3)}\rangle$ obeys the equation
\begin{eqnarray}
|U^{(3)}\rangle&=&(V_{2N}^{12}+V_{3N}^{(3)})GJ^{\mu}_{PT}(q)|\psi_{{}^3{\rm He}}\rangle\nonumber\\
&=&(V_{2N}^{12}+V_{3N}^{(3)})G_0J^{\mu}_{PT}(q)|\psi_{{}^3{\rm He}}\rangle\nonumber\\
&&+(V_{2N}^{12}+V_{3N}^{(3)})G_0(\mathds{1}+P)(V_{2N}^{12}+V_{3N}^{(3)})GJ^{\mu}_{PT}(q)|\psi_{{}^3{\rm He}}\rangle\nonumber\\
&=&(V_{2N}^{12}+V_{3N}^{(3)})G_0J^{\mu}_{PT}(q)|\psi_{{}^3{\rm He}}\rangle+(V_{2N}^{12}+V_{3N}^{(3)})G_0(\mathds{1}+P)|U^{(3)}\rangle\,,
\end{eqnarray}
where the resolvent identity Eq.\,(\ref{eq:residty}) has been utilized. However, this equation does not have a compact kernel. Since the term $V_{2N}^{12}G_0|U^{(3)}\rangle$ contains a $\delta$-function for nucleon (3), the kernel is not fully connected even after a finite number of iterations. The troubling term $V_{2N}^{12}G_0|U^{(3)}\rangle$ can be re-summed independently by the application of an appropriately chosen operator to this equation. We therefore write
\begin{eqnarray}\label{eq:fadint2}
(\mathds{1}-V_{2N}^{12}G_0)|U^{(3)}\rangle&=&(V_{2N}^{12}+V_{3N}^{(3)})G_0J^{\mu}_{PT}(q)|\psi_{{}^3{\rm He}}\rangle+V_{2N}^{12}G_0P|U^{(3)}\rangle\nonumber\\
&&+V_{3N}^{(3)}\,G_0(\mathds{1}+P)|U^{(3)}\rangle\,,
\end{eqnarray}
and exploit the identity
\begin{equation}
(\mathds{1}+t_{12}G_0)(\mathds{1}-V_{2N}^{12}G_0)=\mathds{1}\,.
\end{equation}
The operator $t_{12}$ is the T-matrix in the subsystem of the nucleons labelled by 1 and 2. 
With the help of a multiplication of Eq.\,(\ref{eq:fadint2}) by $(\mathds{1}+t_{12}G_0)$, the Faddeev component $|U^{(3)}\rangle$ is found to obey
\begin{eqnarray}
|U^{(3)}\rangle&=&t_{12}G_0J^{\mu}_{PT}(q)|\psi_{{}^3{\rm He}}\rangle+(\mathds{1}+t_{12}G_0)V_{3N}^{(3)}G_0J^{\mu}_{PT}(q)|\psi_{{}^3{\rm He}}\rangle\nonumber\\
&&+t_{12}G_0P|U^{(3)}\rangle+(\mathds{1}+t_{12}G_0)V_{3N}^{(3)}G_0(\mathds{1}+P)|U^{(3)}\rangle\,,
\end{eqnarray}
which has a connected kernel upon iteration and, therefore, has a well-defined solution. This Faddeev equation has to be solved for the binding energy of the helion (and triton) but for negative parity states. This implies that singularities do not have to be considered here.

\subsection{Results of the numerical EDM computation}\label{sec:results}
The computation of the EDMs of the deuteron, helion and triton with phenomenological $P$- and $T$-conserving wave functions have been performed by various groups over the past years \cite{Khriplovich:1999qr,Liu_Timmermans,Lebedev_Olive,stetcu,Afnan:2010xd,deVries:2011re,deVries:2011an,song,jbepja}. However, the application of 
phenomenological potentials does not lead to results with well-defined uncertainties. Our EDM computations for light nuclei are the first {\em complete} and {\em consistent} computations within $\chi$EFT. This treatment enabled us to provide EDM results with well-defined uncertainties \cite{edmoln}.

The EDMs of the deuteron, helion and triton have been computed  in  \cite{edmoln} by using the next-to-next-to-leading-order (\NtwoLO) $\chi$EFT potentials of \cite{chiralpotentials2,chiralpotentials,Epelbaum:2008ga} for the $P$- and $T$-conserving component of the nuclear potential. As explained in detail in \cite{chiralpotentials2,chiralpotentials,Epelbaum:2008ga}, the regularization of the $P$- and $T$-conserving $\chi$EFT potential requires two kinds of cutoffs:  (i) for loop contributions to the potential, {\it e.g.} the two-pion exchange $NN$ potential, the \textit{Spectral Function Regularization} scheme is employed with a cutoff $\Lambda_{\rm SFR}$ to render the expressions finite. 
(ii) Divergences of the Lippmann-Schwinger equation are removed by the application of an additional cutoff function parametrized by $\Lambda_{\rm LS}$, see \cite{chiralpotentials2,chiralpotentials,Epelbaum:2008ga} for details.
The five cutoff combinations usually chosen are given by \cite{chiralpotentials2,chiralpotentials,Epelbaum:2008ga,edmoln}
\begin{equation}\label{eq:cutoffs}
 (\Lambda_{{\rm LS}},\Lambda_{{\rm SFR}})=\Bigl\{ (0.45,0.5);  (0.6,0.5); 
 (0.55,0.6); (0.45,0.7);
 (0.6,0.7) \Bigr\}\, {\rm GeV}\,.
 \end{equation}
 
The application of $\chi$EFT  to nuclear systems in the $P$- and $T$-conserving sector necessarily requires a non-perturbative 
approach. For the calculations here, we used a perturbatively constructed chiral potential  and 
employ it non-perturbatively as originally suggested by Weinberg \cite{Weinberg:1991um}. Since the iteration of such 
interactions introduces new divergences that cannot be absorbed by the counter terms available \cite{Nogga:2005hy}, 
there has been a vivid discussion on the correct application of $\chi$EFT  to nuclear systems 
\cite{PavonValderrama:2005wv, Epelbaum:2006pt,Birse:2005um}.
Valderrama showed that the problem can be solved by adding counter terms to the leading order and treating higher-order corrections perturbatively \cite{Valderrama:2011hz,Valderrama:2011hw}. Numerically,  however, 
the results obtained in this way 
are  similar as long as cutoffs less than approximately 0.5\,GeV are used for the chiral potentials. Furthermore, a  non-perturbative 
treatment of all orders  is technically much simpler. Therefore, we followed here the traditional non-perturbative approach which 
necessarily constrains the range of cutoffs used in the calculations. However, within this range of cutoffs, several LECs 
of the $P$- and $T$-conserving \NtwoLO\ chiral potential change sign. Therefore, we are confident that the variation is large enough 
for reliable uncertainty estimates of the $P$- and $T$-violating contributions. 
Alternatively, error estimates could also be performed by comparing results of different orders of the expansion
\cite{Epelbaum:2014efa} or adding higher-order contributions  that are only  constrained by naturalness of the 
LECs to account for  theoretical uncertainties  \cite{Furnstahl:2014xsa}.  This goes beyond the scope of this work, but using these ideas 
will be interesting in future to confirm the estimates obtained here. 

In the $P$- and $T$-violating sector, the expansion is strictly perturbative and we do not have to deal with similar issues. 
Nevertheless, a careful analysis is required to estimate the sizes of short-range contributions.  
It has recently been found that naive dimensional analysis of perturbative short-range contributions 
to transition-current operators is not sufficient for a reliable estimate of their contributions \cite{Valderrama:2014vra}. It was shown there
that the short-distance behavior of the nuclear wave functions will be different to the one of plane-wave states which 
in some cases lead to an enhancement. Below we will report explicit calculations for these short-range terms showing that 
their contribution can be quite model- and cutoff-dependent. The explicit calculation will allows us to avoid 
a naive dimensional 
analysis and to obtain a more reliable assessment of these contributions.  We will argue that estimates based on some of the phenomenological 
models severely underestimate the contribution of short-range operators since they lead to strongly suppressed wave functions 
at short distances (see Appendix of \cite{edmoln}). 
As emphasized above, we are confident that the chiral potentials,
even for the limited range of cutoffs, allow 
for reliable estimates of these contributions.

The results of our numerical computation have been presented and discussed in \cite{edmoln}.  Each contribution obtained by employing the $P$- and $T$-conserving \NtwoLO\ $\chi$EFT potential is chosen as the center value of the interval defined by the results for the five cutoff combinations. The uncertainty, to which we refer as the {\it nuclear} uncertainty, is given by one half of the width of the interval.\footnote{Based on the new series of chiral nucleon-nucleon interactions recently published in \cite{Epelbaum:2014efa}, it will be interesting to obtain in future more accurate EDM predictions for the different chiral orders considered in \cite{Epelbaum:2014efa}.} The EDMs of the deuteron, helion and triton as functions of the coefficients defined in Eq.\,(\ref{eq:impcoup}) up to and including \NLO\ equal \cite{edmoln} ($e<0$)
\begin{eqnarray}
d_{^2{\rm H}} &=&
{(0.939\pm 0.009)}(d_n + d_p) - \bigl [ (0.183 \pm 0.017) \,g_1
- (0.748\pm 0.138) \,\Delta_3\bigr] \,e \,{\rm fm} \,\,,
 \label{eq:h2edm} 
 \\
d_{{}^3{\rm He}} &=& (0.90 \pm 0.01)\, d_n-(0.03\pm 0.01 )\,d_p\nn \\
&&\mbox{}+
\bigl\{ (0.017\pm0.006)\,\Delta_3 - (0.11\pm 0.01)\,g_0  -(0.14\pm 0.02)\,g_1  +(0.61\pm 0.14)\,\Delta_3\nn\\
&&\quad\mbox{}+ \left[ (0.04\pm 0.02)C_1 -(0.09\pm 0.02)C_2 \right]\times  {\rm fm}^{-3} \bigr\}
\, e\, {\rm fm} \,, \label{eq:he3edm}
\\
d_{{}^3{\rm H}}&=& -(0.03\pm0.01)\, d_n + (0.92\pm 0.01)\,d_p \nn \\
&&\mbox{}+
\bigl\{  (0.017\pm0.006)\,\Delta_3 + (0.11\pm 0.01)\,g_0  - (0.14\pm 0.02)\,g_1\mbox{} +(0.60\pm 0.14)\,\Delta_3 \nn\\
&&\quad\mbox{}- \left[(0.04 \pm 0.02) C_1 -(0.09\pm 0.02) C_2\right] \times  {\rm fm}^{-3} \bigr\}
\, e\, {\rm fm}\,, \label{eq:h3edm}
\end{eqnarray}
where $d_{n,p}=(d_0\mp d_1)/2$ are the complete neutron/proton EDMs, respectively.\footnote{Note the altered weight factor (including nuclear uncertainty) of 
the total single-nucleon contribution to the deuteron EDM as compared to the one of Ref.\,\cite{edmoln}. In fact, the deviation
of this weight factor from unity, resulting from the subtraction of the small 
$D$-wave contribution of the deuteron,
was first observed in Ref.\,\cite{Yamanaka:2015qfa}.}
The potential operators in Eq.\,(\ref{eq:ptvpotnn}) proportional to $g_0$ and $C_{1,2}$ vanish in the deuteron case due to isospin selection rules \cite{Liu_Timmermans}. It should be stressed that Eqs.\,(\ref{eq:h2edm})-(\ref{eq:h3edm}) are the results of a model-independent computation. While the predictions involving the coefficient $\Delta_3$ are novel and the $C_{1,2}$ contributions exceed the ones based on phenomenological $P$- and $T$-conserving potentials by more than roughly one order in magnitude, the single-nucleon and $g_{0,1}$-induced contributions  are largely consistent with those in \cite{Khriplovich:1999qr,Liu_Timmermans,Lebedev_Olive,stetcu,Afnan:2010xd,deVries:2011re,deVries:2011an,song,jbepja} based on phenomenological $P$- and $T$-conserving potentials. The contributions induced by the $C_{1,2}$ vertices are highly model-dependent. The differences between the $C_{1,2}$ contributions in Eqs.\,(\ref{eq:he3edm})-(\ref{eq:h3edm}) and the corresponding ones from phenomenological potentials can  be attributed to an artificially enhanced short-range repulsion in some of the latter cases,  as mentioned in \cite{edmoln}. However, whereas our results for the $g_{0,1}$ contributions to the helion and triton EDMs are in agreement with those of \cite{song}, they are a factor of two smaller than those in \cite{stetcu,deVries:2011an}. The reason for this discrepancy is currently unknown.

For completeness we also present our results for the leading EDM contributions induced by the $C_3$ and $C_4$ vertices in Eq.\,(\ref{eq:impcoup}) and the isospin-two vertex
\begin{equation}\label{eq:g2c34}
g_2N^{\dagger}\left(3\tau_3\pi_3-\vec{\pi}\cdot\vec{\tau}\right)N\,.
\end{equation}
The leading $NN$ operator induced by the $g_2$ vertex is the iso-tensor component of one-pion-exchange potential, while those induced by the $C_{3,4}$ vertices correspond to the $P$- and $T$-violating and isospin-violating $NN$ contact terms. The $g_2$-induced potential operator does not contribute to the deuteron EDM due to isospin selection rules. The leading EDM contributions solely induced by $g_2$ and $C_{3,4}$ read ({\it cf.} \cite{edmoln} for the $C_{3,4}$ results, $e<0$)
\begin{eqnarray}
\Delta d_{^2{\rm H}} &=&-(0.05\pm 0.05)\, (C_3-C_4)\,e\,{\rm fm}^{-2}\,,\label{eq:g2h2}\\
\Delta d_{{}^3{\rm He}} &=&-(0.238\pm0.026) \,g_2\,e\,{\rm fm}+[(0.04\pm0.03)\, C_3-(0.07\pm0.03)\,C_4]\,e\,{\rm fm}^{-2}\,,\label{eq:g2he3}\\
\Delta d_{{}^3{\rm H}}&=&-(0.233\pm0.026)\, g_2\,e\,{\rm fm}+[(0.04\pm0.03)\, C_3-(0.07\pm0.03)\, C_4]\,e\,{\rm fm}^{-2}\,.\label{eq:g2h3}
\end{eqnarray}
The $g_2$ $\chi$EFT results in Eqs.\,(\ref{eq:g2h2})-(\ref{eq:g2h3}) are consistent with those we obtained from the phenomenological $P$- and $T$-conserving potentials of 
Refs.\,\cite{av18,Pudliner:1997ck,cdbonn,Coon:2001pv}.  They are again a factor of two smaller in magnitude 
than those presented in \cite{stetcu,deVries:2011an} but are larger than those of 
Ref.\,\cite{song}
by roughly  a factor of 2.5. The origin of this discrepancy remains unknown so far.
The signs of our predictions and those of Refs.\,\cite{stetcu,deVries:2011an,song} agree, though. 
The $g_2$-induced EDM contributions can be neglected for all considered sources of $P$ and $T$ violation up to and including \NLO. The ones induced by the $C_3$ and $C_4$ vertices, however, are taken into consideration when the nuclear EDMs for the FQLR-term case are discussed below.

The estimates of the coefficients in Eq.\,(\ref{eq:impcoup}) for the sources of $P$ and $T$ violation in Eqs.\,(\ref{eq:lagtheta})-(\ref{eq:4q8}) were presented in Subsection \ref{sec:selection}. The hierarchies of the different EDM contributions are revealed upon insertion of these estimates into Eqs.\,(\ref{eq:h2edm})-(\ref{eq:h3edm}) and Eqs.\,(\ref{eq:g2h2})-(\ref{eq:g2h3}). We subsequently discuss these hierarchies by considering each source of $P$ and $T$ violation listed in Eqs.\,(\ref{eq:4qlr})-(\ref{eq:4q8}) individually.

\subsubsection{EDMs of light nuclei from the QCD \boldmath{$\theta$}-term}
The single-nucleon EDMs generated by the $\theta$-term receive one-loop contributions \cite{Baluni,Crewther:1979pi} of the same order as the tree-level contributions $d_0$ and $d_1$ \cite{Pich,Ottnad:2009jw,Guo:2012vf} mentioned in Subsection \ref{sec:selection}. Since these tree-level contributions are quantitatively unknown, supplementary Lattice QCD input is required to compute the single-nucleon EDMs. In fact, $d_n$ and $d_p$ have been computed recently \cite{Guo:2012vf,Akan:2014yha} by fitting to currently available Lattice QCD data \cite{Shintani:2008nt,Shintani:2012zca,Shintani:2014}:
\begin{equation}
d_n^{\theta}= \bar\theta\cdot(2.7\pm1.2)\cdot 10^{-16}\, e\,{\rm cm}\,,\quad d_p^{\theta}=-\bar\theta\cdot(2.1\pm 1.2)\cdot10^{-16}\, e\, {\rm cm} \,.
 \label{eq:edm_n_theta}
\end{equation}
The signs were adjusted to our convention \mbox{$e<0$}.
These predictions for the single-nucleon EDMs and the results for the other coupling constants of leading $P$- and $T$-violating vertices, \mbox{Eqs.\,(\ref{eq:sumg0theta})-(\ref{eq:delta3final})} and Eq.\,(\ref{eq:sumctheta}), can be inserted into Eqs.\,(\ref{eq:h2edm})-(\ref{eq:h3edm}) to obtain predictions for the EDMs of the deuteron, helion and triton with well-defined uncertainties \cite{edmoln}:
\begin{eqnarray}
d^\theta_{{}^2{\rm H}} &=&
   \bar\theta\cdot \Bigl\{ \bigl[(0.56\pm 0.01\pm1.59)\bigr] 
  - (0.62 \pm 0.06\pm 0.28) - (0.28\pm 0.05\pm 0.07)
   \Bigr\}\cdot 10^{-16}\,  e\, {\rm cm}\nn\\
   &=&-\bar{\theta}\cdot (0.3\pm1.6)\cdot 10^{-16}\,  e\, {\rm cm} \,,\\
   [2mm]
d_{{}^3{\rm He}}^{\theta}&=&
\bar\theta\cdot \bigl\{  \bigl [(2.44 \pm 0.04 \pm 1.08) + (0.06 \pm 0.01 \pm  0.03) \bigr] \nn \\
&&\quad\mbox{}
 -(0.006\pm 0.002 \pm 0.001)+  (1.72\pm 0.20 \pm 0.21) 
 -(0.48 \pm 0.06\pm 0.22) \nn \\
&&\quad\mbox{} -(0.22\pm 0.05\pm 0.05) \pm 0.2 \bigr\}
\cdot 10^{-16}\, e\, {\rm cm} \nn \\
&=&\bar\theta\cdot( 3.5 \pm 1.2)\cdot 10^{-16}\, e\, {\rm cm}\,,\\
[2mm]
d_{{}^3{\rm H}}^{\theta}&=&
\bar\theta\cdot \bigl\{  \bigl [ - (0.08 \pm 0.02 \pm  0.04)-(1.93 \pm 0.03 \pm 1.10) \bigr] \nn \\
&&\quad\mbox{}
-(0.006\pm0.002\pm0.001)-  (1.68\pm 0.20 \pm 0.21) 
 -(0.47 \pm 0.06\pm 0.21) \nn \\
&&\quad\mbox{} -( 0.22\pm 0.05\pm 0.05) \pm 0.2 \bigr\}
\cdot10^{-16}\, e\, {\rm cm} \nn \\
&=&-\bar\theta\cdot(4.4 \pm 1.2)\cdot10^{-16}\, e\,{\rm cm}\,.
\end{eqnarray}
The sequences of terms in these equations are the same as in Eqs.\,(\ref{eq:h2edm})-(\ref{eq:h3edm}). In each set of parentheses, the first uncertainty is the nuclear uncertainty, while the second uncertainty 
is the hadronic one. 
The contributions of $P$- and $T$-violating $NN$ contact interactions have been added in quadrature and regarded as an additional uncertainty.

The dominating nuclear contributions to the deuteron EDM are induced by $\gone$ and its one-loop correction 
$\Delta_3^{\theta}\, f_{g_1}$. In the cases of the helion and triton, the leading nuclear EDM contribution is generated by $\gzero$, whereas $\gone$ and its one-loop correction yield contributions of a factor of 1/4 and 1/8 smaller, respectively. Furthermore, the isospin-conserving and isospin-violating $NN$ contributions interfere destructively in the helion case, while they add up constructively in the triton case. The effect of the irreducible three-nucleon $P$- and $T$-violating potential is unexpectedly small and may be disregarded.

The uncertainties of the total EDM results are driven by the hadronic uncertainties of the single-nucleon EDMs. 
 The uncertainties of the pure nuclear contributions, especially in the deuteron case, are thus significantly smaller \cite{edmoln}:
\begin{eqnarray}\label{purenucEDMtheta}
d^\theta_{{}^2{\rm H}} -0.94\bigl(d_p^\theta +d_n^\theta\bigr) &=&- \bar\theta\cdot (0.89 \pm 0.30) \cdot 10^{-16}\,  e\, {\rm cm}\,,\\
d_{{}^3{\rm He}}^{\theta}-0.90\, d_n^{\theta}+0.03 \,d_p^{\theta}
&=&\phantom{-} \bar\theta\cdot (1.01\pm 0.42)\cdot 10^{-16}\, e\, {\rm cm}\,,\\
d_{{}^3{\rm H}}^{\theta} -0.92\,d_p^{\theta} +0.03\,d_n^{\theta}
&=&-\bar\theta\cdot (2.37 \pm 0.42)\cdot10^{-16}\, e\, {\rm cm}\,.
\end{eqnarray}
Since  the spins of the two protons in the hellion and of the two neutrons in the triton  couple to
zero  for the case of relative $S$- or $D$-waves,
the total {\em single}-nucleon contributions to the hellion and triton EDMs approximately equal the neutron and proton EDMs, respectively.

\subsubsection{EDMs of light nuclei from the FQLR-term}
The FQLR-term is the only effective {\it dimension-six} source for which the coefficients $g_0$ and $g_1$ can be expressed as functions of just one common parameter, $\Delta_3$, at leading order. Furthermore, the $NN$ potential operators induced by the $C_{3,4}$ vertices define in this case the leading EDM contributions from $NN$-contact interactions, since
the $C_{1,2}$ contributions are suppressed by explicit isospin-violation. If the functions for $g_{0,1}$ and $C_{3,4}$ of Eq.\,(\ref{eq:4qlrg0g1}) and Eq.\,(\ref{eq:4qlrsum}), respectively, are inserted into Eqs.\,(\ref{eq:h2edm})-(\ref{eq:h3edm})
and (\ref{eq:g2h2})-(\ref{eq:g2h3}), 
the nuclear contributions to the EDMs of the deuteron, helion and triton equal \cite{edmoln}
\begin{eqnarray}
d^{LR}_{^2{\rm H}} -0.94\bigl(d_p^{LR} +d_n^{LR}\bigr)
 &=& \Delta_3 \cdot [(1.37 \pm 0.13\pm 0.41) + (0.75\pm 0.14) \pm 0.1
]  
\,e\, {\rm fm}\nn  \\
  &=& \Delta_3\cdot(2.1 \pm 0.5)  \,  e\, {\rm fm}\,,\\
  [1.0mm]
d_{{}^3{\rm He}}^{LR}-0.90\, d_n^{LR}+0.03\,d_p^{LR} &=& \Delta_3\cdot\bigl\{ 
  (0.017\pm 0.006) -  (0.013\pm 0.002 \pm 0.002) 
  \nn \\
  &&\qquad+(1.07 \pm 0.14\pm 0.32)
+(0.61\pm 0.14) \pm 0.1 \bigr\}\, e\, {\rm fm} \nonumber\\
&=& \Delta_3\cdot( 1.7 \pm 0.5 \bigr)\, e\, {\rm fm}\,,\\
[1mm]
d_{{}^3{\rm H}}^{LR} -0.92\,d_p^{LR} +0.03\,d_n^{LR}
&=& \Delta_3\cdot\bigl\{ 
  (0.017\pm 0.006) + (0.013\pm 0.002 \pm 0.002) 
   \nn \\&&\qquad+(1.04 \pm 0.14\pm 0.31)
+(0.60\pm 0.14) \pm 0.1 \bigr\}\, e\, {\rm fm} \nn\\
&=& \Delta_3\cdot ( 1.7\pm 0.5 \bigr)\, e\, {\rm fm}\,.
\end{eqnarray}
The leading nuclear contributions, which here are of  the same sign,
 are now generated by $g_1$ and its one-loop vertex correction, $\Delta_3f_{g_1}$, in all three cases. The nuclear contributions generated by $g_0$ as well as the one induced by the $P$- and $T$-violating $3N$ potential can both be neglected.

\subsubsection{EDMs of light nuclei from the qCEDM}
Due to the existence of comparable isospin-conserving as well as isospin-violating components of the qCEDM effective dimension-six source, the leading coefficients in Eq.\,(\ref{eq:impcoup}) are no longer expressible as functions of just one parameter in this case. In fact, the dominating coefficients induced by the qCEDM are $g_0$ and $g_1$. If comparable strengths of the parameters $\tilde{\delta}_G^{0,3}$ in Eq.\,(\ref{eq:qcedm}) at the scale $\Lambda_{{\rm had}}$ are assumed, they both define the leading coupling constants of $P$- and $T$-violating hadronic vertices. EDM contributions induced by $g_0$ and $g_1$ might thus be relatively larger than their $\theta$-term counterparts. This would naively imply a significant enhancement of the nuclear contributions to the EDMs of the deuteron, helion and triton with respect to their single-nucleon EDM contributions. However, explicit computations show, as noted in \cite{deVries:2011an,edmoln},
that the EDM contributions from $P$- and $T$-violating one-pion exchanges turn out to be smaller than expected. The enhancement of $g_{0,1}$ contributions is, therefore, probably less profound than power counting and NDA suggest. 

The $P$- and $T$-violating $3\pi$ vertex only enters through the second axial rotation during the adjustment of the ground state. The $3N$ contributions that this vertex induces are of subleading orders. The $C_{1,2}$- and $C_{3,4}$-induced $NN$ contact interactions are both of the same order and are expected to scale as the suppressed $g_{0,1}$-induced two-pion exchanges, {\it i.e.} \mbox{$C_{1,2 (3,4)}{\sim}\mathcal{O}(g_{0(1)}/(\fpi m_N^2)$}. Therefore, the dominating EDM contributions most likely emerge from the $g_{0,1}$-induced one-pion exchange diagrams and the $d_{0,1}$-induced single-nucleon EDM contributions. The qCEDM-induced single-nucleon and light nuclei EDMs are then functions of these four coupling constants up to \NLO. 

\subsubsection{EDMs of light nuclei from the qEDM}

Since in the case of the qEDM all coupling constants of $P$- and $T$-violating vertices without the photon field are suppressed by at least a factor of $\alpha_{em}/(4\pi)$, the EDMs of the deuteron, helion and triton approximately equal their single-nucleon contributions. Assuming that the absolute values of the parameters of the isospin-conserving and isospin-violating components of the qEDM, namely $\tilde{\delta}_F^{1,3}$ in Eq.\,(\ref{eq:qedm}), are comparable, the isoscalar and isovector single-nucleon contributions are expected to be of the same order. The EDMs of single nucleons and light nuclei are, therefore, determined by just two coupling constants, $d_n$ and $d_p$,  up to \NLO.

\subsubsection{EDMs of light nuclei from the gCEDM and 4q-term}

Since the gCEDM and the 4q-term are both chiral singlets, these two effective dimension-six sources are indistinguishable by chiral symmetry considerations. The difficulties in assessing their hierarchies of single-nucleon and nuclear EDM contributions are compounded by the emergence of $P$- and $T$-violating $4N$ {\it and} $\gamma NN$ interactions at leading order. The nuclear contributions of the one-pion-exchange $NN$ potential operators are suppressed by a factor of $\mpi^2/m_N^2$ according to Goldstone's theorem. In total, most of the first seven vertices in Eq.\,(\ref{eq:impcoup}) contribute at the same order to single-nucleon and light nuclei EDMs up to \NLO\ for these sources of $P$ and $T$ violation. 

Lattice QCD predictions for the unknown LECs that emerge in amended $\chi$EFT might in the future allow for improved assessments of the hierarchies of single-nucleon and nuclear EDM contributions~\cite{Bhattacharya:2014cla,Shindler:2014oha}.


\section{Conclusion and Outlook}\label{sec:concl}
BSM theories such as supersymmetry, multi-Higgs scenarios and left-right symmetric models give rise to effective quark multilinears at the energy scale \mbox{$\Lambda_{{\rm had}}\gtrsim 1\,{\rm GeV}$}. In order to study the impact of BSM physics on low-energy systems in full generality without limiting the scope to one particular theory, the set of all possible quark multilinears at $\Lambda_{{\rm had}}$ and their induced hadronic interactions below the energy scale \mbox{$\Lambda_{{\rm QCD}}\sim 200$} MeV have to be considered. The focus of this paper has been twofold: we investigated the connection between quark multilinears at the energy scale $\Lambda_{{\rm had}}$ and their induced hadronic operators in Section \ref{sec:chptlag}. We presented a general scheme to derive the terms in the effective low-energy Lagrangian induced by arbitrary quark multilinears within the Gasser-Leutwyler formulation of $\chi$EFT. Effective quark multilinears that do not appear in the SM Lagrangian induce effective hadronic operators which are analogously beyond the realm of standard $\chi$EFT. Since new and independent quark multilinears can be regarded as an amendment of the QCD Lagrangian, we call the modified low-energy chiral effective field theory {\it amended} $\chi$EFT. The amended $\chi$EFT contains in general a set of new and independent LECs that are quantitatively unknown, which has also been pointed out in \cite{deVries:2010ah,deVries:2012ab}. These LECs yield only minor contributions to measurable $P$- and $T$-conserving observables, such as the strong part of the proton-neutron mass difference. This renders any attempt to disentangle them from the conventional LECs impossible. We demonstrated that all chiral structures (effective terms modulo their source fields and LECs) that are induced by arbitrary quark multilinears are in principle already encountered in the {\it standard} $\chi$EFT Lagrangian. The terms generated by a specific effective quark term can thus be extracted from the standard $\chi$EFT Lagrangian by a suitable replacement of LECs and source fields. The orders at which particular chiral structures occur are in general different for different effective quark multilinears. We stress that our techniques will be of general use in the study of BSM physics in hadronic systems.

Terms of the amended $\chi$EFT Lagrangian derived within our scheme can be exploited to compute single-nucleon and nuclear form factors which are induced by arbitrary quark multilinears. To this end, we also presented in Section \ref{sec:edmcomp} the derivation of scattering (two-nucleon systems) and Faddeev equations (three-nucleon systems) on which our numerical computation of these form factors of two- and three-nucleon systems is based. These equations will also be useful for other observables that require to sum reducible contributions in intermediate states.

One application for which these concepts have been primarily developed and which is the focus of this paper are the EDMs of single nucleons and light nuclei, {\it i.e.} the deuteron, helion and triton. As EDMs of single nucleons and light nuclei are $P$- and $T$-violating observables, the focus of this paper was on $P$- and $T$-violating quark multilinears and their induced hadronic interactions. The $P$- and $T$-violating QCD $\theta$-term, FQLR-term, qCEDM, qEDM, gCEDM and 4q-term defined in Eq.\,(\ref{eq:lagtheta}) and Eqs.\,(\ref{eq:4qlr})-(\ref{eq:4q8}), respectively, served as the starting point of our analysis, which is therefore completely general and does not depend on a particular BSM model. We presented our investigation of the connection between quark multilinears and their induced hadronic interaction by focussing on the above-stated sources of $P$ and $T$ violation and gave explicit expressions for the coefficients of induced hadronic vertices in Eq.\,(\ref{eq:impcoup}). The hadronic interactions modify the nuclear potential and we derived the resulting scattering and Faddeev equations. 

The final results of our independent analysis of the induced $P$- and $T$-violating effective operators are in complete agreement with those obtained within the Weinberg formulations of $\chi$EFT \cite{Mereghetti:2010tp,deVries:2010ah,deVries:2012ab}. The advantage of our derivation is the efficient compilation of higher-order effective terms and the straightforward generalization to $SU(3)$ $\chi$EFT once particular quark multilinears with strange quark content have been specified. The findings derived in this paper are the basis of the results of the complete and consistent $\chi$EFT computation of the nuclear contributions to the EDMs of the deuteron, helion and triton presented in \cite{edmoln}, which we briefly summarized at the end of of the previous
section.  

For the cases of the QCD $\theta$-term and the FQLR-term, the three leading induced coupling constants, $\Delta_3$, $g_0$ and $g_1$, are functions of only one but in $\chi$EFT itself unknown parameter at leading order in the standard chiral counting. The unknown parameter
is specific either to the $\theta$-term or the FQLR scenario. This observation led to quantitative predictions of the nuclear contributions to light-nuclei EDMs as functions of one common parameter for both of these cases, respectively. The consistent treatment within $\chi$EFT yielded controlled nuclear and hadronic uncertainties of our results as shown in \cite{edmoln}, which provides the basis for experimental tests of these two sources of $P$ and $T$ violation. 

In the case of all the other dimension-six sources of $P$ and $T$ violation,
the coupling constants in Eq.\,(\ref{eq:impcoup}) depend on more than one LEC at leading order. These LECs are in general the counterparts of the LECs $l_i$, $c_i$, $e_i$, $\bar{C}_i$ etc.\ in the standard $\chi$EFT Lagrangian. Lattice QCD has the capacity to accurately compute these LECs and close the connection between effective quark multilinears at the energy scale $\Lambda_{{\rm had}}\sim 1$ GeV and their induced hadronic interactions. In the absence of any such Lattice QCD results, order-of-magnitude estimates may be obtained by Naive Dimensional Analysis (NDA). These estimates are already sufficient to reveal distinct hierarchies among single-nucleon and nuclear EDM contributions for some sources of $P$ and $T$ violation. The qEDM yields EDM contributions which are functions of the two generically independent single-nucleon EDMs, $d_n$ and $d_p$. The leading coupling constants generated by the qCEDM are $g_{0}$ and $g_1$. Since the one-pion-exchange contributions induced by $g_{0,1}$ prove to be smaller than expected by explicit computation, $d_{0,1}$-induced contributions might also be significant for light-nuclei EDMs. The indistinguishable (by chiral-symmetry considerations) chiral-singlet sources, the gCEDM and the 4q-term, give rise to almost all vertices in Eq.\,(\ref{eq:impcoup}) at the same level, including the short-range $C_{1,2}$ interactions. All these conclusions have also been pointed out in \cite{deVries:2011an}. The specific hierarchies can be falsified by EDM experiments, but the EDM measurements of single nucleons, the deuteron and helion do in general not suffice to learn more about these sources of $P$ and $T$ violation. EDM measurements of more systems such as the triton and atoms or measurements of other electromagnetic moments are required in these cases to fix the relevant coupling constants in Eq.\,(\ref{eq:impcoup}).

Im summary, this paper provides a thorough investigation of hadronic interaction generated by the QCD $\theta$-term and $P$- and $T$-violating effective multi-quark terms within the framework of SU(2) $\chi$EFT. Our investigation is easily extendible to other effective multi-quark terms that conserve or violate discrete symmetries and to $SU(3)$ $\chi$EFT. We explicitly considered the induced EDMs of the deuteron, helion and triton as an application and derived the scattering and Faddeev equations required to compute  photon-nucleus form factors to this end. Our general findings can  be  used to study manifestations of specific $P$- and $T$-violating models in light nuclei.

\section*{Acknowledgements}
We are especially grateful to J.~de~Vries, E.~Epelbaum, C.~Hanhart, T.~L\"ahde, S.~Liebig, T.~Luu, D.~Minossi, N.~N.~Nikolaev and A.~Shindler for many useful discussions. We also express our gratitude to J.~de~Vries, S.~Liebig and T.~Luu for carefully reading the manuscript. Furthermore, we would like to thank C.-P.~Liu, E.~Mereghetti, U.~van~Kolck and R.~Timmermans for useful communications. This work is supported in part by the DFG and the NSFC through funds provided to the Sino-German CRC 110 ``Symmetries and the Emergence of 
Structure in QCD''. The resources of the J{\"u}lich Supercomputing Center at the 
Forschungszentrum J{\"u}lich, namely the supercomputers JUQUEEN and JUROPA, have been instrumental for the computations reported here.

\setcounter{equation}{0}
\renewcommand{\theequation}{\thesection.\arabic{equation}}

\newpage
\appendix



\section{Quark multilinears}\label{app:o4rep}
The quark multilinears considered in Section \ref{sec:chptlag} are quark bilinears and quark quadrilinears. A chiral $SU(2)_L\!\times\! SU(2)_R$ transformation of the left-handed and right-handed quark fields in quark multilinears induces a transformation in the space of quark multilinears. It will be shown in this appendix that a quark multilinear in general admits a decomposition into quark multilinears which transform as basis states of particular irreducible representations of $O(4)$, for which $SU(2)_L\!\times\!SU(2)_R$ is the double covering group. In order to provide a systematic study of quark bilinears and quadrilinears, the connection between the representation theory of $O(4)$ and quark multilinears is explained and the set of all quark bilinears and quark quadrilinears (relevant for this work) which transform as basis states of irreducible representations of $O(4)$ is compiled. 

We will explain that the relationship between quark multilinears and the representation theory of $O(4)$ is established by the set of $SU(2)_L\!\times\!SU(2)_R$ group actions on (symmetric) tensor products of elements of the quaternion algebra $\mathds{H}_4$. There is a one-to-one correspondence between the quaternion algebra $\mathds{H}_4$ and the real vector space of matrices spanned by the set $\{\mathds{1},i\tau_1,i\tau_2,i\tau_3\}$ with $\tau_i$ being the Pauli matrices. A group action $F$ of a group $G$ and a space $X$ is defined as the group homomorphism of $G$ into the group of homeomorphic maps from $X$ onto itself:
\begin{alignat}{1}
&F:G\times X\rightarrow X\,,\quad (g,x)\mapsto F_g(x)\,,\nonumber\\
& F_{g_1}\circ F_{g_2}(x)=F_{g_1\cdot g_2}(x)\quad\forall g_1,g_2\in G\,,\forall x\in X\,.
\end{alignat}

The connection between $SU(2)_L\!\times\!SU(2)_R$ group actions on $\mathds{H}_4$ and quark multilinears is based on the following observation:
when the quark field is defined to be the $SU(2)$ flavor doublet of the two lightest quark flavors $(u,d)$, the space of quark bilinears is given by the symmetric tensor product of the Clifford algebra of Dirac matrices ($\mathcal{A}_D$) and the algebra of the isospin Pauli matrices ($\mathcal{A}_I$): 
\begin{equation}\label{eq:diracquat}
\mathcal{A}_D=\{\mathds{1},\gamma^{\mu},\gamma^{\mu}\gamma_5,i\gamma_5,\sigma^{\mu\nu}\},\quad \mathcal{A}_I=\{\mathds{1},\tau_1,\tau_2,\tau_3\}\,.
\end{equation}
Whereas the matrices in $\mathcal{A}_D$ define the chiralities of the quark fields in a quark bilinear, the isospin matrices in $\mathcal{A}_I$ determine the quark flavors in a quark bilinear. Since quark bilinears with elements of $\mathcal{A}_D$ in Eq.\,(\ref{eq:diracquat}) are hermitian, they have to be eigenstates of the parity transformation $P:\,(L,R)\mapsto(R,L)$ which converts a right-handed quark field into a left-handed quark field and vice versa. This is also true for quark quadrilinears and quark multilinears in general which demonstrates the connection of quark multilinears to the representation theory of $O(4)$: a matrix in $\mathcal{A}_D$ defines a particular $SU(2)_L\!\times\!SU(2)_R$ group action on the isospin algebra $\mathcal{A}_I$, which can be identified with the quaternion algebra $\mathds{H}_4$ (the above defined basis of the $\mathcal{A}_I$ is obtained from the standard basis of $\mathds{H}_4$ by multiplication of each element by $-i$). The set $\mathcal{A}_D$ therefore corresponds to all possible $SU(2)_L\!\times\!SU(2)_R\!\times\!\mathds{Z}_2$ group actions on $\mathcal{A}_I\sim \mathds{H}_4$, where $\mathds{Z}_2$ denotes the group of parity transformations. This is, however, not a one-to-one correspondence, since two different matrices in $\mathcal{A}_D$ can in general define the same $SU(2)_L\!\times\!SU(2)_R\!\times\!\mathds{Z}_2$ group action, {\it e.g.} $\mathds{1}$ and $\sigma^{\mu\nu}$. To illustrate this statement, consider the quark bilinear
\begin{equation}
i\bar{q}\tau_k\gamma_5q=i\bar{q}_L\tau_kq_R-i\bar{q}_R\tau_kq_L\,.
\end{equation}
The Dirac matrix $i\gamma_5$ defines the $SU(2)_L\!\times\!SU(2)_R$ group action on the element $(i)\tau_k$ of $\mathds{H}_4$ which is associated with a basis state of the $(1/2,1/2)^+$ representation (see Eq.\,(\ref{eq:o4-1212}) below).

One has to emphasize the subtle difference between the parity transformation in relativistic quantum field theory and the parity group action on $\mathds{H}_4$: as shown below, the parity operation on $\mathds{H}_4$ amounts to a combination of the exchange \mbox{$(L,R)\mapsto(R,L)$} of the elements of $SU(2)_L$ and $SU(2)_R$ and the hermitian conjugation of the basis elements, such that all states of {\it e.g.} the $(1/2,1/2)^+$ irreducible representation have the same parity eigenvalue. In the field theory case, the requirement of hermiticity causes quark bilinears that transform under $SU(2)_L\!\times\! SU(2)_R\!\times\!\mathds{Z}_2$ as basis states of a particular $O(4)$ representation to have in general different parity eigenvalues.

This appendix is organized as follows: the well-known algebra of quaternions $\mathds{H}_4$ is briefly introduced in the first section of this appendix. The second section is concerned with the investigation of $SU(2)_L\!\times\!SU(2)_R$ group actions on $\mathds{H}_4$ and the representation theory of $SO(4)$. The third section of this appendix explains the connection between the representation theory of $O(4)$ and group actions on (tensor products of) $\mathds{H}_4$, from which the set of all quark bilinears and quark quadrilinears which transform as basis states of irreducible representations of $O(4)$ can be derived.

\subsection{Quaternions}
A detailed explanation of the quaternion algebra can be found in {\it e.g.}\cite{quaternion3,quaternion1,quaternion2}. A few well-known aspects are briefly summarized here.
The quaternion algebra $\mathds{H}_4$ is a four dimensional, non-commutative, associative algebra over the real numbers. $\mathds{H}_4$ is generated by $\{1,i,j,k\}$ where $i$, $j$ and $k$ obey
\begin{equation}
i^2=j^2=k^2=ijk=-1\,.
\end{equation}
For a general element $q\in\mathds{H}_4$, the conjugate $q^{\ast}\in\mathds{H}_4$ is defined by
\begin{equation}
q=a+bi+cj+dk\quad\rightarrow\quad q^{\ast}=a-bi-cj-dk\,.
\end{equation}
The quaternion algebra $\mathds{H}_4$ is isomorphic to the associative algebra generated by $\{\mathds{1},i\tau_1,i\tau_2,i\tau_3\}$, where $\tau_i$ denote Pauli matrices. The norm of a quaternion $q\in\mathds{H}_4$ is defined by 
\begin{equation}
||q||=\sqrt{q^{\ast}q}=\sqrt{a^2+b^2+c^2+d^2}\,.
\end{equation}

Exploiting the isomorphism to the associative algebra generated by the set $\{\mathds{1},i\tau_1,i\tau_2,i\tau_3\}$, the norm of a quaternion can be expressed in terms of the determinant of a complex $2\times 2$ matrix:
\begin{equation}
\det
\begin{pmatrix}
 a+ib & id+c\\
id-c & a-ib
\end{pmatrix}
=a^2+b^2+c^2+d^2\,.
\end{equation}
This proves that the group of quaternions of norm equal to one -- called unit quaternions -- is isomorphic to $SU(2)$:
\begin{equation}
\{q\in\mathds{H}_4|\,\,||q||=1\}\simeq SU(2)\,.
\end{equation}
Let $M$ be the isomorphism
\begin{equation}\label{eq:isom}
M: a+ib+jc+kd\mapsto 
\begin{pmatrix}
 a+ib & id+c\\
id-c & a-ib
\end{pmatrix}
\,,
\end{equation}
and $F$ be the group action of $SU(2)_L\!\times\! SU(2)_R$ on $\mathds{H}_4$ defined by: 
\begin{equation}\label{eq:fundgroupact}
F:\,SU(2)_L\!\times\!SU(2)_R\times \mathds{H}_4\rightarrow \mathds{H}_4\,,\quad q\mapsto L\,M(q)\,R^{\dagger}\,.
\end{equation}
Since $F$ preserves the norm for all $q\in\mathds{H}_4$, this group action is an automorphism of $SU(2)$ if restricted to unit quaternions. For all $g\in SU(2)$ the negative counter part, $-g$, is also an element of $SU(2)$, which implies that $F$ is not injective on $SU(2)_L\!\times\! SU(2)_R$ since the pairs $(L,R),\,(-L,-R)\in SU(2)_L\!\times\!SU(2)_R$ map onto the same homeomorphism of $\mathds{H}_4$.

\subsection{The representation theory of \boldmath{$SO(4)$}}\label{app:so4repquat}
The connection between the group $SU(2)_L\!\times\! SU(2)_R$ and the Lie group $SO(4)$ is drawn by the well-known Cayley-Klein decomposition of $SO(4)$ matrices (a detailed explanation of this connection between the quaternion algebra and $SO(4)$ is given in {\it e.g.}\cite{quaternion5,quaternion4,quaternion3,quaternion1,quaternion2,quaternion7}). Let $N$ denote the isomorphism between $\mathds{R}^4$ and $\mathds{H}_4$:
\begin{equation}\label{eq:niso}
N: \mathds{R}^4\rightarrow \mathds{H}_4\,,\quad (a,b,c,d)\mapsto a+ib+jc+kd\,.
\end{equation}
An $SO(4)$ rotation matrix $A$ can be re-expressed as (Cayley-Klein)
\begin{equation}\label{eq:cayley}
A\in SO(4)\,,x\in\mathds{R}^4:\,A\,x=(M\circ N)^{-1}(L\,M\circ N(x)\,R^{\dagger})\,,\quad L,R\in SU(2)_{L,R}\,,
\end{equation}
with the isomorphism $M$ as defined in Eq.\,(\ref{eq:isom}). This demonstrates that the group action $F$ of Eq.\,(\ref{eq:fundgroupact}) defines a 2-1 homomorphism $SU(2)_L\!\times\!SU(2)_R\rightarrow SO(4)$, since the pairs 
\mbox{$(L,R)\in SU(2)_L\!\times\!SU(2)_R$} and $(-L,-R)\in SU(2)_L\!\times\!SU(2)_R$ map onto the same element \mbox{of SO(4)}.

 Let $D_{(j_1,j_2)}$ be a representation of $SU(2)_L\times SU(2)_R$ of dimension $(2j_1\!+\!1)(2j_2\!+\!1)$ for integers or half integers  $j_1,j_2$. In order for $D_{(j_1,j_2)}$ to be also a representation of $SO(4)$, it has to obey
\begin{equation}
D_{(j_1,j_2)}(L,R)=D_{(j_1,j_2)}(-L,-R)\,,\quad \forall\,(L,R)\in SU(2)_L\!\times\!SU(2)_R\,,
\end{equation}
which requires $j_1+j_2$ to equal integer values. In general, multiple tensor products of (fundamental) two-dimensional representations of $SU(2)_L\times SU(2)_R$,
\begin{eqnarray}
&&\underbrace{D_{(1/2,0)}\!\otimes\!\cdots\!\otimes\! D_{(1/2,0)}}_{m-\text{times}}\otimes\underbrace{D_{(0,1/2)}\!\otimes\!\cdots\!\otimes\! D_{(0,1/2)}}_{n-\text{times}}:\nonumber\\
&&SU(2)_L\times SU(2)_R\times\underbrace{\mathds{C}^2\otimes\cdots\otimes\mathds{C}^2}_{m-\text{times}}\otimes\underbrace{\mathds{C}^2\otimes\cdots\otimes\mathds{C}^2}_{n-\text{times}}\nonumber\\
&&\rightarrow\underbrace{\mathds{C}^2\otimes\cdots\otimes\mathds{C}^2}_{m-\text{times}}\otimes\underbrace{\mathds{C}^2\otimes\cdots\otimes\mathds{C}^2}_{n-\text{times}}\,, 
\end{eqnarray}
are (not necessarily irreducible) representations of $SO(4)$ if $m+n$ is even.
The group actions corresponding to irreducible representations of $SO(4)$ with $m+n\leq1$ can be defined by
\begin{alignat}{3}
&\text{dim}=4\qquad&&D_{(1/2,1/2)}&&: M(q)\mapsto LM(q)R^{\dagger}\,,\label{eq:so4d4}\\
&\text{dim}=4\qquad&&D_{(1/2,1/2)}&&: RM(q)L^{\dagger}\,,\\
&\text{dim}=3&&D_{(1,0)}\qquad&&: M(q)\mapsto LM(q)L^{\dagger}\,,\\
&\text{dim}=3&&D_{(0,1)}&&: M(q)\mapsto RM(q)R^{\dagger}\,,\\
&\text{dim}=1&&D_{(0,0)}&&:M(1)=\mathds{1}\mapsto LM(1)L^{\dagger}\,,\\
&\text{dim}=1&&D_{(0,0)}&&:M(1)=\mathds{1}\mapsto RM(1)R^{\dagger}\,,\label{eq:so4d12}
\end{alignat}
for $q\in\mathds{H}_4$ and $(L,R)\in SU(2)_L\!\times\!SU(2)_R$.
Note that the group action on $\mathds{H}_4$ corresponding to a particular irreducible representation of $SO(4)$ (and $SU(2)_L\!\times\! SU(2)_R $) is not necessarily unique. All group actions defined in Eqs.\,(\ref{eq:so4d4})-(\ref{eq:so4d12}) are formally different group actions.

Group actions $F$ on the tensor product space of multiple copies of $\mathds{H}_4$,
\begin{equation}
F: SU(2)_L\!\times\! SU(2)_R\times (\mathds{H}_4\otimes\cdots\otimes\mathds{H}_4)\rightarrow (\mathds{H}_4\otimes\cdots\otimes\mathds{H}_4)\,,
\end{equation}
can constitute higher dimensional irreducible representations of $SO(4)$. Let the notation $(i\tau_j)_L$ and $(i\tau_j)_R$, $j=1,2,3$\,, imply the following definitions of $SU(2)_L\times SU(2)_R$ group actions on a subspace of $\mathds{H}_4$:
\begin{alignat}{2}
&\text{dim}=3:\quad (i\tau_j)_L&&\mapsto (Li\tau_jL^{\dagger})_L\,,\label{eq:so4ga31}\\
&\text{dim}=3:\quad (i\tau_j)_R&&\mapsto (Ri\tau_jR^{\dagger})_R\,.\label{eq:so4ga32}
\end{alignat}
The group actions on $\mathds{H}_4\otimes\mathds{H}_4$ -- subsequently referred to as the (rank 2) tensor product of two elementary (rank 1) $SU(2)_L\!\times\! SU(2)_R$ group actions -- corresponding to tensor products of the representations $D_{(1,0)}$ and  $D_{(0,1)}$ are then given by (the '$\otimes$' is omitted for convenience below, {\it e.g.} $(i\tau_j)_{L,R}(i\tau_k)_{L,R}$ is meant to imply $(i\tau_j)_{L,R}\otimes(i\tau_k)_{L,R}$)
\begin{alignat}{4}
&(0,1)\otimes (0,1)&&:\quad &&(i\tau_j)_R(i\tau_k)_R&&\mapsto (Ri\tau_j R^{\dagger})_R(Ri\tau_k R^{\dagger})_R\,,\\
&(1,0)\otimes (1,0)&&: &&(i\tau_j)_L(i\tau_k)_L&&\mapsto (Li\tau_j L^{\dagger})_L(Li\tau_k L^{\dagger})_L\,,\\
&(1,0)\otimes (0,1)&&: &&(i\tau_j)_L(i\tau_k)_R&&\mapsto (Li\tau_j L^{\dagger})_L(Ri\tau_k R^{\dagger})_R\,,
\end{alignat}
which do a priori not constitute irreducible representations of $SO(4)$ and may decompose into direct sums of irreducible representation of $SO(4)$. The basis states of the lowest-dimensional irreducible representations are obtained by identifying linear combinations of the basis vectors $(i\tau_j)_{L/R}(i\tau_k)_{L/R}$ with well-defined properties under exchanges of $j$ and $k$ (summation of identical indices implied):
\begin{alignat}{4}
&\text{dim}=1:\quad&&(0,0)\quad &&:\,\quad&&(i\tau_m)_L(i\tau_m)_L\,,\\
&\text{dim}=1:\quad&&(0,0)\quad &&:\,\quad&&(i\tau_m)_R(i\tau_m)_R\,,\\
&\text{dim}=3:&&(1,0) &&:&&(i\tau_j)_L(i\tau_k)_L-(i\tau_k)_L(i\tau_j)_L\,,\\
&\text{dim}=3:&&(0,1) &&:&&(i\tau_j)_R(i\tau_k)_R-(i\tau_k)_R(i\tau_j)_R\,,\\
&\text{dim}=5:&&(2,0)&&:&&(i\tau_j)_L(i\tau_k)_L+(i\tau_k)_L(i\tau_j)_L-2(i\tau_m)_L(i\tau_m)_L\,,\\
&\text{dim}=5:&&(0,2)&&:&&(i\tau_j)_R(i\tau_k)_R+(i\tau_k)_R(i\tau_j)_R-2(i\tau_m)_R(i\tau_m)_R\,,\\
&\text{dim}=9:&&(1,1)&&:&&(i\tau_j)_L(i\tau_k)_R\,,\\
&\text{dim}=9:&&(1,1)&&:&&(i\tau_j)_R(i\tau_k)_L\,.
\end{alignat}

In order to explore the $SU(2)_L\!\times\! SU(2)_R$ group actions on $\mathds{H}_4\otimes \mathds{H}_4$ corresponding to the tensor product of representations $D_{(1/2,1/2)}\otimes D_{(1/2,1/2)}$, the quantities $(t_{\alpha})$ and $(t_{\alpha}^{\dagger})$ are introduced to imply the group actions on $\mathds{H}_4$ defined by
\begin{alignat}{3}
&\text{dim}=4:\qquad &&(t_{\alpha})&&\mapsto (Lt_{\alpha}R^{\dagger})\,,\label{eq:so4ga41}\\
&\text{dim}=4:&&(t_{\alpha}^{\dagger})&&\mapsto(Rt_{\alpha}^{\dagger}L^{\dagger})\label{eq:so4ga42}
\end{alignat}
with 
\begin{equation}
t_0=\mathds{1}\,,\quad t_1=i\tau_1\,,\quad t_2=i\tau_2\,,\quad t_3=i\tau_3\,.
\end{equation}
For $g_{\alpha\beta}=\rm{diag}(1,-1,-1,-1),$\footnote{We do not distinguish between covariance and contravariance here.} the basis states of the lowest-dimensional irreducible representations of $SO(4)$ expressed in terms of tensor products of the four-dimensional group action Eq.\,(\ref{eq:so4ga41}) are given by:
\begin{alignat}{4}
&\text{dim}=1& &(0,0) &&:\,&&(t_{\alpha})(t_{\beta})\,g_{\alpha\beta}\,,\\
&\text{dim}=3+3 \quad&&(1,0)\oplus(0,1) &&: &&(t_{\alpha})(t_{\beta})-(t_{\beta})(t_{\alpha})\,,\\
&\text{dim}=9 &&(1,1) &&:  &&(t_{\alpha})(t_{\beta})+(t_{\beta})(t_{\alpha})-2(t_{\alpha})(t_{\beta})\,g_{\alpha\beta}\,.
\end{alignat}
By replacing $(t_{\alpha,\beta})$ by $(t_{\alpha,\beta}^{\dagger})$ the lowest-dimensional irreducible representations of $SO(4)$ expressed in terms of tensor products of the group action defined in Eq.\,(\ref{eq:so4ga42}) are obtained.

\subsection{The representation theory of \boldmath{$O(4)$}}
The well-known connection between the quaternion algebra $\mathds{H}_4$ and the Lie group $O(4)$ is the following \cite{quaternion4,quaternion2,quaternion6,quaternion7}: the group $O(4)=SO(4)\times \mathds{Z}_2$ consists of two copies of $SO(4)$ and is thus not connected. The parity transformation $P$ defined as the spacial inversion of three components of a general real four-vector,
\begin{equation}
P: \mathds{R}^4\rightarrow \mathds{R}^4\,,\quad(x_1,x_2,x_3,x_4)\mapsto (x_1,-x_2,-x_3,-x_4)\,,
\end{equation}
transforms an element of one connected component into an element of the other connected component. By exploiting the isomorphism $N$ of Eq.\,(\ref{eq:niso}) between $\mathds{R}^4$ and $\mathds{H}_4$, the parity transformation $P$ is found to correspond to the following $\mathds{Z}_2$ group action on $\mathds{H}_4$:
\begin{equation}
\tilde{P}=NPN^{-1}:\,\mathds{Z}_2\times \mathds{H}_4\rightarrow\mathds{H}_4\,,\quad\mathds{1}\,,i\tau_1\,,i\tau_2\,,i\tau_3\mapsto\mathds{1}\,,-i\tau_1\,,-i\tau_2\,,-i\tau_3\,,
\end{equation}
{\it i.e.} to be equivalent to the conjugation of quaternions (or to hermitian conjugation of $\mathds{1}$ and $i\tau_{1,2,3}$). The composition of the parity transformation with an $SO(4)$ matrix $A$ then amounts to an exchange of the left and right unit quaternion in the Cayley-Klein 2-1 isomorphism of Eq.\,(\ref{eq:cayley}):
\begin{eqnarray}\label{eq:o4rel}
P\,A\,x&=&(M\circ N)^{-1}([L\,M\circ N(x)\,R^{\dagger}]^{\dagger})\nonumber\\
&=&(M\circ N)^{-1}(R\,[M\circ N(x)]^{\dagger}\,L^{\dagger})=(M\circ N)^{-1}(R\,M\circ N(Px)\,L^{\dagger})\,.
\end{eqnarray}

Let $D_{(j,j)}$ be a representation of $O(4)$ and $v_1\otimes v_2$ an element of the base space of $D_{(j,j)}$. Equation~(\ref{eq:o4rel}) demonstrates that the representation of the element $P$ has to obey 
\begin{eqnarray}\label{eq:o4rel2}
D_{(j,j)}( P)\,D_{(j,j)}(L,R)\,v_1\otimes v_2&=&D_{(j,j)}( P)\,(D_{(j)}(L)\,v_1\otimes D_{(j)}( R)\,v_2)\nonumber\\
&=&D_{(j,j)}(R,L)\,D_{(j,j)}( P)v_1\otimes v_2\,.
\end{eqnarray}
Therefore, a representation $D_{(j_1,j_2)}$ of $SO(4)$ is also a representation of $O(4)$ if and only if it is symmetric under an exchange of the indices $j_1$ and $j_2$.   For arbitrary $j_1$ and $j_2$, the irreducible representation $D_{(j_1,j_2)}$ of $SO(4)$ induces the symmetrized representation $D_{(j_1,j_2)\oplus(j_2,j_1)}$ of $O(4)$, which is irreducible if $j_1\neq j_2$ and decomposes into two irreducible representation of equal dimensions if $j_1=j_2$. This observation implies that irreducible representations $D_{(j_1,j_2)}$ of $SO(4)$ with $j_1\neq j_2$ induce exactly one irreducible $O(4)$ representation, whereas irreducible representations $D_{(j,j)}$ of $SO(4)$ induce two separate representations of $O(4)$. 

For arbitrary half integers or integers $j_1,j_2$ with $j_1+j_2\in\mathds{N}$, the generalization of Eq.\,(\ref{eq:o4rel2}) reads:
\begin{equation}
D_{(j_1,j_2)\oplus (j_2,j_1)}(P)D_{(j_1,j_2)\oplus (j_2,j_1)}(L,R)=D_{(j_1,j_2)\oplus (j_2,j_1)}(R,L)D_{(j_1,j_2)\oplus (j_2,j_1)}(P)\,.
\end{equation}
The action of $D_{(j_1,j_2)\oplus (j_2,j_1)}(P)$ on an element of the base space $v_1\otimes v_2\oplus v_3\otimes v_4$ is required to be
\begin{equation}
D_{(j_1,j_2)\oplus (j_2,j_1)}(P)(v_1\otimes v_2\oplus v_3\otimes v_4)=v_4\otimes v_3\oplus v_2\otimes v_1\,,
\end{equation}
which can be proven by the following brief computation:
\begin{eqnarray}
&&D_{(j_1,j_2)\oplus (j_2,j_1)}(L,R)\,(v_1\otimes v_2\oplus v_3\otimes v_4)\nonumber\\
&=&[D_{(j_1,j_2)\oplus (j_2,j_1)}(P)]^2D_{(j_1,j_2)\oplus (j_2,j_1)}(L,R)\,(v_1\otimes v_2\oplus v_3\otimes v_4)\nonumber\\
&=&D_{(j_1,j_2)\oplus (j_2,j_1)}(P)D_{(j_1,j_2)\oplus (j_2,j_1)}(R,L)D_{(j_1,j_2)\oplus (j_2,j_1)}(P)\,(v_1\otimes v_2\oplus v_3\otimes v_4)\nonumber\\
&=&D_{(j_1,j_2)\oplus (j_2,j_1)}(P)D_{(j_1,j_2)\oplus (j_2,j_1)}(R,L)\,(v_4\otimes v_3\oplus v_2\otimes v_1)\nonumber\\
&=&D_{(j_1,j_2)\oplus (j_2,j_1)}(P)\,(D_{(j_1)}(R)v_4\otimes D_{(j_2)}(L)v_3\oplus D_{(j_2)}(R)v_2\otimes D_{(j_1)}(L)v_1)\nonumber\\
&=&D_{(j_1)}(L)v_1\otimes D_{(j_2)}(R)v_2\oplus D_{(j_2)}(L)v_3\otimes D_{(j_1)}(R)v_4)\nonumber\\
&=&D_{(j_1,j_2)\oplus (j_2,j_1)}(L,R)\,(v_1\otimes v_2\oplus v_3\otimes v_4)\,.
\end{eqnarray}
Due to
\begin{equation}
[D_{(j_1,j_2)\oplus (j_2,j_1)}(P)]^2=\mathds{1}\,,
\end{equation}
$D_{(j_1,j_2)\oplus (j_2,j_1)}(P)$ must have eigenvalues $\pm 1$ and the basis states of the irreducible representations have definite parity, $p=\pm 1$. It is convenient to refer to the two irreducible representations of $O(4)$ in $D_{(j,j)\oplus (j,j)}$ by $D_{(j,j)^{\pm}}$.

\subsubsection{Representations of \boldmath{$O(4)$} and quaternions}
Utilizing the notation introduced by Eq.\,(\ref{eq:so4ga31}), Eq.\,(\ref{eq:so4ga32}), Eq.\,(\ref{eq:so4ga41}) and Eq.\,(\ref{eq:so4ga42}) and defining
\begin{alignat}{4}
&(0,0)\quad&&\dim=1:\qquad &&(\mathds{1})_L&&\mapsto (L\mathds{1}L^{\dagger})_R=(\mathds{1})_L\,,\\
&(0,0)&&\dim=1:&& (\mathds{1})_R&&\mapsto (R\mathds{1}R^{\dagger})_R=(\mathds{1})_R\,,
\end{alignat}
the group actions of $SU(2)_L\!\times\!SU(2)_R$ on $\mathds{H}_4$ which correspond to the lowest-dimensional irreducible representations of $O(4)$ are given by:
\begin{alignat}{4}
&(0,0)^{\pm}\quad&&\dim=1\,&&:\qquad &&(\mathds{1})_L\pm(\mathds{1})_R\,,\\
&(1/2,1/2)^{\pm}&&\dim=4+4&&: && (t_0)\pm (t_{0}^{\dagger})\,,(t_i)\mp (t_{i}^{\dagger})\,,\label{eq:o4-1212}\\
&(1,0)\oplus (0,1)\quad&&\dim=6&&: && (i\tau_j)_L+(i\tau_j)_R\,,(i\tau_j)_L-(i\tau_j)_R\,.
\end{alignat}

For convenience, the tensor products $(i\tau_j)_{L,R}\otimes(i\tau_k)_{L,R}$ and $(t_{\alpha}^{(\dagger)})\otimes(t_{\beta}^{(\dagger)})$ are denoted by $(i\tau_j)_{L,R}(i\tau_k)_{L,R}$ and $(t_{\alpha}^{(\dagger)})(t_{\beta}^{(\dagger)})$, respectively. The basis states of the lowest-dimensional irreducible representations of $O(4)$ expressed in terms of tensor products of $(i\tau_j)$ read (which can also be proven by a direct computation):
\begin{alignat}{3}
&(0,0)^{\pm}&&:&\,&(i\tau_m)_R(i\tau_m)_R\pm(i\tau_m)_L(i\tau_m)_L\,,\label{eq:o4-100}\\
&(1,0)\oplus(0,1)&&:&&(\delta^{jl}\delta^{km}-\delta^{jm}\delta^{kl})[(i\tau_l)_L(i\tau_m)_L\pm(i\tau_l)_R(i\tau_m)_R]\,,\label{eq:o4-110}\\
&(2,0)\oplus(0,2)&&:&&(\delta^{jl}\delta^{km}+\delta^{jm}\delta^{kl}-2\delta^{jk}\delta^{lm})[(i\tau_k)_L(i\tau_l)_L\pm(i\tau_k)_R(i\tau_l)_R]\,,\label{eq:o4-22}\\
&(1,1)^{+}&&: &&(i\tau_j)_L(i\tau_k)_R-(i\tau_j)_R(i\tau_k)_L-(j\leftrightarrow k)\,,\label{eq:o4-11p1}\\
&&&&&(i\tau_j)_L(i\tau_k)_R+(i\tau_j)_R(i\tau_k)_L+(j\leftrightarrow k)\,,\label{eq:o4-11p2}\\
&(1,1)^{-}&&: &&(i\tau_j)_L(i\tau_k)_R-(i\tau_j)_R(i\tau_k)_L+(j\leftrightarrow k)\,,\label{eq:o4-11m1}\\
&&&&&(i\tau_j)_L(i\tau_k)_R+(i\tau_j)_R(i\tau_k)_L-(j\leftrightarrow k)\,.\label{eq:o4-11m2}
\end{alignat}
The tensor product of the two six-dimensional representations of $O(4)$ decomposes into:
\begin{eqnarray}
&&[(1,0)\oplus(0,1)]\otimes[(1,0)\oplus(0,1)]\nonumber\\
&=&[(2,0)\oplus(0,2)]\oplus(1,1)^+\oplus(1,1)^-\oplus[(1,0)\oplus(0,1)]\oplus(0,0)^+\oplus(0,0)^-\,,
\end{eqnarray}
{\it i.e.} into one 10-dimensional, two 9-dimensional, one 6-dimensional and two one-dimensional irreducible representations of $O(4)$, whose basis states with their implied group actions are given by Eq.\,(\ref{eq:o4-100}) and Eqs.\,(\ref{eq:o4-110})-(\ref{eq:o4-11m2}).

The other relevant tensor products of irreducible representations of $O(4)$ decompose into
\begin{align}
(1/2,1/2)^{\pm}\otimes(1/2,1/2)^{\pm}&=(1,1)^{+}\oplus[(1,0)\oplus(0,1)]\oplus(0,0)^{+}\,,\\
(1/2,1/2)^{\pm}\otimes(1/2,1/2)^{\mp}&=(1,1)^{-}\oplus[(1,0)\oplus(0,1)]\oplus(0,0)^{-}\,.
\end{align}
Utilizing the notation for group actions defined by Eq.\,(\ref{eq:so4ga41}) and Eq.\,(\ref{eq:so4ga42}), the basis states with implied group actions of the lowest-dimensional irreducible representations of $O(4)$ expressed in terms of tensor products of $(t_{\alpha})$ and $(t^{\dagger}_{\alpha})$ are given by:\footnote{The basis states of the two $(1,1)^{\pm}$ representations have been combined here for compactness.}
\begin{alignat}{3}
&(0,0)^{\pm}&&\text{dim}=1& &:\,[(t_{\alpha})(t_{\beta})\pm(t_{\alpha}^{\dagger})(t_{\beta}^{\dagger})]g_{\alpha\beta}\,,\label{eq:o4-00}\\
&(1,0)\oplus(0,1)\,&&\text{dim}=6& &:\,[(t_{\alpha})(t_{\beta})\pm(t_{\alpha}^{\dagger})(t_{\beta}^{\dagger})]-[(t_{\beta})(t_{\alpha})\pm(t_{\beta}^{\dagger})(t_{\alpha}^{\dagger})]\,,\label{eq:1001twoidrep1}\\
&(1,1)^{\pm}&&\text{dim}=9+9& &:\,(g_{\alpha\rho}g_{\beta\sigma}+g_{\beta\rho}g_{\alpha\sigma}-2g_{\alpha\beta}g_{\rho\sigma})[(t_{\rho})(t_{\sigma})\pm(t_{\rho}^{\dagger})(t_{\sigma}^{\dagger})]\,.
\end{alignat}
Note that the different relative signs in Eq.\,(\ref{eq:1001twoidrep1}) define two identical (Fierz-equivalent) sets of tensors (which are given explicitly by Eqs.\,(\ref{eq:1001idrep1})-(\ref{eq:1001idrep2}) and Eqs.\,(\ref{eq:1001idrep3})-(\ref{eq:1001idrep4}), respectively). The basis states of the $(1,1)^+$ and $(1,1)^-$ representations are given by Eq.\,(\ref{eq:o4-11p3})-(\ref{eq:o4-11p4}) and Eq.\,(\ref{eq:o4-11m3})-(\ref{eq:o4-11m4}) below, respectively. The basis states of the irreducible representations of $O(4)$ in the decompositions of specific tensor products of four-dimensional irreducible representations of $O(4)$ read\newline
$(1/2,1/2)^{\pm}\otimes(1/2,1/2)^{\pm}$ with $i,j=1,2,3$, $i\neq j$:
\begin{alignat}{4}
&(1,0)\oplus(0,1)\quad&&\text{dim}=6 &&:&\quad&(t_i)(t_j)+(t_i^{\dagger})(t_j^{\dagger})-(i\leftrightarrow j)\,,\label{eq:1001idrep1}\\
&&&&&&&(t_0)(t_i)-(t_0^{\dagger})(t_i^{\dagger})-(0\leftrightarrow i)\,,\label{eq:1001idrep2}\\
&(1,1)^+&&\text{dim}=9& &:& &(t_i)(t_j)+(t_i^{\dagger})(t_j^{\dagger})+(i\leftrightarrow j)\,,\label{eq:o4-11p3}\\
&&&&&&&(t_0)(t_i)-(t_0^{\dagger})(t_i^{\dagger})+(0\leftrightarrow i)\,,\label{eq:o4-11p4}
\end{alignat}
$(1/2,1/2)^{\pm}\otimes(1/2,1/2)^{\mp}$ with $i,j=1,2,3$, $i\neq j$:
\begin{alignat}{4}
&(1,0)\oplus(0,1)\quad&&\text{dim}=6& &:&\quad&(t_i)(t_j)-(t_i^{\dagger})(t_j^{\dagger})-(i\leftrightarrow j)\,,\label{eq:1001idrep3}\\
&&&&&&&(t_0)(t_i)+(t_0^{\dagger})(t_i^{\dagger})-(0\leftrightarrow i)\,,\label{eq:1001idrep4}\\
&(1,1)^-&&\text{dim}=9& &:& &(t_i)(t_j)-(t_i^{\dagger})(t_j^{\dagger})+(i\leftrightarrow j)\,,\label{eq:o4-11m3}\\
&&&&&&&(t_0)(t_i)+(t_0^{\dagger})(t_i^{\dagger})+(0\leftrightarrow i)\,,\label{eq:o4-11m4}
\end{alignat}
where only the six ``off-diagonal" ($\alpha\neq\beta$) basis states for the $(1,1)^{\pm}$ representations are shown. 

The ``off-diagonal" basis states of the $(1,1)^{\pm}$ representations in terms of tensor products of $(t_{\alpha})$ and $(t_{\beta}^{\dagger})$ are given by ($i,j=1,2,3$ and $i\neq j$):
\begin{alignat}{4}
&(1,1)^+\quad&&\text{dim}=9& &:&\quad&(t_0)(t_i^{\dagger})-(t_0^{\dagger})(t_i)+(t_i^{\dagger})(t_0)-(t_i)(t_0^{\dagger})\,,\label{eq:o4-11p5}\\
&&&&&&&(t_i)(t_j^{\dagger})+(t_i^{\dagger})(t_j)+(t_j^{\dagger})(t_i)+(t_j)(t_i^{\dagger})\label{eq:o4-11p6}\,,\\
&(1,1)^-&&\text{dim}=9& &:&\quad&(t_0)(t_i^{\dagger})+(t_0^{\dagger})(t_i)+(t_i^{\dagger})(t_0)+(t_i)(t_0^{\dagger})\,,\label{eq:o4-11m5}\\
&&&&&&&(t_i)(t_j^{\dagger})-(t_i^{\dagger})(t_j)+(t_j^{\dagger})(t_i)-(t_j)(t_i^{\dagger})\,.\label{eq:o4-11m6}
\end{alignat}
The basis states of the $(1,1)^+$ representation in Eqs.\,(\ref{eq:o4-11p5})-(\ref{eq:o4-11p6}) are identical (Fierz-equivalent) to the basis states given in Eqs.\,(\ref{eq:o4-11p1})-(\ref{eq:o4-11p2}) since they define exactly the same $SU(2)_L\!\times\!SU(2)_R$ tensors. Phrased differently, representations of $O(4)$ allow for different expressions in terms of group actions on (multiple) tensor products of copies of $\mathds{H}_4$. In fact, the existence of Fierz identities is a direct consequence of this statement. By the means of a Fierz-type projection \cite{Nieves:2003in} a connection between them can be drawn: let
\begin{equation}
(X_s\otimes X'_t)_{ijkl},\,(Y_u\otimes Y'_v)_{ijkl}\,,\quad X_{s},X_{t}',Y_{u},Y_{v}'\in \mathds{H}_4\,,
\end{equation}
be two bases of the same space which can be expressed as a tensor product of two spaces in more than one way. A scalar product $B$ is defined by
\begin{equation}\label{eq:repscal}
B(X_s\otimes X'_t,Y_u\otimes Y'_v)=\sum_{ijkl}(X_s\otimes X'_t)_{ijkl}(Y_u\otimes Y'_v)_{ijkl}^{\dagger}\,.
\end{equation}
For instance, two different expressions of the basis states of the irreducible (1,0) $SO(4)$ representation are given by
\begin{equation}
(i\tau_s)_R=(\mathds{1})_R(i\tau_s)_R\,,\quad \epsilon^{s'tu}(i\tau_t)_R(i\tau_u)_R\,.
\end{equation}
The scalar product Eq.\,(\ref{eq:repscal}) of  basis states from two different expressions in terms of tensor products yields:
\begin{align}
-i[(\mathds{1})_R]_{ij}[(\tau_s)_R]_{kl}[(\tau_t)_R]_{ij}[(\tau_u)_R]_{kl}\epsilon^{s'tu}&/4=0\,,\\
-i[(\mathds{1})_R]_{kl}[(\tau_s)_R]_{ij}[(\tau_t)_R]_{ij}[(\tau_u)_R]_{kl}\epsilon^{s'tu}&/4=0\,,\\
-i[(\mathds{1})_R]_{li}[(\tau_s)_R]_{jk}[(\tau_t)_R]_{ij}[(\tau_u)_R]_{kl}\epsilon^{s'tu}&/4=-\delta_{ss'}\,,\\
-i[(\mathds{1})_R]_{jk}[(\tau_s)_R]_{li}[(\tau_t)_R]_{ij}[(\tau_u)_R]_{kl}\epsilon^{s'tu}&/4=\delta_{ss'}\,.
\end{align}

\subsubsection{\boldmath{$O(4)$} representations and quark multilinears}
So far only the regular tensor product of basis states of $SO(4)$ representations with implied $SU(2)_L\!\times SU(2)_R$ group actions have been considered. However, quark multilinears which include more than two quarks constitute {\em symmetric} tensor products of basis states with  implied \mbox{$SU(2)_L\!\times SU(2)_R$} group actions, since a general quark multilinear is the {\em commutative} product of multiple quark bilinears. The symmetric tensor product, denoted by '$\odot$', is defined for $n$ vectors $v_i$ by
\begin{equation}\label{eq:o4-symten}
v_1\odot\cdots\odot v_n=\sum_{\sigma}\frac{1}{n!}v_{\sigma(1)}\otimes\cdots\otimes v_{\sigma(n)}\,,
\end{equation}
where the sum is over all permutations $\sigma(1),\cdots,\sigma(n)$ of the index set $\{1,\cdots,n\}$. The {\it symmetric} tensor product of two irreducible representations of dimension $d$ yields a representation of dimension $D=d+(d-1)+\cdots+1$.

The interest of this paper is in the decompositions of the symmetric tensor products of the lowest-dimensional irreducible representations of $O(4)$. The symmetric tensor product of the two 6-dimensional irreducible representations yields a reducible representation of dimension 21 which decomposes into the irreducible representations
\begin{equation}
[(1,0)\oplus(0,1)]\odot[(1,0)\oplus(0,1)]=(1,1)^+\oplus(2,0)\oplus(0,2)\oplus(0,0)^+\oplus(0,0)^-\,,
\end{equation}
where the basis states of the irreducible representations $(1,1)^+$, $(2,0)\oplus(0,2)$ and $(0,0)^{\pm}$ are given by Eq.\,(\ref{eq:o4-11p1}), Eq.\,(\ref{eq:o4-22}) and Eq.\,(\ref{eq:o4-100}) with $(i\tau_j)_{L}(i\tau_k)_{R}=(i\tau_k)_{R}(i\tau_j)_{L}$.
Note that the basis states of the $(1,1)^-$ can not be expressed in terms of symmetric tensor products of $(i\tau_i)_{L,R}$. The symmetric tensor product of two identical 4-dimensional irreducible representations yields a 10-dimensional reducible representation which decomposes into
\begin{equation}\label{eq:o4-prod12pp}
(1/2,1/2)^{\pm}\odot (1/2,1/2)^{\pm}=(1,1)^+\oplus(0,0)^+\,.
\end{equation}
The basis states of $(0,0)^+$ and $(1,1)^+$ are defined by Eq.\,(\ref{eq:o4-00}), Eqs.\,(\ref{eq:o4-11p3})-(\ref{eq:o4-11p4}) and Eqs.\,(\ref{eq:o4-11p5})-(\ref{eq:o4-11p6}) with $AB=BA$ for $A,B\in\{(t_{\alpha}),(t_{\alpha}^{\dagger})\,|\,\alpha=0,1,2,3\}$. The decomposition of the symmetric tensor product of two different 4-dimensional irreducible representations reads
\begin{equation}
(1/2,1/2)^{\pm}\odot(1/2,1/2)^{\mp}=(1,1)^-\oplus(0,0)^-\,,
\end{equation}
where the basis states of the of $(0,0)^-$ and $(1,1)^-$ are defined by Eq.\,(\ref{eq:o4-00}), Eqs.\,(\ref{eq:o4-11m3})-(\ref{eq:o4-11m4}) and Eqs.\,(\ref{eq:o4-11m5})-(\ref{eq:o4-11m6}) with $A,B\in\{(t_{\alpha}),(t_{\alpha}^{\dagger})\,|\,\alpha=0,1,2,3\}$.

Based on the findings of this appendix, a list of all quark quadrilinears which transform as basis states of particular irreducible $O(4)$ representations can be compiled. The set of $P$- and/or $T$-violating quark quadrilinears, the irreducible $O(4)$ representation to which they belong (first column) and the number of $P$- and/or $T$-violating quark quadrilinears each representation contains (second column) is given by ($i,j,k,l=1,2,3$, summation over $i$ and $j$ implied, $k$ and $l$ are fixed):
\begin{alignat}{6}
&(0,0)^-\quad&&1\,&&\slashed{P}\quad&&\text{-state}\quad&&:\quad&&\bar{q}\gamma_{\mu}\tau_iq\bar{q}\gamma^{\mu}\gamma_5\tau_iq\,,\label{eq:o4-qq1}\\
&(1,1)^+&&3&&\slashed{P}\slashed{T}&&\text{-states}&&:&&\epsilon^{kij}\bar{q}\gamma_{\mu}\tau_iq\bar{q}\gamma^{\mu}\gamma_5\tau_jq\,,\\
&(2,0)\oplus(0,2)\quad&&5&&\slashed{P}&&\text{-states}&&:&&(\delta_{ik}\delta_{jl}\!+\!\delta_{jk}\delta_{il}\!-\!2\delta_{ij}\delta_{kl})\bar{q}\gamma_{\mu}\tau_iq\bar{q}\gamma^{\mu}\gamma_5\tau_jq\,,\\
&(0,0)^-&&1&&\slashed{P}\slashed{T}&&\text{-state}&&:&&\bar{q}q\bar{q}i\gamma_5q-\bar{q}\tau_iq\bar{q}i\gamma_5\tau_iq\,,\label{eq:o4-qq4}\\
&(1,1)^{+}&&3&&\slashed{P}\slashed{T}&&\text{-states}&&:&&\bar{q}q\bar{q}i\gamma_5\tau_kq\pm\bar{q}\tau_kq\bar{q}i\gamma_5q \,,\label{eq:o4-qq5}\\
&(1,1)^-&&6&&\slashed{P}\slashed{T}&&\text{-states}&&:&&\bar{q}\tau_kq\bar{q}i\gamma_5\tau_lq\pm\bar{q}\tau_lq\bar{q}i\gamma_5\tau_kq\quad k\neq l\,,\nonumber \\
&&&&&&&&&&&\bar{q}\tau_{k}q\bar{q}i\gamma_5\tau_{k}q+\bar{q}q\bar{q}i\gamma_5q\quad\text{with}\quad k=1,2,3\,.\label{eq:o4-qq7}
\end{alignat}
Note that the different relative signs in Eq.\,(\ref{eq:o4-qq5}) and Eq.\,(\ref{eq:o4-qq7}) define two different sets of tensors which transform as two equivalent realizations of the same respective basis states of the same representation.

The derivation of quark multilinears presented in this appendix immediately reveals Fierz identities among quark multilinears. A Fierz identity between two quark multilinears exists when they transform as the same basis state of the same irreducible representation and as identical tensors. This is, {\it e.g.}, the case for (summation over $i,j$ implied, $k$ is fixed)
\begin{equation}
 \bar{q}q\bar{q}i\gamma_5\tau_kq-\bar{q}\tau_kq\bar{q}i\gamma_5q\quad \text{and}\quad \epsilon^{kij}\bar{q}\gamma_{\mu}\tau_iq\bar{q}\gamma^{\mu}\gamma_5\tau_jq\,,
 \end{equation}
which transform as the same basis state of the $(1,1)^+$ representation and as identical tensors.

\setcounter{equation}{0}

\section{The Weinberg formulation of \boldmath{$SU(2)$} ChPT}\label{GassWein}
The aim of this appendix is to demonstrate the connection between the two equivalent formulations of $SU(2)$ ChPT, by Gasser and Leutwyler \cite{GassLeut1} and by Weinberg \cite{WeinbergII}, respectively. The equivalence of these two formulations is based on the observation made in Appendix \ref{app:o4rep}: a chiral  $SU(2)_L\!\times\! SU(2)_R$ transformation of the quark fields in a quark bilinear induces a vector or tensor $SO(4)$ rotation on the set of quark bilinears. The chiral $SU(2)_L\!\times\! SU(2)_R$ transformation of the quark fields and the induced $SO(4)$ transformation are just two sides of the same coin. The relationship between quark fields in a quark bilinears and the vector space of quark bilinears is resemblant of the relationship between vector spaces and their dual vector spaces. The same is true for quark multilinears in general.

In order establish the relationship between the two formulations of $SU(2)$ ChPT, the derivation of the transformation properties of the Goldstone boson fields \cite{GassLeut1,Leut,LeutProceedings,WeinbergII} has to be revisited. Let $v_0\in\mathcal{B}$ be the ground state in the field space $\mathcal{B}$ ({\it i.e.} the state without Goldstone bosons) and let $F: G\times\mathcal{B}\rightarrow\mathcal{B}$ be a group action of the group $G$ on this space, with the ground state $v_0$ being invariant under the subgroup $H\subset G$:
\begin{equation}
F_{g_1g_2}(v)=F_{g_1}(F_{g_2}(v))\quad \forall v\in\mathcal{B}\,,\quad v_0\mapsto F_h(v_0)=v_0\quad\forall h\in H\,.
\end{equation}
$F$ defines a one-to-one map from the set of left cosets $G/H$ onto $\mathcal{B}$:
\begin{equation}
F: G\times\mathcal{B}\rightarrow G/H\times \mathcal{B}\rightarrow\mathcal{B}\,,\quad F|_{G/H\times\{v_0\}}:G/H\times\{v_0\}\leftrightarrow \mathcal{B}\,.
\end{equation}
Since $H\subset G$ is a subgroup of G, any product $g_1\cdot g_2$, $g_1,g_2\in G$, equals $g'\cdot h$ for some $h\in H$ and $g'\in G$ and the group action thus obeys
\begin{equation}\label{eq:gbtrans1}
F_{g_1\cdot g_2}=F_{g'\circ h}\,,
\end{equation}
for all $g_1,g_2\in G$ and suitable $g'\in G$, $h\in H$. Let $\{\pi^{i}\}$ be the set of Goldstone boson fields, which are essentially the parameters of $G/H$ whose elements act on the ground state $v_0$. According to Eq.\,(\ref{eq:gbtrans1}), the Goldstone boson fields transform under $G$ by
\begin{equation}\label{eq:gbtrans2}
F_{\tilde{g}}\circ F_{g(\{\pi^i\})}(v_0)=F_{g'(\{\pi^i\},\tilde{g})}\circ F_{h(\{\pi^i\},\tilde{g})}(v_0)=F_{g'(\{\pi^i\},\tilde{g})}(v_0)\,,
\end{equation}
where $g'$ and $h$ depend on $\tilde{g}$ and $g(\{\pi^{i}\})$.

Eq.\,(\ref{eq:gbtrans2}) is the starting point for both formulations of $SU(2)$ ChPT. Both formulations are now discussed in the exponential parameterization \cite{WessZumino1,WessZumino2} first. The realization of the $SU(2)_L\!\times\!SU(2)_R$ group action on the ground state $v_0$ chosen by Gasser and Leutwyler \cite{GassLeut1} is defined by an axial rotation of the ground state $v_0=\mathds{1}$:
\begin{equation}
F_{g(\{\pi^i\})}(v_0)=u(\vec{\pi})\,\mathds{1}\,u(\vec{\pi})=U(\vec{\pi})\,,\quad u,U\in SU(2)\,,
\end{equation}
where $U=u^2$ is the standard $U$ matrix (the group action $F$ for a general $g\!=\!(L,R)$ is given by $\mathds{1}\!\mapsto\!R\mathds{1}L^{\dagger}$). The matrix $U=\exp(i\vec{\pi}\cdot\vec{\tau})$ is essentially an axial $SU(2)_A$ rotation of the ground state $v_0=\mathds{1}$ parametrized by the Goldstone boson fields $\{\pi^{i}\}$. For this definition of $F$ -- the Gasser-Leutwyler realization -- the transformation law  of Eq.\,(\ref{eq:gbtrans2}) reads
\begin{alignat}{3}\label{eq:glga}
F_{\tilde{g}}\circ F_{g(\{\pi^i\})}(v_0)&=Ru(\vec{\pi})\mathds{1}u(\vec{\pi})L^{\dagger}&&=u(\vec{\pi}\,')V\mathds{1}V^{\dagger}u(\vec{\pi}\,')&&=U(\vec{\pi}\,')\nonumber\\
&&&=F_{g'(\{\pi^{i}\},\tilde{g})}\circ F_{h(\{\pi^i\},\tilde{g})}(v_0)&&=F_{g'(\{\pi^i\},\tilde{g})}(v_0)\,,
\end{alignat}
for $V\in SU(2)_V=H$.

 The other realization of the group action employed by Weinberg utilizes the induced $SO(4)$ transformation: since the Goldstone bosons map bijectively onto the quotient group
$G/H=SU(2)_L\!\times\!SU(2)_R/SU(2)_V$ where $SU(2)_V$ is the diagonal vector subgroup $\{(V,V)|V\in SU(2)\}$, the elements $\{\mathds{1},\mathds{1}\}$ and $\{-\mathds{1},-\mathds{1}\}$ are in the same left coset. The spontaneous symmetry breaking pattern  $SU(2)_L\!\times\! SU(2)_R\rightarrow SU(2)_V$ is then equivalent to $SO(4)\rightarrow SO(3)$ because of
\begin{equation}
SU(2)/\{\mathds{1},-\mathds{1}\}=\mathds{R}P^3=SO(3)\,,
 \end{equation}
 where $\mathds{R}P^3$ is the real projective 3-space.
Let the group action $F$ on $G$ now be explicitly defined by the $SO(4)$ rotations $g\in SO(4)$ and $h\in SO(3)\subset SO(4)$ which act on the ground state denoted by the real four-vector $v_0$.
The Goldstone boson fields map bijectively onto $SO(4)/SO(3)$ and can be represented by an $SO(4)$ matrix $U_4$ (the Goldstone bosons parametrize the $S^3\sim SO(4)/SO(3)$ manifold). The generators of $SO(4)$ are readily obtained from the group action of $SU(2)_L\!\times\! SU(2)_R$ on $\mathds{H}_4$ defined in Eq.\,(\ref{eq:cayley}) and read
\begin{alignat}{2}
(T^k)_{i4}&=-(T^k)_{4i}=-i\delta_{ki},\quad&&(T^k)_{ij}=(T^k)_{44}=0\,,\\
(H^l)_{ij}&=-i\epsilon_{ijl},&&(H^l)_{i4}=(H^l)_{4i}=(H^l)_{44}=0\,,
\end{alignat}
where $H^k$ are the generators of $H\!\subset\! G$. A possible parameterization of the $SO(4)$ matrix $U_4$ corresponding to an axial transformation of the ground state ({\it i.e.} parameterizing $SO(4)/SO(3)$) is then given by:
\begin{equation}\label{eq:u4riem}
U_4=\exp\left[i
\begin{pmatrix}
0&0&0&-i\pi_1/F_{\pi}\\
0&0&0&-i\pi_2/F_{\pi}\\
0&0&0&-i\pi_3/F_{\pi}\\
i\pi_1/F_{\pi}&i\pi_2/F_{\pi}&i\pi_3/F_{\pi}&0
\end{pmatrix}\right]\,.
\end{equation}
The four-vector $v_0=(0,0,0,n)$, $n\in\mathds{R}$, represents a ground state which is invariant under $H$. The transformation law of Eq.\,(\ref{eq:gbtrans2}) for the Goldstone bosons then reads
\begin{alignat}{3}
F^{\ast}_{\tilde{g}}\circ F^{\ast}_{g(\{\pi^i\})}(v_0)&=\tilde{g}\,U_4(\vec{\pi})\,v_0&&=U_4(\vec{\pi}\,')\,h(\vec{\pi},\tilde{g})\,v_0&&=U_4(\vec{\pi}')v_0\nonumber\\
&&&=F^{\ast}_{g'(\{\pi^{i}\},\tilde{g})}\circ F^{\ast}_{h(\{\pi^i\},\tilde{g})}(v_0)&&=F^{\ast}_{g'(\{\pi\},\tilde{g})}(v_0)\,,
\end{alignat}
where $\tilde{g}\in G$ and $h\in H\subset G$ are $SO(4)$ matrices. The map $F$ in Eq.\,(\ref{eq:gbtrans2}) has been replaced by $F^{\ast}$ here since the Weinberg realization of the group action can be regarded as the group action induced by the Gasser-Leutwyler realization of $F$ in the same manner as a map of the quark fields in a quark bilinear induces a map on the vector space of quark bilinears. Note that with the choice of the group action $F$ defined in Eq.\,(\ref{eq:glga}), compositions $F_{g_2}\circ F_{g_1}$ and their induced compositions $F^{\ast}_{g_2}\circ F^{\ast}_{g_1}$ exhibit the same ordering for all $g_1,g_2\in G$.

While in the Gasser-Leutwyler formulation the leading kinetic term, for instance, of the effective Lagrangian is given by
\begin{equation}
c_{GL}\langle \partial_{\mu}U\partial^{\mu}U^{\dagger}\rangle=c_{GL}2(\partial_{\mu}\vec{\pi})(\partial^{\mu}\vec{\pi})+\cdots\,,
\end{equation}
the leading kinetic term of the effective Lagrangian in the Weinberg formulation would read (see section 19.5 of \cite{WeinbergII})
\begin{equation}
2c_W(\partial_{\mu}U_4v_0)^T(\partial_{\mu}U_4v_0)=2c_Wn^2(\partial_{\mu}\vec{\pi})(\partial^{\mu}\vec{\pi})+\cdots\,,
\end{equation}
for some real constants $c_{GL}$ and $c_W$. The connection between both realizations of the group action is established by the 2-1 homomorphism of $SU(2)_L\times SU(2)_R$ onto $SO(4)$ Eq.\,(\ref{eq:cayley}). 

There exists another more convenient parameterization for the Weinberg realization of the group action which is based on the standard stereographic projection of the 3-sphere $S^3$ (the north pole is taken to lie on the fourth axis and $\pi^2\!=\!\pi_1^2\!+\!\pi_2^2\!+\!\pi_3^2$):
\begin{equation}
\left(\frac{2\pi_1/F_0}{1+\pi^2/F_0},\frac{2\pi_2/F_0}{1+\pi^2/F_0},\frac{2\pi_3/F_0}{1+\pi^2/F_0},\frac{1-(\pi_1^2+\pi_2^2+\pi_3^2)/F_0}{1+\pi^2/F_0}\right)\,,\quad F_0=2\fpi\,.
\end{equation}
This parameterization can be expressed as an (axial) $SO(4)$ rotation given by the matrix
\begin{equation}\label{eq:u4stereo}
U_4=\begin{pmatrix} 1-\frac{2\pi_1^2/F_0^2}{1+\pi^2/F_0^2} & -\frac{2\pi_1\pi_2/F_0^2}{1+\pi^2/F_0^2} & -\frac{2\pi_1\pi_3/F_0^2}{1+\pi^2/F_0^2}
&\frac{2\pi_1/F_0}{1+\pi^2/F_0^2} \\
-\frac{2\pi_2\pi_1/F_0^2}{1+\pi^2/F_0^2} &1-\frac{2\pi_2^2/F_0^2}{1+\pi^2/F_0^2} & -\frac{2\pi_2\pi_3/F_0^2}{1+\pi^2/F_0^2}
&\frac{2\pi_2/F_0}{1+\pi^2/F_0^2} \\
-\frac{2\pi_3\pi_1/F_0^2}{1+\pi^2/F_0^2} & -\frac{2\pi_3\pi_2/F_0^2}{1+\pi^2/F_0^2} &1-\frac{2\pi_3^2/F_0^2}{1+\pi^2/F_0^2} &
\frac{2\pi_3/F_0}{1+\pi^2/F_0^2} \\
-\frac{2\pi_1/F_0}{1+\pi^2/F_0^2} & -\frac{2\pi_2/F_0}{1+\pi^2/F_0^2}& -\frac{2\pi_3/F_0}{1+\pi^2/F_0^2}& \frac{1-\pi^2/F_0^2}{1+\pi^2/F_0^2}
\end{pmatrix}\,,
\end{equation} 
which acts on the north pole $(0,0,0,n)$.
This is the preferred parameterization in the Weinberg formulation of $SU(2)$ ChPT \cite{WeinbergII}.

It has been demonstrated in Appendix \ref{app:o4rep} that quark multilinears can be regarded as basis states of irreducible representations of $O(4)$. In general, a chiral-symmetry breaking quark structure $\mathcal{S}_q$ ({\it i.e.} a quark multilinear) exhibits the same transformation properties under $G$ as a corresponding structure $\mathcal{S}_{\rm eff}$ in the effective Lagrangian ($e$ is the identity element):
\begin{eqnarray}\label{eq:csbglwein}
\mathcal{S}_q[\{q_i\}]&\mapsto&\mathcal{S}_q[F_{\tilde{g}}^q(\{q_i\})]=F_{\tilde{g}}^{\ast}(\mathcal{S}_q[\{q_i\}])\,,\nonumber\\
\mathcal{S}_{\rm eff}[F_{g(\{\pi^i\})},v_0]&\mapsto& \mathcal{S}_{\rm eff}[F_{\tilde{g}}\circ F_{g(\{\pi^i\})},v_0]=F^{\ast}_{\tilde g}\circ F^{\ast}_{g(\{\pi^i\})}(\mathcal{S}_{\rm eff}[F_e,v_0])\,,
\end{eqnarray}
where $F^q$ is the group action on the quark fields, $F$ is the corresponding group action in the effective theory (on the ground state or on the Goldstone boson fields) and $F^{\ast}$ is -- as before -- the induced map on the (vector-)space of chiral structures, which is identical for quark structures and chiral structures in the effective field theory. The group action $F$ is in general a group action on an element of the set $\{\mathds{1},\tau_1,\tau_2,\tau_3\}$. The equation in the second line of Eq.\,(\ref{eq:csbglwein}) reveals the two equivalent methods of constructing chiral-symmetry breaking structures in the effective field theory Lagrangian. In the Gasser-Leutwyler formulation, a chiral-symmetry breaking term in the effective Lagrangian can be constructed from the Goldstone boson matrix $U$ and its hermitian conjugate to transform under $SU(2)_L\!\times\! SU(2)_R$ (or equivalently $O(4)$) identically to the corresponding quark multilinear. As an example one may consider the isospin-violating component of the quark-mass matrix and its leading counterpart in the pion-sector effective Lagrangian, which read in the Gasser-Leutwyler formulation:
\begin{alignat}{3}
&\mathcal{S}[\{q_i\}]=\bar{q}_L\tau_3q_R+\bar{q}_R\tau_3q_L&&\mapsto\,&& \mathcal{S}[F_{\tilde{g}}^{q}(\{q_i\})]=\bar{q}_LL^{\dagger}\tau_3Rq_R+\bar{q}_RR^{\dagger}\tau_3Lq_L\,,\\
&\mathcal{S}_{\rm eff}[F_{g(\{\pi^i\})},v_0]=\langle U\tau_3+U^{\dagger}\tau_3\rangle&&\mapsto&& \mathcal{S}_{\rm eff}[F_{\tilde{g}}\!\circ\!F_{g(\{\pi^i\})},v_0]=\langle RUL^{\dagger}\tau_3\!+\!LU^{\dagger}R^{\dagger}\tau_3\rangle\,.
\end{alignat}

In the Weinberg formulation of $SU(2)$ ChPT, a term without Goldstone bosons is constructed which exhibits the same transformation properties under $O(4)$ (or equivalently $SU(2)_L\times SU(2)_R$) as its corresponding quark multilinear. The Goldstone bosons are then introduced by an axial $SO(4)$ rotation ({\it i.e.} expressible in terms of $U_4$ of Eq.\,(\ref{eq:u4riem}) or Eq.\,(\ref{eq:u4stereo})) which is paramterized by the Goldstone boson fields: 
\begin{equation}
\mathcal{S}_{\rm eff}[F_{e},v_0]\mapsto F^{\ast}_{g(\{\pi^{i}\})}(\mathcal{S}_{\rm eff}[F_e,v_0])\,.
\end{equation}

We demonstrate the Weinberg procedure of constructing chiral-symmetry breaking terms in the pion- and pion-nucleon-sector effective Lagrangian for the quark bilinears in the \mbox{4-dimensional} $(1/2,1/2)^+$ representation of $O(4)$. The fourth component of a vector in this representation transforms as the isospin conserving component of quark-mass term, {\it i.e.} it transforms as a scalar under $H\!\subset\! G$. The other components of a vector in this representation transform as a $P$- and $T$-violating three-vector under $H\!\subset\! G$. The application of $U_4$ of Eq.\,(\ref{eq:u4riem}) to the four-vectors in the pion sector and the pion-nucleon sector with the correct transformation properties yields 
\begin{eqnarray}
(U_4(0,0,0,1)^T)_4&=&(\vec{\pi}/\fpi+\cdots, (1-\vec{\pi}^2/(2\fpi^2))+\cdots)\,,\\
(U_4(0,0,0,N^{\dagger}N)^T)_4&=&  (\vec{\pi}/\fpi+\cdots, (1-\vec{\pi}^2/(2\fpi^2))+\cdots)N^{\dagger}N\,,
\end{eqnarray}
which is identical to the expressions obtained within the Gasser-Leutwyler formulation. A list of a few further structures in the Weinberg formulation \cite{Mereghetti:2010tp,WeinbergII} with their corresponding structures in the Gasser-Leutwyler formulation reads
\begin{alignat}{3}
 &N^{\dagger}N\quad&&\leftrightarrow\quad &&N^{\dagger}N\,,\\
&(U_4(0,0,0,N^{\dagger}N)^T)_4&&\leftrightarrow &&\langle\chi_+\rangle N^{\dagger}N\,,\\
&(U_4(N^{\dagger}\vec{\tau}N,0)^T)_3&&\leftrightarrow &&N^{\dagger}\hat{\chi}_+N=N^{\dagger}(\chi_+-\langle\chi_+\rangle)N\,,
\\
&eF_{\mu\nu}\left(((U_4)_{ik}(U_4)_{jl}-(U_4)_{jk}(U_4)_{il})N^{\dagger}T_{kl}N\right)_{34}
&&\leftrightarrow &&N^{\dagger}\hat{f}^+_{\mu\nu}N=N^{\dagger}(f^+_{\mu\nu}-\langle f^+_{\mu\nu}\rangle)N\,,
\end{alignat}
where the antisymmetric matrix $T$ is given by
\begin{equation}
T=
\begin{pmatrix}
0 &0 &0 &\tau_1\\
0 &0 &0 &\tau_2\\
0 &0 &0 &\tau_3\\
-\tau_1 &-\tau_2& -\tau_3& 0
\end{pmatrix}\,,
\end{equation}
and $\chi_+$ and $f_{\mu\nu}^+$ are defined by Eq.\,(\ref{eq:chi}) and Eq.\,(\ref{eq:fmunupm}), respectively.

Higher-order terms in the effective Lagrangian are tensor products of the lowest-dimensional representations of $O(4)$. They can be obtained in exactly the same manner by constructing tensors without Goldstone bosons and with the correct transformation properties under $O(4)$ transformations. Whereas in the Gasser-Leutwyler formulation such structures in the pion sector and pion-nucleon sector are easily obtained by the composition of fundamental building blocks, the construction of higher-order structures in the Weinberg formulation proves to be increasingly tedious. Furthermore, the simple extension of $SU(2)$ ChPT to $SU(3)$ ChPT in the Gasser-Leutwyler formulation is not possible for the Weinberg formulation of $SU(2)$ ChPT.


\bibliography{bibliography}

\begin{thebibliography}{100}
\expandafter\ifx\csname url\endcsname\relax
  \def\url#1{\texttt{#1}}\fi
\expandafter\ifx\csname urlprefix\endcsname\relax\def\urlprefix{URL }\fi
\expandafter\ifx\csname href\endcsname\relax
  \def\href#1#2{#2} \def\path#1{#1}\fi

\bibitem{Bigi:2000yz}
I.~I.~Y. Bigi, A.~I. Sanda, {CP violation}, Camb. Monogr. Part. Phys. Nucl.
  Phys. Cosmol. 9 (2000) 1--382.

\bibitem{sakharov}
A.~D. Sakharov, {Violation of CP Invariance, C Asymmetry, and Baryon Asymmetry
  of the Universe}, Pisma Zh. Eksp. Teor. Fiz. 5 (1967) 32--35.
\newblock \href {http://dx.doi.org/10.1070/PU1991v034n05ABEH002497}
  {\path{doi:10.1070/PU1991v034n05ABEH002497}}.

\bibitem{Bennett:2012zja}
C.~L. Bennett, et~al., {Nine-Year Wilkinson Microwave Anisotropy Probe (WMAP)
  Observations: Final Maps and Results}, Astrophys. J. Suppl. 208 (2013) 20.
\newblock \href {http://arxiv.org/abs/1212.5225} {\path{arXiv:1212.5225}},
  \href {http://dx.doi.org/10.1088/0067-0049/208/2/20}
  {\path{doi:10.1088/0067-0049/208/2/20}}.

\bibitem{Ade:2013zuv}
P.~A.~R. Ade, et~al., {Planck 2013 results. XVI. Cosmological parameters},
  Astron. Astrophys. 571 (2014) A16.
\newblock \href {http://arxiv.org/abs/1303.5076} {\path{arXiv:1303.5076}},
  \href {http://dx.doi.org/10.1051/0004-6361/201321591}
  {\path{doi:10.1051/0004-6361/201321591}}.

\bibitem{Pospelov_Ritz}
M.~Pospelov, A.~Ritz, {Electric dipole moments as probes of new physics},
  Annals Phys. 318 (2005) 119--169.
\newblock \href {http://arxiv.org/abs/hep-ph/0504231}
  {\path{arXiv:hep-ph/0504231}}, \href
  {http://dx.doi.org/10.1016/j.aop.2005.04.002}
  {\path{doi:10.1016/j.aop.2005.04.002}}.

\bibitem{Baker:2006ts}
C.~A. Baker, D.~D. Doyle, P.~Geltenbort, K.~Green, M.~G.~D. van~der Grinten,
  et~al., {An improved experimental limit on the electric dipole moment of the
  neutron}, Phys. Rev. Lett. 97 (2006) 131801.
\newblock \href {http://arxiv.org/abs/hep-ex/0602020}
  {\path{arXiv:hep-ex/0602020}}, \href
  {http://dx.doi.org/10.1103/PhysRevLett.97.131801}
  {\path{doi:10.1103/PhysRevLett.97.131801}}.

\bibitem{Griffith:2009zz}
W.~C. Griffith, M.~D. Swallows, T.~H. Loftus, M.~V. Romalis, B.~R. Heckel,
  et~al., {Improved Limit on the Permanent Electric Dipole Moment of Hg-199},
  Phys. Rev. Lett. 102 (2009) 101601.
\newblock \href {http://dx.doi.org/10.1103/PhysRevLett.102.101601}
  {\path{doi:10.1103/PhysRevLett.102.101601}}.

\bibitem{Baron:2013eja}
J.~Baron, et~al., {Order of Magnitude Smaller Limit on the Electric Dipole
  Moment of the Electron}, Science 343~(6168) (2014) 269--272.
\newblock \href {http://arxiv.org/abs/1310.7534} {\path{arXiv:1310.7534}},
  \href {http://dx.doi.org/10.1126/science.1248213}
  {\path{doi:10.1126/science.1248213}}.

\bibitem{dim4CKM1}
A.~Czarnecki, B.~Krause, {Neutron electric dipole moment in the standard model:
  Valence quark contributions}, Phys. Rev. Lett. 78 (1997) 4339--4342.
\newblock \href {http://arxiv.org/abs/hep-ph/9704355}
  {\path{arXiv:hep-ph/9704355}}, \href
  {http://dx.doi.org/10.1103/PhysRevLett.78.4339}
  {\path{doi:10.1103/PhysRevLett.78.4339}}.

\bibitem{Mannel:2012qk}
T.~Mannel, N.~Uraltsev, {Loop-Less Electric Dipole Moment of the Nucleon in the
  Standard Model}, Phys. Rev. D 85 (2012) 096002.
\newblock \href {http://arxiv.org/abs/1202.6270} {\path{arXiv:1202.6270}},
  \href {http://dx.doi.org/10.1103/PhysRevD.85.096002}
  {\path{doi:10.1103/PhysRevD.85.096002}}.

\bibitem{Mannel:2012hb}
T.~Mannel, N.~Uraltsev, {Charm CP Violation and the Electric Dipole Moments
  from the Charm Scale}, JHEP 1303 (2013) 064.
\newblock \href {http://arxiv.org/abs/1205.0233} {\path{arXiv:1205.0233}},
  \href {http://dx.doi.org/10.1007/JHEP03(2013)064}
  {\path{doi:10.1007/JHEP03(2013)064}}.

\bibitem{Semertzidis:2003iq}
Y.~K. Semertzidis, et~al., {A new method for a sensitive deuteron EDM
  experiment}, AIP Conf. Proc. 698 (2004) 200--204.
\newblock \href {http://arxiv.org/abs/hep-ex/0308063}
  {\path{arXiv:hep-ex/0308063}}, \href {http://dx.doi.org/10.1063/1.1664226}
  {\path{doi:10.1063/1.1664226}}.

\bibitem{Orlov:2006su}
Y.~F. Orlov, W.~M. Morse, Y.~K. Semertzidis, {Resonance method of
  electric-dipole-moment measurements in storage rings}, Phys. Rev. Lett. 96
  (2006) 214802.
\newblock \href {http://arxiv.org/abs/hep-ex/0605022}
  {\path{arXiv:hep-ex/0605022}}, \href
  {http://dx.doi.org/10.1103/PhysRevLett.96.214802}
  {\path{doi:10.1103/PhysRevLett.96.214802}}.

\bibitem{Semertzidis:2011qv}
Y.~K. Semertzidis, {A Storage Ring Proton Electric Dipole Moment Experiment:
  Most Sensitive Experiment to CP-Violation beyond the Standard Model} (2011).
\newblock \href {http://arxiv.org/abs/1110.3378} {\path{arXiv:1110.3378}}.

\bibitem{Semertzidis:2011zz}
Y.~K. Semertzidis, {Review of EDM experiments}, J. Phys. Conf. Ser. 335 (2011)
  012012.
\newblock \href {http://dx.doi.org/10.1088/1742-6596/335/1/012012}
  {\path{doi:10.1088/1742-6596/335/1/012012}}.

\bibitem{Lehrach}
A.~Lehrach, B.~Lorentz, W.~Morse, N.~Nikolaev, F.~Rathmann, {Precursor
  Experiments to Search for Permanent Electric Dipole Moments (EDMs) of Protons
  and Deuterons at COSY} (2012).
\newblock \href {http://arxiv.org/abs/1201.5773} {\path{arXiv:1201.5773}}.

\bibitem{Pretz:2013us}
J.~Pretz, {Measurement of Permanent Electric Dipole Moments of Charged Hadrons
  in Storage Rings}, Hyperfine Interact. 214~(1-3) (2013) 111--117.
\newblock \href {http://arxiv.org/abs/1301.2937} {\path{arXiv:1301.2937}},
  \href {http://dx.doi.org/10.1007/s10751-013-0799-4}
  {\path{doi:10.1007/s10751-013-0799-4}}.

\bibitem{Rathmann:2013rqa}
F.~Rathmann, A.~Saleev, N.~N. Nikolaev, {The search for electric dipole moments
  of light ions in storage rings}, J. Phys. Conf. Ser. 447 (2013) 012011.
\newblock \href {http://dx.doi.org/10.1088/1742-6596/447/1/012011}
  {\path{doi:10.1088/1742-6596/447/1/012011}}.

\bibitem{BNL_deuteron}
D.~Anastassopoulos, et~al., {Search for a permanent electric dipole moment of
  the deuteron nucleus at the $10^{-29}$ e cm level, AGS Proposal, April 2008,
  available from http://www.bnl.gov/edm/} (2008).

\bibitem{jbepja}
J.~Bsaisou, C.~Hanhart, S.~Liebig, U.-G. Mei{\ss}ner, A.~Nogga, A.~Wirzba, {The
  electric dipole moment of the deuteron from the QCD $\theta$-term}, Eur.
  Phys. J. A 49~(3) (2013) 31.
\newblock \href {http://arxiv.org/abs/1209.6306} {\path{arXiv:1209.6306}},
  \href {http://dx.doi.org/10.1140/epja/i2013-13031-x}
  {\path{doi:10.1140/epja/i2013-13031-x}}.

\bibitem{deVries:2011an}
J.~de~Vries, R.~Higa, C.-P. Liu, E.~Mereghetti, I.~Stetcu, et~al., {Electric
  Dipole Moments of Light Nuclei From Chiral Effective Field Theory}, Phys.
  Rev. C 84 (2011) 065501.
\newblock \href {http://arxiv.org/abs/1109.3604} {\path{arXiv:1109.3604}},
  \href {http://dx.doi.org/10.1103/PhysRevC.84.065501}
  {\path{doi:10.1103/PhysRevC.84.065501}}.

\bibitem{Engel:2013lsa}
J.~Engel, M.~J. Ramsey-Musolf, U.~van Kolck, {Electric Dipole Moments of
  Nucleons, Nuclei, and Atoms: The Standard Model and Beyond}, Prog. Part.
  Nucl. Phys. 71 (2013) 21--74.
\newblock \href {http://arxiv.org/abs/1303.2371} {\path{arXiv:1303.2371}},
  \href {http://dx.doi.org/10.1016/j.ppnp.2013.03.003}
  {\path{doi:10.1016/j.ppnp.2013.03.003}}.

\bibitem{Wirzba:2014mka}
A.~Wirzba, {Electric dipole moments of the nucleon and light nuclei}, Nucl.
  Phys. A 928 (2014) 116--127.
\newblock \href {http://arxiv.org/abs/1404.6131} {\path{arXiv:1404.6131}},
  \href {http://dx.doi.org/10.1016/j.nuclphysa.2014.04.003}
  {\path{doi:10.1016/j.nuclphysa.2014.04.003}}.

\bibitem{Yamanaka:2014nba}
N.~Yamanaka, T.~Sato, T.~Kubota, {Linear Programming Analysis of the $R$-Parity
  Violation within EDM-Constraints}, JHEP 1412 (2014) 110.
\newblock \href {http://arxiv.org/abs/1406.3713} {\path{arXiv:1406.3713}},
  \href {http://dx.doi.org/10.1007/JHEP12(2014)110}
  {\path{doi:10.1007/JHEP12(2014)110}}.

\bibitem{edmoln}
J.~Bsaisou, J.~de~Vries, C.~Hanhart, S.~Liebig, U.-G. Mei{\ss}ner, D.~Minossi,
  A.~Nogga, A.~Wirzba, {Nuclear Electric Dipole Moments in Chiral Effective
  Field Theory}, JHEP 1503 (2015) 104.
\newblock \href {http://arxiv.org/abs/1411.5804} {\path{arXiv:1411.5804}},
  \href {http://dx.doi.org/10.1007/JHEP03(2015)104}
  {\path{doi:10.1007/JHEP03(2015)104}}.

\bibitem{dissertation}
J.~Bsaisou, \href{http://hss.ulb.uni-bonn.de/2014/3585/3585.htm}{{Electric
  Dipole Moments of Light Nuclei}}, Dissertation, University of Bonn (April
  2014).
\newline\urlprefix\url{http://hss.ulb.uni-bonn.de/2014/3585/3585.htm}

\bibitem{'tHooft:1976up}
G.~'t~Hooft, {Symmetry Breaking Through Bell-Jackiw Anomalies}, Phys. Rev.
  Lett. 37 (1976) 8--11.
\newblock \href {http://dx.doi.org/10.1103/PhysRevLett.37.8}
  {\path{doi:10.1103/PhysRevLett.37.8}}.

\bibitem{Baluni}
V.~Baluni, {CP Violating Effects in QCD}, Phys. Rev. D 19 (1979) 2227--2230.
\newblock \href {http://dx.doi.org/10.1103/PhysRevD.19.2227}
  {\path{doi:10.1103/PhysRevD.19.2227}}.

\bibitem{Crewther:1979pi}
R.~J. Crewther, P.~Di~Vecchia, G.~Veneziano, E.~Witten, {Chiral Estimate of the
  Electric Dipole Moment of the Neutron in Quantum Chromodynamics}, Phys. Lett.
  88B (1979) 123--127, [{\em Erratum ibid.} {91B}, 487 (1980)].
\newblock \href {http://dx.doi.org/10.1016/0370-2693(79)90128-X}
  {\path{doi:10.1016/0370-2693(79)90128-X}}.

\bibitem{Mereghetti:2010tp}
E.~Mereghetti, W.~H. Hockings, U.~van Kolck, {The Effective Chiral Lagrangian
  From the Theta Term}, Annals Phys. 325 (2010) 2363--2409.
\newblock \href {http://arxiv.org/abs/1002.2391} {\path{arXiv:1002.2391}},
  \href {http://dx.doi.org/10.1016/j.aop.2010.03.005}
  {\path{doi:10.1016/j.aop.2010.03.005}}.

\bibitem{RamseyMusolf:2006vr}
M.~J. Ramsey-Musolf, S.~Su, {Low Energy Precision Test of Supersymmetry}, Phys.
  Rept. 456 (2008) 1--88.
\newblock \href {http://arxiv.org/abs/hep-ph/0612057}
  {\path{arXiv:hep-ph/0612057}}, \href
  {http://dx.doi.org/10.1016/j.physrep.2007.10.001}
  {\path{doi:10.1016/j.physrep.2007.10.001}}.

\bibitem{Weinberg:1989dx}
S.~Weinberg, {Larger Higgs Exchange Terms in the Neutron Electric Dipole
  Moment}, Phys. Rev. Lett. 63 (1989) 2333.
\newblock \href {http://dx.doi.org/10.1103/PhysRevLett.63.2333}
  {\path{doi:10.1103/PhysRevLett.63.2333}}.

\bibitem{Pati:1974yy}
J.~C. Pati, A.~Salam, {Lepton Number as the Fourth Color}, Phys. Rev. D 10
  (1974) 275--289, [{\em Erratum ibid.} D {11}, 703 (1975)].
\newblock \href {http://dx.doi.org/10.1103/PhysRevD.10.275}
  {\path{doi:10.1103/PhysRevD.10.275}}.

\bibitem{Mohapatra:1974hk}
R.~N. Mohapatra, J.~C. Pati, {Left-Right Gauge Symmetry and an Isoconjugate
  Model of CP Violation}, Phys. Rev. D 11 (1975) 566--571.
\newblock \href {http://dx.doi.org/10.1103/PhysRevD.11.566}
  {\path{doi:10.1103/PhysRevD.11.566}}.

\bibitem{Mohapatra:1974gc}
R.~N. Mohapatra, J.~C. Pati, {A Natural Left-Right Symmetry}, Phys. Rev. D 11
  (1975) 2558.
\newblock \href {http://dx.doi.org/10.1103/PhysRevD.11.2558}
  {\path{doi:10.1103/PhysRevD.11.2558}}.

\bibitem{Senjanovic:1975rk}
G.~Senjanovic, R.~N. Mohapatra, {Exact Left-Right Symmetry and Spontaneous
  Violation of Parity}, Phys. Rev. D 12 (1975) 1502.
\newblock \href {http://dx.doi.org/10.1103/PhysRevD.12.1502}
  {\path{doi:10.1103/PhysRevD.12.1502}}.

\bibitem{Minkowski:1977sc}
P.~Minkowski, {$\mu \to e \gamma$ at a Rate of One Out of 1-Billion Muon
  Decays?}, Phys. Lett. 67B (1977) 421.
\newblock \href {http://dx.doi.org/10.1016/0370-2693(77)90435-X}
  {\path{doi:10.1016/0370-2693(77)90435-X}}.

\bibitem{Senjanovic:1978ev}
G.~Senjanovic, {Spontaneous Breakdown of Parity in a Class of Gauge Theories},
  Nucl. Phys. B 153 (1979) 334--364.
\newblock \href {http://dx.doi.org/10.1016/0550-3213(79)90604-7}
  {\path{doi:10.1016/0550-3213(79)90604-7}}.

\bibitem{Mohapatra:1979ia}
R.~N. Mohapatra, G.~Senjanovic, {Neutrino Mass and Spontaneous Parity
  Violation}, Phys. Rev. Lett. 44 (1980) 912.
\newblock \href {http://dx.doi.org/10.1103/PhysRevLett.44.912}
  {\path{doi:10.1103/PhysRevLett.44.912}}.

\bibitem{Mohapatra:1980yp}
R.~N. Mohapatra, G.~Senjanovic, {Neutrino Masses and Mixings in Gauge Models
  with Spontaneous Parity Violation}, Phys. Rev. D 23 (1981) 165.
\newblock \href {http://dx.doi.org/10.1103/PhysRevD.23.165}
  {\path{doi:10.1103/PhysRevD.23.165}}.

\bibitem{Buchmuller:1985jz}
W.~Buchm{\"u}ller, D.~Wyler, {Effective Lagrangian Analysis of New Interactions
  and Flavor Conservation}, Nucl. Phys. B 268 (1986) 621.
\newblock \href {http://dx.doi.org/10.1016/0550-3213(86)90262-2}
  {\path{doi:10.1016/0550-3213(86)90262-2}}.

\bibitem{dim6Weinberg}
S.~Weinberg, {Baryon and Lepton Nonconserving Processes}, Phys. Rev. Lett. 43
  (1979) 1566--1570.
\newblock \href {http://dx.doi.org/10.1103/PhysRevLett.43.1566}
  {\path{doi:10.1103/PhysRevLett.43.1566}}.

\bibitem{DeRujula:1990db}
A.~De~Rujula, M.~B. Gavela, O.~Pene, F.~J. Vegas, {Signets of CP violation},
  Nucl. Phys. B 357 (1991) 311--356.
\newblock \href {http://dx.doi.org/10.1016/0550-3213(91)90472-A}
  {\path{doi:10.1016/0550-3213(91)90472-A}}.

\bibitem{Grzadkowski:2010es}
B.~Grzadkowski, M.~Iskrzynski, M.~Misiak, J.~Rosiek, {Dimension-Six Terms in
  the Standard Model Lagrangian}, JHEP 1010 (2010) 085.
\newblock \href {http://arxiv.org/abs/1008.4884} {\path{arXiv:1008.4884}},
  \href {http://dx.doi.org/10.1007/JHEP10(2010)085}
  {\path{doi:10.1007/JHEP10(2010)085}}.

\bibitem{dim6ngtulin}
J.~Ng, S.~Tulin, {D versus d: CP Violation in Beta Decay and Electric Dipole
  Moments}, Phys. Rev. D 85 (2012) 033001.
\newblock \href {http://arxiv.org/abs/1111.0649} {\path{arXiv:1111.0649}},
  \href {http://dx.doi.org/10.1103/PhysRevD.85.033001}
  {\path{doi:10.1103/PhysRevD.85.033001}}.

\bibitem{deVries:2012ab}
J.~de~Vries, E.~Mereghetti, R.~G.~E. Timmermans, U.~van Kolck, {The Effective
  Chiral Lagrangian From Dimension-Six Parity and Time-Reversal Violation},
  Annals Phys. 338 (2013) 50--96.
\newblock \href {http://arxiv.org/abs/1212.0990} {\path{arXiv:1212.0990}},
  \href {http://dx.doi.org/10.1016/j.aop.2013.05.022}
  {\path{doi:10.1016/j.aop.2013.05.022}}.

\bibitem{Dekens:2013zca}
W.~Dekens, J.~de~Vries, {Renormalization Group Running of Dimension-Six Sources
  of Parity and Time-Reversal Violation}, JHEP 1305 (2013) 149.
\newblock \href {http://arxiv.org/abs/1303.3156} {\path{arXiv:1303.3156}},
  \href {http://dx.doi.org/10.1007/JHEP05(2013)149}
  {\path{doi:10.1007/JHEP05(2013)149}}.

\bibitem{newmodelspaper}
W.~Dekens, J.~de~Vries, J.~Bsaisou, W.~Bernreuther, C.~Hanhart, U.-G.
  Mei{\ss}ner, A.~Nogga, A.~Wirzba, {Unraveling models of CP violation through
  electric dipole moments of light nuclei}, JHEP 1407 (2014) 069.
\newblock \href {http://arxiv.org/abs/1404.6082} {\path{arXiv:1404.6082}},
  \href {http://dx.doi.org/10.1007/JHEP07(2014)069}
  {\path{doi:10.1007/JHEP07(2014)069}}.

\bibitem{GassLeut1}
J.~Gasser, H.~Leutwyler, {Chiral Perturbation Theory to One Loop}, Annals Phys.
  158 (1984) 142.
\newblock \href {http://dx.doi.org/10.1016/0003-4916(84)90242-2}
  {\path{doi:10.1016/0003-4916(84)90242-2}}.

\bibitem{GassLeut2}
J.~Gasser, H.~Leutwyler, {Chiral Perturbation Theory: Expansions in the Mass of
  the Strange Quark}, Nucl. Phys. B 250 (1985) 465.
\newblock \href {http://dx.doi.org/10.1016/0550-3213(85)90492-4}
  {\path{doi:10.1016/0550-3213(85)90492-4}}.

\bibitem{GasserSvarc}
J.~Gasser, M.~Sainio, A.~Svar\'c, {Nucleons with Chiral Loops}, Nucl. Phys. B
  307 (1988) 779.
\newblock \href {http://dx.doi.org/10.1016/0550-3213(88)90108-3}
  {\path{doi:10.1016/0550-3213(88)90108-3}}.

\bibitem{MeissnerFettes3}
N.~Fettes, U.-G. Mei{\ss}ner, M.~Moj{\v z}i{\v s}, S.~Steininger, {The Chiral
  effective pion nucleon Lagrangian of order $p^4$}, Annals Phys. 283 (2000)
  273--302.
\newblock \href {http://arxiv.org/abs/hep-ph/0001308}
  {\path{arXiv:hep-ph/0001308}}, \href
  {http://dx.doi.org/10.1006/aphy.2000.6059}
  {\path{doi:10.1006/aphy.2000.6059}}.

\bibitem{deVries:2010ah}
J.~de~Vries, R.~G.~E. Timmermans, E.~Mereghetti, U.~van Kolck, {The Nucleon
  Electric Dipole Form Factor From Dimension-Six Time-Reversal Violation},
  Phys. Lett. B 695 (2011) 268--274.
\newblock \href {http://arxiv.org/abs/1006.2304} {\path{arXiv:1006.2304}},
  \href {http://dx.doi.org/10.1016/j.physletb.2010.11.042}
  {\path{doi:10.1016/j.physletb.2010.11.042}}.

\bibitem{Bhattacharya:2014cla}
T.~Bhattacharya, V.~Cirigliano, R.~Gupta, {Neutron Electric Dipole Moments from
  Beyond the Standard Model Physics}, PoS LATTICE\,2013 (2014) 299.
\newblock \href {http://arxiv.org/abs/1403.2445} {\path{arXiv:1403.2445}}.

\bibitem{Shindler:2014oha}
A.~Shindler, J.~de~Vries, T.~Luu, {Beyond-the-Standard-Model matrix elements
  with the gradient flow}, PoS LATTICE\,2014 (2014) 251.
\newblock \href {http://arxiv.org/abs/1409.2735} {\path{arXiv:1409.2735}}.

\bibitem{Manohar:1983md}
A.~Manohar, H.~Georgi, {Chiral Quarks and the Nonrelativistic Quark Model},
  Nucl. Phys. B 234 (1984) 189.
\newblock \href {http://dx.doi.org/10.1016/0550-3213(84)90231-1}
  {\path{doi:10.1016/0550-3213(84)90231-1}}.

\bibitem{Pitschmann:2014jxa}
M.~Pitschmann, C.-Y. Seng, C.~D. Roberts, S.~M. Schmidt, {Nucleon tensor
  charges and electric dipole moments}, Phys. Rev. D 91~(7) (2015) 074004.
\newblock \href {http://arxiv.org/abs/1411.2052} {\path{arXiv:1411.2052}},
  \href {http://dx.doi.org/10.1103/PhysRevD.91.074004}
  {\path{doi:10.1103/PhysRevD.91.074004}}.

\bibitem{WeinbergII}
S.~Weinberg, {The Quantum Theory of Fields. Vol. 2: Modern Applications},
  {Cambridge University Press}, 1996.

\bibitem{BKM1995}
V.~Bernard, N.~Kaiser, U.-G. Mei{\ss}ner, {Chiral dynamics in nucleons and
  nuclei}, Int. J. Mod. Phys. E 4 (1995) 193--346.
\newblock \href {http://arxiv.org/abs/hep-ph/9501384}
  {\path{arXiv:hep-ph/9501384}}, \href
  {http://dx.doi.org/10.1142/S0218301395000092}
  {\path{doi:10.1142/S0218301395000092}}.

\bibitem{Scherer:2002tk}
S.~Scherer, {Introduction to chiral perturbation theory}, Adv. Nucl. Phys. 27
  (2003) 277.
\newblock \href {http://arxiv.org/abs/hep-ph/0210398}
  {\path{arXiv:hep-ph/0210398}}.

\bibitem{Bernard:2006gx}
V.~Bernard, U.-G. Mei{\ss}ner, {Chiral perturbation theory}, Ann. Rev. Nucl.
  Part. Sci. 57 (2007) 33--60.
\newblock \href {http://arxiv.org/abs/hep-ph/0611231}
  {\path{arXiv:hep-ph/0611231}}, \href
  {http://dx.doi.org/10.1146/annurev.nucl.56.080805.140449}
  {\path{doi:10.1146/annurev.nucl.56.080805.140449}}.

\bibitem{Kubis:2007iy}
B.~Kubis, {An Introduction to chiral perturbation theory} (2007).
\newblock \href {http://arxiv.org/abs/hep-ph/0703274}
  {\path{arXiv:hep-ph/0703274}}.

\bibitem{veroniquereview}
V.~Bernard, {Chiral Perturbation Theory and Baryon Properties}, Prog. Part.
  Nucl. Phys. 60 (2008) 82--160.
\newblock \href {http://arxiv.org/abs/0706.0312} {\path{arXiv:0706.0312}},
  \href {http://dx.doi.org/10.1016/j.ppnp.2007.07.001}
  {\path{doi:10.1016/j.ppnp.2007.07.001}}.

\bibitem{Leut}
H.~Leutwyler, {On the foundations of chiral perturbation theory}, Annals Phys.
  235 (1994) 165--203.
\newblock \href {http://arxiv.org/abs/hep-ph/9311274}
  {\path{arXiv:hep-ph/9311274}}, \href
  {http://dx.doi.org/10.1006/aphy.1994.1094}
  {\path{doi:10.1006/aphy.1994.1094}}.

\bibitem{LeutProceedings}
H.~Leutwyler, {Chiral effective Lagrangians}, Lect. Notes Phys. 396 (1991)
  97--138.
\newblock \href {http://dx.doi.org/10.1007/3-540-54978-1_8}
  {\path{doi:10.1007/3-540-54978-1_8}}.

\bibitem{PDG}
K.~A. Olive, et~al., {Review of Particle Physics}, Chin. Phys. C 38 (2014)
  090001.
\newblock \href {http://dx.doi.org/10.1088/1674-1137/38/9/090001}
  {\path{doi:10.1088/1674-1137/38/9/090001}}.

\bibitem{WeinbergChPTEarly}
S.~Weinberg, {Dynamical approach to current algebra}, Phys. Rev. Lett. 18
  (1967) 188--191.
\newblock \href {http://dx.doi.org/10.1103/PhysRevLett.18.188}
  {\path{doi:10.1103/PhysRevLett.18.188}}.

\bibitem{WeinbergChPTEarly2}
S.~Weinberg, {Precise relations between the spectra of vector and axial vector
  mesons}, Phys. Rev. Lett. 18 (1967) 507--509.
\newblock \href {http://dx.doi.org/10.1103/PhysRevLett.18.507}
  {\path{doi:10.1103/PhysRevLett.18.507}}.

\bibitem{WeinbergChPTEarly3}
S.~Weinberg, {Nonlinear realizations of chiral symmetry}, Phys. Rev. 166 (1968)
  1568--1577.
\newblock \href {http://dx.doi.org/10.1103/PhysRev.166.1568}
  {\path{doi:10.1103/PhysRev.166.1568}}.

\bibitem{LiPagelsChPTEarly}
L.-F. Li, H.~Pagels, {Perturbation theory about a Goldstone symmetry}, Phys.
  Rev. Lett. 26 (1971) 1204--1206.
\newblock \href {http://dx.doi.org/10.1103/PhysRevLett.26.1204}
  {\path{doi:10.1103/PhysRevLett.26.1204}}.

\bibitem{GellMannOakesRenner}
M.~Gell-Mann, R.~Oakes, B.~Renner, {Behavior of current divergences under SU(3)
  x SU(3)}, Phys. Rev. 175 (1968) 2195--2199.
\newblock \href {http://dx.doi.org/10.1103/PhysRev.175.2195}
  {\path{doi:10.1103/PhysRev.175.2195}}.

\bibitem{Georgi:1985kw}
H.~Georgi, {Weak Interactions and Modern Particle Theory}, {Benjamin/Cummings
  Pub. Co.}, 1984.

\bibitem{Jenkins:1990jv}
E.~E. Jenkins, A.~V. Manohar, {Baryon chiral perturbation theory using a heavy
  fermion Lagrangian}, Phys. Lett. B 255 (1991) 558--562.
\newblock \href {http://dx.doi.org/10.1016/0370-2693(91)90266-S}
  {\path{doi:10.1016/0370-2693(91)90266-S}}.

\bibitem{Bernard:1992qa}
V.~Bernard, N.~Kaiser, J.~Kambor, U.-G. Mei{\ss}ner, {Chiral structure of the
  nucleon}, Nucl. Phys. B 388 (1992) 315--345.
\newblock \href {http://dx.doi.org/10.1016/0550-3213(92)90615-I}
  {\path{doi:10.1016/0550-3213(92)90615-I}}.

\bibitem{Bernard:1996gq}
V.~Bernard, N.~Kaiser, U.-G. Mei{\ss}ner, {Aspects of chiral pion-nucleon
  physics}, Nucl. Phys. A 615 (1997) 483--500.
\newblock \href {http://arxiv.org/abs/hep-ph/9611253}
  {\path{arXiv:hep-ph/9611253}}, \href
  {http://dx.doi.org/10.1016/S0375-9474(97)00021-3}
  {\path{doi:10.1016/S0375-9474(97)00021-3}}.

\bibitem{MeissnerFettes1}
N.~Fettes, U.-G. Mei{\ss}ner, S.~Steininger, {Pion-nucleon scattering in chiral
  perturbation theory. 1. Isospin symmetric case}, Nucl. Phys. A 640 (1998)
  199--234.
\newblock \href {http://arxiv.org/abs/hep-ph/9803266}
  {\path{arXiv:hep-ph/9803266}}, \href
  {http://dx.doi.org/10.1016/S0375-9474(98)00452-7}
  {\path{doi:10.1016/S0375-9474(98)00452-7}}.

\bibitem{Walker-Loud:2014iea}
A.~Walker-Loud, {Nuclear Physics Review}, PoS LATTICE\,2013 (2014) 013.
\newblock \href {http://arxiv.org/abs/1401.8259} {\path{arXiv:1401.8259}}.

\bibitem{Borsanyi:2014jba}
S.~Borsanyi, S.~D{\"{u}}rr, Z.~Fodor, C.~Hoelbling, S.~D. Katz, et~al., {Ab
  initio calculation of the neutron-proton mass difference}, Science 347 (2015)
  1452--1455.
\newblock \href {http://arxiv.org/abs/1406.4088} {\path{arXiv:1406.4088}},
  \href {http://dx.doi.org/10.1126/science.1257050}
  {\path{doi:10.1126/science.1257050}}.

\bibitem{BaruHanhartc1-1}
V.~Baru, C.~Hanhart, M.~Hoferichter, B.~Kubis, A.~Nogga, et~al., {Precision
  calculation of threshold $\pi^-$ d scattering, $\pi N$ scattering lengths,
  and the GMO sum rule}, Nucl. Phys. A 872 (2011) 69--116.
\newblock \href {http://arxiv.org/abs/1107.5509} {\path{arXiv:1107.5509}},
  \href {http://dx.doi.org/10.1016/j.nuclphysa.2011.09.015}
  {\path{doi:10.1016/j.nuclphysa.2011.09.015}}.

\bibitem{c1-2}
T.~Becher, H.~Leutwyler, {Low energy analysis of $\pi N \to \pi N$}, JHEP 0106
  (2001) 017.
\newblock \href {http://arxiv.org/abs/hep-ph/0103263}
  {\path{arXiv:hep-ph/0103263}}, \href
  {http://dx.doi.org/10.1088/1126-6708/2001/06/017}
  {\path{doi:10.1088/1126-6708/2001/06/017}}.

\bibitem{c1-3}
J.~Gasser, M.~Ivanov, E.~Lipartia, M.~Moj{\v z}i{\v s}, A.~Rusetsky, {Ground
  state energy of pionic hydrogen to one loop}, Eur. Phys. J. C 26 (2002)
  13--34.
\newblock \href {http://arxiv.org/abs/hep-ph/0206068}
  {\path{arXiv:hep-ph/0206068}}, \href
  {http://dx.doi.org/10.1007/s10052-002-1013-z}
  {\path{doi:10.1007/s10052-002-1013-z}}.

\bibitem{c1-4}
P.~B{\"u}ttiker, U.-G. Mei{\ss}ner, {Pion nucleon scattering inside the
  Mandelstam triangle}, Nucl. Phys. A 668 (2000) 97--112.
\newblock \href {http://arxiv.org/abs/hep-ph/9908247}
  {\path{arXiv:hep-ph/9908247}}, \href
  {http://dx.doi.org/10.1016/S0375-9474(99)00813-1}
  {\path{doi:10.1016/S0375-9474(99)00813-1}}.

\bibitem{WeinbergPowerCounting}
S.~Weinberg, {Phenomenological Lagrangians}, Physica A 96 (1979) 327--340.
\newblock \href {http://dx.doi.org/10.1016/0378-4371(79)90223-1}
  {\path{doi:10.1016/0378-4371(79)90223-1}}.

\bibitem{weinbergpiNN}
S.~Weinberg, {Three body interactions among nucleons and pions}, Phys. Lett. B
  295 (1992) 114--121.
\newblock \href {http://arxiv.org/abs/hep-ph/9209257}
  {\path{arXiv:hep-ph/9209257}}, \href
  {http://dx.doi.org/10.1016/0370-2693(92)90099-P}
  {\path{doi:10.1016/0370-2693(92)90099-P}}.

\bibitem{Hanhart:2007mu}
C.~Hanhart, A.~Wirzba, {Remarks on $NN \to NN\pi$ beyond leading order}, Phys.
  Lett. B 650 (2007) 354--361.
\newblock \href {http://arxiv.org/abs/nucl-th/0703012}
  {\path{arXiv:nucl-th/0703012}}, \href
  {http://dx.doi.org/10.1016/j.physletb.2007.05.038}
  {\path{doi:10.1016/j.physletb.2007.05.038}}.

\bibitem{Mereghetti:2013bta}
C.-P. Liu, J.~de~Vries, E.~Mereghetti, R.~G.~E. Timmermans, U.~van Kolck,
  {Deuteron Magnetic Quadrupole Moment From Chiral Effective Field Theory},
  Phys. Lett. B 713 (2012) 447--452.
\newblock \href {http://arxiv.org/abs/1203.1157} {\path{arXiv:1203.1157}},
  \href {http://dx.doi.org/10.1016/j.physletb.2012.06.024}
  {\path{doi:10.1016/j.physletb.2012.06.024}}.

\bibitem{Maekawa:2011vs}
C.~M. Maekawa, E.~Mereghetti, J.~de~Vries, U.~van Kolck, {The Time-Reversal-
  and Parity-Violating Nuclear Potential in Chiral Effective Theory}, Nucl.
  Phys. A 872 (2011) 117--160.
\newblock \href {http://arxiv.org/abs/1106.6119} {\path{arXiv:1106.6119}},
  \href {http://dx.doi.org/10.1016/j.nuclphysa.2011.09.020}
  {\path{doi:10.1016/j.nuclphysa.2011.09.020}}.

\bibitem{Dashen:1970et}
R.~F. Dashen, {Some features of chiral symmetry breaking}, Phys. Rev. D 3
  (1971) 1879--1889.
\newblock \href {http://dx.doi.org/10.1103/PhysRevD.3.1879}
  {\path{doi:10.1103/PhysRevD.3.1879}}.

\bibitem{Borasoy:2000pq}
B.~Borasoy, {The Electric dipole moment of the neutron in chiral perturbation
  theory}, Phys. Rev. D 61 (2000) 114017.
\newblock \href {http://arxiv.org/abs/hep-ph/0004011}
  {\path{arXiv:hep-ph/0004011}}, \href
  {http://dx.doi.org/10.1103/PhysRevD.61.114017}
  {\path{doi:10.1103/PhysRevD.61.114017}}.

\bibitem{GassLeut_eta}
J.~Gasser, H.~Leutwyler, {$\eta \to 3\pi$ to One Loop}, Nucl. Phys. B 250
  (1985) 539.
\newblock \href {http://dx.doi.org/10.1016/0550-3213(85)90494-8}
  {\path{doi:10.1016/0550-3213(85)90494-8}}.

\bibitem{Aoki:2013ldr}
S.~Aoki, Y.~Aoki, C.~Bernard, T.~Blum, G.~Colangelo, et~al., {Review of lattice
  results concerning low-energy particle physics}, Eur. Phys. J. C 74~(9)
  (2014) 2890.
\newblock \href {http://arxiv.org/abs/1310.8555} {\path{arXiv:1310.8555}},
  \href {http://dx.doi.org/10.1140/epjc/s10052-014-2890-7}
  {\path{doi:10.1140/epjc/s10052-014-2890-7}}.

\bibitem{Lebedev_Olive}
O.~Lebedev, K.~A. Olive, M.~Pospelov, A.~Ritz, {Probing CP violation with the
  deuteron electric dipole moment}, Phys. Rev. D 70 (2004) 016003.
\newblock \href {http://arxiv.org/abs/hep-ph/0402023}
  {\path{arXiv:hep-ph/0402023}}, \href
  {http://dx.doi.org/10.1103/PhysRevD.70.016003}
  {\path{doi:10.1103/PhysRevD.70.016003}}.

\bibitem{CSBpn2dpi}
A.~Filin, V.~Baru, E.~Epelbaum, J.~Haidenbauer, C.~Hanhart, A.~Kudryavstev,
  U.-G. Mei{\ss}ner, {Extraction of the strong neutron-proton mass difference
  from the charge symmetry breaking in $pn \to d \pi^0$}, Phys. Lett. B 681
  (2009) 423--427.
\newblock \href {http://arxiv.org/abs/0907.4671} {\path{arXiv:0907.4671}},
  \href {http://dx.doi.org/10.1016/j.physletb.2009.10.069}
  {\path{doi:10.1016/j.physletb.2009.10.069}}.

\bibitem{Ottnad:2009jw}
K.~Ottnad, B.~Kubis, U.-G. Mei{\ss}ner, F.-K. Guo, {New insights into the
  neutron electric dipole moment}, Phys. Lett. B 687 (2010) 42--47.
\newblock \href {http://arxiv.org/abs/0911.3981} {\path{arXiv:0911.3981}},
  \href {http://dx.doi.org/10.1016/j.physletb.2010.03.005}
  {\path{doi:10.1016/j.physletb.2010.03.005}}.

\bibitem{Mereghetti:2010kp}
E.~Mereghetti, J.~de~Vries, W.~H. Hockings, C.~M. Maekawa, U.~van Kolck, {The
  Electric Dipole Form Factor of the Nucleon in Chiral Perturbation Theory to
  Sub-leading Order}, Phys. Lett. B 696 (2011) 97--102.
\newblock \href {http://arxiv.org/abs/1010.4078} {\path{arXiv:1010.4078}},
  \href {http://dx.doi.org/10.1016/j.physletb.2010.12.018}
  {\path{doi:10.1016/j.physletb.2010.12.018}}.

\bibitem{Guo:2012vf}
F.-K. Guo, U.-G. Mei{\ss}ner, {Baryon electric dipole moments from strong CP
  violation}, JHEP 1212 (2012) 097.
\newblock \href {http://arxiv.org/abs/1210.5887} {\path{arXiv:1210.5887}},
  \href {http://dx.doi.org/10.1007/JHEP12(2012)097}
  {\path{doi:10.1007/JHEP12(2012)097}}.

\bibitem{Seng:2014pba}
C.-Y. Seng, J.~de~Vries, E.~Mereghetti, H.~H. Patel, M.~Ramsey-Musolf, {Nucleon
  electric dipole moments and the isovector parity- and time-reversal-odd
  pionﾐnucleon coupling}, Phys. Lett. B 736 (2014) 147--153.
\newblock \href {http://arxiv.org/abs/1401.5366} {\path{arXiv:1401.5366}},
  \href {http://dx.doi.org/10.1016/j.physletb.2014.07.014}
  {\path{doi:10.1016/j.physletb.2014.07.014}}.

\bibitem{Akan:2014yha}
T.~Akan, F.-K. Guo, U.-G. Mei{\ss}ner, {Finite-volume corrections to the CP-odd
  nucleon matrix elements of the electromagnetic current from the QCD vacuum
  angle}, Phys. Lett. B 736 (2014) 163.
\newblock \href {http://arxiv.org/abs/1406.2882} {\path{arXiv:1406.2882}},
  \href {http://dx.doi.org/10.1016/j.physletb.2014.07.022}
  {\path{doi:10.1016/j.physletb.2014.07.022}}.

\bibitem{Khriplovich:1999qr}
I.~B. Khriplovich, R.~A. Korkin, {P and T odd electromagnetic moments of
  deuteron in chiral limit}, Nucl. Phys. A 665 (2000) 365--373.
\newblock \href {http://arxiv.org/abs/nucl-th/9904081}
  {\path{arXiv:nucl-th/9904081}}, \href
  {http://dx.doi.org/10.1016/S0375-9474(99)00403-0}
  {\path{doi:10.1016/S0375-9474(99)00403-0}}.

\bibitem{Liebig:2010ki}
S.~Liebig, V.~Baru, F.~Ballout, C.~Hanhart, A.~Nogga, {Towards a high precision
  calculation for the pion-nucleus scattering lengths}, Eur. Phys. J. A 47
  (2011) 69.
\newblock \href {http://arxiv.org/abs/1003.3826} {\path{arXiv:1003.3826}},
  \href {http://dx.doi.org/10.1140/epja/i2011-11069-4}
  {\path{doi:10.1140/epja/i2011-11069-4}}.

\bibitem{Gloeckle:1995jg}
W.~Gl{\"o}ckle, H.~Witala, D.~Huber, H.~Kamada, J.~Golak, {The three nucleon
  continuum: achievements, challenges and applications}, Phys. Rept. 274 (1996)
  107--285.
\newblock \href {http://dx.doi.org/10.1016/0370-1573(95)00085-2}
  {\path{doi:10.1016/0370-1573(95)00085-2}}.

\bibitem{Liu_Timmermans}
C.-P. Liu, R.~G.~E. Timmermans, {P- and T-odd two-nucleon interaction and the
  deuteron electric dipole moment}, Phys. Rev. C 70 (2004) 055501.
\newblock \href {http://arxiv.org/abs/nucl-th/0408060}
  {\path{arXiv:nucl-th/0408060}}, \href
  {http://dx.doi.org/10.1103/PhysRevC.70.055501}
  {\path{doi:10.1103/PhysRevC.70.055501}}.

\bibitem{stetcu}
I.~Stetcu, C.-P. Liu, J.~L. Friar, A.~C. Hayes, P.~Navratil, {Nuclear Electric
  Dipole Moment of He-3}, Phys. Lett. B 665 (2008) 168--172.
\newblock \href {http://arxiv.org/abs/0804.3815} {\path{arXiv:0804.3815}},
  \href {http://dx.doi.org/10.1016/j.physletb.2008.06.019}
  {\path{doi:10.1016/j.physletb.2008.06.019}}.

\bibitem{Afnan:2010xd}
I.~R. Afnan, B.~F. Gibson, {Model Dependence of the 2H Electric Dipole Moment},
  Phys. Rev. C 82 (2010) 064002.
\newblock \href {http://arxiv.org/abs/1011.4968} {\path{arXiv:1011.4968}},
  \href {http://dx.doi.org/10.1103/PhysRevC.82.064002}
  {\path{doi:10.1103/PhysRevC.82.064002}}.

\bibitem{deVries:2011re}
J.~de~Vries, E.~Mereghetti, R.~G.~E. Timmermans, U.~van Kolck, {Parity- and
  Time-Reversal-Violating Form Factors of the Deuteron}, Phys. Rev. Lett. 107
  (2011) 091804.
\newblock \href {http://arxiv.org/abs/1102.4068} {\path{arXiv:1102.4068}},
  \href {http://dx.doi.org/10.1103/PhysRevLett.107.091804}
  {\path{doi:10.1103/PhysRevLett.107.091804}}.

\bibitem{song}
Y.-H. Song, R.~Lazauskas, V.~Gudkov, {Nuclear electric dipole moment of
  three-body system}, Phys. Rev. C 87 (2013) 015501.
\newblock \href {http://arxiv.org/abs/1211.3762} {\path{arXiv:1211.3762}},
  \href {http://dx.doi.org/10.1103/PhysRevC.87.015501}
  {\path{doi:10.1103/PhysRevC.87.015501}}.

\bibitem{chiralpotentials2}
E.~Epelbaum, W.~Gl{\"o}ckle, U.-G. Mei{\ss}ner, {The two-nucleon system at
  next-to-next-to-next-to-leading order}, Nucl. Phys. A 747 (2005) 362--424.
\newblock \href {http://arxiv.org/abs/nucl-th/0405048}
  {\path{arXiv:nucl-th/0405048}}, \href
  {http://dx.doi.org/10.1016/j.nuclphysa.2004.09.107}
  {\path{doi:10.1016/j.nuclphysa.2004.09.107}}.

\bibitem{chiralpotentials}
E.~Epelbaum, {Four-nucleon force in chiral effective field theory}, Phys. Lett.
  B 639 (2006) 456--461.
\newblock \href {http://arxiv.org/abs/nucl-th/0511025}
  {\path{arXiv:nucl-th/0511025}}, \href
  {http://dx.doi.org/10.1016/j.physletb.2006.06.046}
  {\path{doi:10.1016/j.physletb.2006.06.046}}.

\bibitem{Epelbaum:2008ga}
E.~Epelbaum, H.-W. Hammer, U.-G. Mei{\ss}ner, {Modern Theory of Nuclear
  Forces}, Rev. Mod. Phys. 81 (2009) 1773--1825.
\newblock \href {http://arxiv.org/abs/0811.1338} {\path{arXiv:0811.1338}},
  \href {http://dx.doi.org/10.1103/RevModPhys.81.1773}
  {\path{doi:10.1103/RevModPhys.81.1773}}.

\bibitem{Weinberg:1991um}
S.~Weinberg, {Effective chiral Lagrangians for nucleon--pion interactions and
  nuclear forces}, Nucl. Phys. B 363 (1991) 3--18.
\newblock \href {http://dx.doi.org/10.1016/0550-3213(91)90231-L}
  {\path{doi:10.1016/0550-3213(91)90231-L}}.

\bibitem{Nogga:2005hy}
A.~Nogga, R.~G.~E. Timmermans, U.~van Kolck, {Renormalization of one-pion
  exchange and power counting}, Phys. Rev. C 72 (2005) 054006.
\newblock \href {http://arxiv.org/abs/nucl-th/0506005}
  {\path{arXiv:nucl-th/0506005}}, \href
  {http://dx.doi.org/10.1103/PhysRevC.72.054006}
  {\path{doi:10.1103/PhysRevC.72.054006}}.

\bibitem{PavonValderrama:2005wv}
M.~Pav\'{o}n~Valderrama, E.~Ruiz~Arriola, {Renormalization of NN interaction
  with chiral two pion exchange potential. central phases and the deuteron},
  Phys. Rev. C 74 (2006) 054001.
\newblock \href {http://arxiv.org/abs/nucl-th/0506047}
  {\path{arXiv:nucl-th/0506047}}, \href
  {http://dx.doi.org/10.1103/PhysRevC.74.054001}
  {\path{doi:10.1103/PhysRevC.74.054001}}.

\bibitem{Epelbaum:2006pt}
E.~Epelbaum, U.-G. Mei{\ss}ner, {On the Renormalization of the One-Pion
  Exchange Potential and the Consistency of Weinberg's Power Counting}, Few
  Body Syst. 54 (2013) 2175--2190.
\newblock \href {http://arxiv.org/abs/nucl-th/0609037}
  {\path{arXiv:nucl-th/0609037}}, \href
  {http://dx.doi.org/10.1007/s00601-012-0492-1}
  {\path{doi:10.1007/s00601-012-0492-1}}.

\bibitem{Birse:2005um}
M.~C. Birse, {Power counting with one-pion exchange}, Phys. Rev. C 74 (2006)
  014003.
\newblock \href {http://arxiv.org/abs/nucl-th/0507077}
  {\path{arXiv:nucl-th/0507077}}, \href
  {http://dx.doi.org/10.1103/PhysRevC.74.014003}
  {\path{doi:10.1103/PhysRevC.74.014003}}.

\bibitem{Valderrama:2011hz}
M.~Pav\'{o}n~Valderrama, {Perturbative Renormalizability of Chiral Two Pion
  Exchange in Nucleon-Nucleon Scattering}, Phys. Rev. C 83 (2011) 024003.
\newblock \href {http://arxiv.org/abs/0912.0699} {\path{arXiv:0912.0699}},
  \href {http://dx.doi.org/10.1103/PhysRevC.83.024003}
  {\path{doi:10.1103/PhysRevC.83.024003}}.

\bibitem{Valderrama:2011hw}
M.~Pav\'{o}n~Valderrama, {Perturbative Renormalizability of Chiral Two Pion
  Exchange in Nucleon-Nucleon Scattering: P- and D-waves}, Phys. Rev. C 84
  (2011) 064002.
\newblock \href {http://arxiv.org/abs/1108.0872} {\path{arXiv:1108.0872}},
  \href {http://dx.doi.org/10.1103/PhysRevC.84.064002}
  {\path{doi:10.1103/PhysRevC.84.064002}}.

\bibitem{Epelbaum:2014efa}
E.~Epelbaum, H.~Krebs, U.-G. Mei{\ss}ner, {Improved chiral nucleon-nucleon
  potential up to next-to-next-to-next-to-leading order} (2014).
\newblock \href {http://arxiv.org/abs/1412.0142} {\path{arXiv:1412.0142}}.

\bibitem{Furnstahl:2014xsa}
R.~J. Furnstahl, D.~R. Phillips, S.~Wesolowski, {A recipe for EFT uncertainty
  quantification in nuclear physics}, J. Phys. G 42~(3) (2015) 034028.
\newblock \href {http://arxiv.org/abs/1407.0657} {\path{arXiv:1407.0657}},
  \href {http://dx.doi.org/10.1088/0954-3899/42/3/034028}
  {\path{doi:10.1088/0954-3899/42/3/034028}}.

\bibitem{Valderrama:2014vra}
M.~Pav\'{o}n~Valderrama, D.~R. Phillips, {Power Counting of Contact-Range
  Currents in Effective Field Theory}, Phys.Rev.Lett. 114~(8) (2015) 082502.
\newblock \href {http://arxiv.org/abs/1407.0437} {\path{arXiv:1407.0437}},
  \href {http://dx.doi.org/10.1103/PhysRevLett.114.082502}
  {\path{doi:10.1103/PhysRevLett.114.082502}}.

\bibitem{Yamanaka:2015qfa}
N.~Yamanaka, E.~Hiyama, {Enhancement of the CP-odd effect in the nuclear
  electric dipole moment of $^6$Li} (2015).
\newblock \href {http://arxiv.org/abs/1503.04446} {\path{arXiv:1503.04446}}.

\bibitem{av18}
R.~B. Wiringa, V.~G.~J. Stoks, R.~Schiavilla, {An accurate nucleon-nucleon
  potential with charge independence breaking}, Phys. Rev. C 51 (1995) 38--51.
\newblock \href {http://arxiv.org/abs/nucl-th/9408016}
  {\path{arXiv:nucl-th/9408016}}, \href
  {http://dx.doi.org/10.1103/PhysRevC.51.38}
  {\path{doi:10.1103/PhysRevC.51.38}}.

\bibitem{Pudliner:1997ck}
B.~S. Pudliner, V.~R. Pandharipande, J.~Carlson, S.~C. Pieper, R.~B. Wiringa,
  {Quantum Monte Carlo calculations of nuclei with A $\leq$ 7}, Phys. Rev. C 56
  (1997) 1720--1750.
\newblock \href {http://arxiv.org/abs/nucl-th/9705009}
  {\path{arXiv:nucl-th/9705009}}, \href
  {http://dx.doi.org/10.1103/PhysRevC.56.1720}
  {\path{doi:10.1103/PhysRevC.56.1720}}.

\bibitem{cdbonn}
R.~Machleidt, {The high precision, charge dependent Bonn nucleon-nucleon
  potential (CD-Bonn)}, Phys. Rev. C 63 (2001) 024001.
\newblock \href {http://arxiv.org/abs/nucl-th/0006014}
  {\path{arXiv:nucl-th/0006014}}, \href
  {http://dx.doi.org/10.1103/PhysRevC.63.024001}
  {\path{doi:10.1103/PhysRevC.63.024001}}.

\bibitem{Coon:2001pv}
S.~A. Coon, H.~K. Han, {Reworking the Tucson-Melbourne three nucleon
  potential}, Few Body Syst. 30 (2001) 131--141.
\newblock \href {http://arxiv.org/abs/nucl-th/0101003}
  {\path{arXiv:nucl-th/0101003}}, \href
  {http://dx.doi.org/10.1007/s006010170022} {\path{doi:10.1007/s006010170022}}.

\bibitem{Pich}
A.~Pich, E.~de~Rafael, {Strong CP violation in an effective chiral Lagrangian
  approach}, Nucl. Phys. B 367 (1991) 313--333.
\newblock \href {http://dx.doi.org/10.1016/0550-3213(91)90019-T}
  {\path{doi:10.1016/0550-3213(91)90019-T}}.

\bibitem{Shintani:2008nt}
E.~Shintani, S.~Aoki, Y.~Kuramashi, {Full QCD calculation of neutron electric
  dipole moment with the external electric field method}, Phys. Rev. D 78
  (2008) 014503.
\newblock \href {http://arxiv.org/abs/0803.0797} {\path{arXiv:0803.0797}},
  \href {http://dx.doi.org/10.1103/PhysRevD.78.014503}
  {\path{doi:10.1103/PhysRevD.78.014503}}.

\bibitem{Shintani:2012zca}
E.~Shintani, T.~Blum, T.~Izubuchi, {Lattice caclulation of neutron and proton
  EDM in full QCD}, PoS ConfinementX (2012) 330.

\bibitem{Shintani:2014}
E.~Shintani,
  \href{{http://crunch.ikp.physik.tu-darmstadt.de/hirschegg/2014/}}{{Lattice
  calculation of nucleon EDM}}, { Talk given at ``Hadrons from Quarks and
  Gluons'', Hirschegg, Austria} ({January} {2014}).
\newline\urlprefix\url{{http://crunch.ikp.physik.tu-darmstadt.de/hirschegg/2014/}}

\bibitem{quaternion3}
T.~Y. Yam, {Hamilton's Quaternions, Handbook of Algebra Vol. 3, 429-454},
  {North-Holland}, 2003.

\bibitem{quaternion1}
J.~Stillwell, {Naive Lie Theory}, {Springer}, 2008.

\bibitem{quaternion2}
A.~F. Beardon, {Algebra and Geometry}, {Cambridge University Press}, 2005.

\bibitem{quaternion5}
M.~Lorente, P.~D. Kramer, {Tensor and spin representations of SO(4) and
  discrete quantum gravity} (2003).
\newblock \href {http://arxiv.org/abs/gr-qc/0402050}
  {\path{arXiv:gr-qc/0402050}}.

\bibitem{quaternion4}
H.~S.~M. Coxeter, {Quaternions and Reflections}, The American Mathematically
  Monthly 53 (1946) 136--146.

\bibitem{quaternion7}
W.~Fulton, J.~Harris, {Representation Theory}, {Springer}, 2004.

\bibitem{quaternion6}
M.~Hamermesh, {Group Theory and its Application to Physical Problems},
  {Addison-Wesley}, 1962.

\bibitem{Nieves:2003in}
J.~F. Nieves, P.~B. Pal, {Generalized Fierz identities}, Am. J. Phys. 72 (2004)
  1100--1108.
\newblock \href {http://arxiv.org/abs/hep-ph/0306087}
  {\path{arXiv:hep-ph/0306087}}, \href {http://dx.doi.org/10.1119/1.1757445}
  {\path{doi:10.1119/1.1757445}}.

\bibitem{WessZumino1}
S.~R. Coleman, J.~Wess, B.~Zumino, {Structure of phenomenological Lagrangians.
  1.}, Phys. Rev. 177 (1969) 2239--2247.
\newblock \href {http://dx.doi.org/10.1103/PhysRev.177.2239}
  {\path{doi:10.1103/PhysRev.177.2239}}.

\bibitem{WessZumino2}
C.~G. Callan~Jr., S.~R. Coleman, J.~Wess, B.~Zumino, {Structure of
  phenomenological Lagrangians. 2.}, Phys. Rev. 177 (1969) 2247--2250.
\newblock \href {http://dx.doi.org/10.1103/PhysRev.177.2247}
  {\path{doi:10.1103/PhysRev.177.2247}}.

\end{thebibliography}

\end{document}